\documentclass[useAMS,usenatbib,usegraphicx]{mn2e}
\newcommand{\der}{d}
\newcommand{\epsx}{\epsilon_\mathrm{X}}
\newcommand{\epsy}{\epsilon_\mathrm{Y}}
\newcommand{\epsz}{\epsilon_\mathrm{Z}}

\newcommand{\vlos}{v_\mathrm{los}}
\newcommand{\vperp}{v_\bot}

\newcommand{\betaphi}{\beta_\varphi}
\newcommand{\betatheta}{\beta_\vartheta}

\newcommand{\reff}{r_\mathrm{eff}}

\newcommand{\mlfit}{\Upsilon_\mathrm{fit}}
\newcommand{\mlin}{\Upsilon_\mathrm{in}}
\newcommand{\mfit}{M^\mathrm{(fit)}}
\newcommand{\mmin}{M^\mathrm{(in)}}
\newcommand{\shape}{\langle H_4 \rangle_0}

\newcommand{\sig}{\langle \sigma \rangle_0}
\newcommand{\sigr}{\sigma_r}
\newcommand{\sigt}{\sigma_\vartheta}
\newcommand{\sigp}{\sigma_\varphi}
\newcommand{\mc}{\chi^2_\mathrm{min}}
\newcommand{\ms}{S_\mathrm{max}}
\newcommand{\inp}{^\mathrm{(in)}}
\newcommand{\fit}{^\mathrm{(fit)}}
\newcommand{\mlrat}{\mlfit/\mlin}
\newcommand{\mrat}{\mfit/\mmin}
\newcommand{\mrateff}{\mfit/\mmin}
\newcommand{\mratcen}{\mfit/\mmin}
\newcommand{\trat}{T\fit/T\inp}
\title[Schwarzschild models of $N$-body merger remnants]{Axisymmetric orbit models of $N$-body merger remnants: a dependency of reconstructed mass on viewing angle}
\author[J. Thomas, R. Jesseit, T. Naab, R.~P. Saglia, A. Burkert  and R. Bender]{J. Thomas$^{1,2}$\thanks{E-mail:jthomas@mpe.mpg.de}, 
R. Jesseit$^{2}$\thanks{E-mail: jesseit@usm.uni-muenchen.de}, T. Naab$^{2}$, R.~P. Saglia$^{1}$, A. Burkert$^{2}$ and R. Bender$^{1,2}$ \\
$^{1}$Max-Planck-Institut f\"ur Extraterrestrische Physik, Giessenbachstrasse, D-85748, Garching, Germany\\
$^{2}$Universit\"atssternwarte, Scheinerstr. 1, 81679 M\"unchen, Germany}
\begin{document}

\date{Submitted to MNRAS ------; Accepted ------------}

\pagerange{\pageref{firstpage}--\pageref{lastpage}} \pubyear{2005}

\maketitle

\label{firstpage}
\begin{abstract}
We model mock observations of collisionless $N$-body disc-disc mergers with the same 
axisymmetric orbit superposition program that has been used to model elliptical galaxies
in Coma. The remnants sample representatively the shape distribution of disc-disc mergers, 
including the most extreme cases, like highly 
prolate, maximally triaxial and dominantly oblate objects. The aim of our study is to better
understand how the assumption of axial symmetry affects reconstructed masses and stellar
motions of systems which are intrinsically not axisymmetric, whether the axisymmetry assumption then
leads to a bias and how such a potential bias can be recognised in models of real galaxies. 
The mass recovery at the half-light radius 
depends on viewing-angle and intrinsic shape: edge-on views allow to
reconstruct total masses with an accuracy between 20 percent (triaxial/prolate remnants) and
3 percent (oblate remnant). Masses of highly flattened, face-on systems
are underestimated by up to 50 percent. Deviations in local
mass densities can be larger where remnants are strongly triaxial
or prolate. Luminous mass-to-light ratios are sensitive to box orbits 
in the remnants. Box orbits cause the 
central value of the Gauss-Hermite parameter $H_4$ to vary with viewing-angle.
Reconstructed luminous mass-to-light ratios, as well as reconstructed central masses,
follow this variation. Luminous mass-to-light ratios are always underestimated (up to a 
factor of $2.5$). Respective dark halos in the models can be overestimated 
by about the same amount, depending again on viewing angle. Reconstructed velocity 
anisotropies $\beta$ depend on viewing angle as well as on the orbital composition 
of the remnant and are mostly accurate to about $\Delta \beta = 0.2$. Larger
deviations can occur towards the centre or the outer regions, respectively. We construct
$N$-body realisations of the Schwarzschild models to discuss chaotic orbits 
and the virial equilibrium in our models. In this study we explore the extreme limits 
of axisymmetric models. Apparently flattened, rotating ellipticals of 
intermediate mass are likely close to both, axial symmetry and edge-on orientation.
Our results imply that Schwarzschild models allow a reconstruction of
their masses and stellar anisotropies with high accuracy.
\end{abstract}

\begin{keywords}
galaxies: kinematics and dynamics -- galaxies: elliptical and lenticular, cD -- galaxies: formation
\end{keywords}

\section{Introduction}
Subject of this paper is the reconstruction of {\it synthetic} dynamical systems --
$N$-body merger remnants -- with orbit models. The motivation behind is to 
better understand models of {\it real} dynamical systems, especially
those of elliptical galaxies.

Elliptical galaxies are optically smooth stellar systems
in approximate dynamical equilibrium. They can result from 
various kinds of merging processes -- 
for example the merging of two discs \citep[e.g.][]{Too72} -- or from
some kind of monolithic collapse \citep[e.g.][]{Egg62,Lar74}. In a 
cosmological context an early-type can go through several distinct phases of the
above prototypical forms \citep[e.g.][]{Nab07}.

Apart from following the cosmic evolution of potential progenitor systems the 
only way to determine elliptical galaxy evolutionary histories is to scan their
present structure for characteristic
fingerprints of different evolutionary events. This concerns
both, scaling relations of ellipticals as a class and internal properties
of individual systems. 
For example, the cold collapse of a stellar system results in a typical gradient
from central isotropy to strong outer radial anisotropy in stellar orbits
\citep{vanAl82}. Galaxy mergers,
on the other hand, can produce a variety of dynamical systems. The 
final structure of disc-disc merger remnants depends, for example, on progenitor 
properties \citep{Bar92,Her92,Her93}, the merging geometry \citep{Wei96,Dub98} 
and on the mass ratio of the progenitors \citep{Nab03,Jes05}. 
Ellipticals as progenitors can be merged as well \citep{Nab06c}.

The difficulty with real galaxies is that their intrinsic properties, like the 
intrinsic shape, the distribution of mass or the 
geometry of stellar orbits, are not directly observable. They
have to be inferred from observations through dynamical modeling.

The state-of-the-art method for such modeling is Schwarzschild's orbit superposition
technique \citep{S79}. 
In very rough terms (1) the photometry of a galaxy is deprojected into the
3d internal light distribution; (2) the light distribution is multiplied with the
stellar mass-to-light ratio and -- depending on the specific application -- a black hole
or dark halo is added to obtain the composite mass distribution; (3) thousands of orbits are
calculated in the resulting gravitational potential; (4) the orbits are added together to
fit the kinematic and photometric observations of the galaxy. Thereby each orbit is 
weighted individually to optimise the match with the data.

Schwarzschild's method can be read as a numerical implementation of Jeans' theorem,
which states that {\it stationary} distribution functions  
(the density of stars in six dimensional phase-space) of collisionless 
systems are {\it necessarily} functions of 
the integrals of motion \citep[e.g.][]{Bin87}. 
In other words -- since integrals of motion label orbits
(and vice versa) --
the phase-space density in a stationary
system is constant along individual orbits. This explains why
the fundamental building blocks of stationary dynamical systems
are entire orbits and no density variation along individual orbits needs to
be considered. In principle then, the only assumption underlying Schwarzschild
modeling is that galaxies are stationary and collisionless.

In practice, however, applications also assume a specific internal
symmetry for each object under study. This is to reduce
the degrees of freedom in the deprojection and to simplify the sampling of phase-space
with orbits. Axial symmetry is the simplest geometry
to account for intrinsic flattening, inclination effects, rotation, and the presence of 
disc-like subsystems in real galaxies. Several implementations of Schwarzschild's 
method for axially symmetric potentials have been developed
\citep{Cre99,Geb00,Haf00,Val04,Tho04,Cap05} and have been used to analyse surveys of elliptical 
galaxy kinematics (\citealt{Geb03,Cap05,Tho07}).

Recovery of synthetic axisymmetric test models from either idealised noiseless data
\citep{Cre99,Kra05} or from realistic noisy kinematics \citep{Tho05} has proven 
an accuracy level of better than 10 percent in these models (in cases where the
reconstruction is reasonably well defined). Concerning applications to real galaxies, however, 
the distribution of apparent ellipticities, isophotal 
twists and/or minor-axis rotation indicate that ellipticals cannot be {\it exactly} 
axisymmetric \citep[e.g.][]{Ber79,Fra89,Jed89,Tre96}. Up to now, it is not clear how 
such intrinsic deviations from rotational symmetry in real galaxies affect the 
results of axisymmetric dynamical modeling.

In this paper, 
we present first results of a project aimed to systematically survey the properties
of axisymmetric Schwarzschild models that are applied to non-axisymmetric test objects. 
Specifically, we imitate
realistic photometric and kinematical observations (realistic 
in terms of spatial coverage and resolution) of collisionless 
$N$-body merger remnants and fit them with the same Schwarzschild code 
that has been used for a study of Coma ellipticals \citep{Tho07}.
We determine internal mass distributions and velocity anisotropies just as for real galaxies,
but since we know the corresponding properties of our test objects, we can examine the models.

The final goal of our project is twofold. Firstly, we 
want to explore possible systematic deviations that are caused by applying axisymmetric models
to objects that do not respect any internal symmetry. Thereby, we want to understand
how such deviations can be recognised when modeling a real galaxy, whose internal structure
is not known a priori. Collisionless disc mergers are ideal for such a study, because they
represent physically motivated dynamical systems that cover a large range of intrinsic
shapes and dynamical structures.

By investigating how intrinsically non-axisymmetric systems are mapped onto 
axisymmetric models we also gain templates for the interpretation of real galaxy models.
A second goal of our study is therefore to compare the resulting Schwarzschild models 
of merger remnants with Schwarzschild models of real galaxies. Since we use the same 
modeling code in both cases, differences are indicative for structural 
differences between galaxies and the analysed merger remnants. Knowing such 
differences allows a deeper understanding of the physical mechanisms involved in 
elliptical galaxy build-up. 

The present paper focusses on the first part of the project. A detailed discussion
of the results with respect to observations and models of real galaxies is planed for a future
publication.
We further plan to extend our survey of Schwarzschild models to samples of
mergers involving gas physics and/or dynamical systems developing from cosmological initial
conditions.

The paper is organised as follows: Sec.~\ref{section:sample} describes the sample of merger
remnants used for this work. Our implementation of Schwarzschild's technique is reviewed in 
Sec.~\ref{section:schwarzschild}. Sec.~\ref{section:hernquist} summarises
tests with a Hernquist sphere. The modeling results are
detailed in Sec.~\ref{section:schwarznotes} (general notes), 
Sec.~\ref{section:mass} (reconstructed masses) and Sec.~\ref{section:intmom} (reconstructed 
velocity anisotropies). In Sec.~\ref{section:virial} we discuss various modeling
uncertainties. Sec.~\ref{sec:totm} deals with the viewing-angle dependency of the total
mass recovery and Sec.~\ref{sec:lumml} discusses the relation between reconstructed 
luminous mass-to-light ratios and the central orbital structure of the merger remnants.
Implications for models of real galaxies are briefly discussed in Sec.~\ref{sec:implications}.
The paper closes with a summary in Sec.~\ref{section:conclusions}.

\section{Merger sample}
\label{section:sample}

A careful selection of the sample 
of merger remnants is crucial for our study. 
The merger remnants for this paper are taken from the collisionless
disc-disc merger sample of \citet{Nab03}. Their progenitor galaxies
consist of exponential discs and Hernquist bulges with a bulge-to-disc ratio of 1:3,
embedded in pseudo-isothermal dark matter halos such that the overall circular
velocity curve is approximately flat in the outer parts.
With respect to their global kinematical and photometrical properties, 
these merger remnants resemble giant ellipticals of intermediate
mass \citep{Nab06a}. For further details on the general properties and numerical details
of how the merger remnants have formed we refer the reader to \citet{Nab03}.

Different merger remnants result from different merging geometries, but here we select them
only according to their shape and do not care how they have formed. 
As the shape of the merger remnants is very closely correlated to their
orbital content \citep{Jes05}, we know that sampling different shapes ensures 
that we explore a range of different orbital makeups as well. 

\subsection{Orbital composition and shape of remnants}
\label{sub:intrinsic}
According to rotational symmetry, all orbits in axisymmetric potentials conserve the
$z$-component $L_z$ of angular momentum and are {\it minor axis tubes}, or {\it Z-tubes}. 
Such Z-tubes can have various shapes between equatorial-radial, equatorial-circular,
shell-like and polar-radial \citep[e.g.][]{Ric82}.

In triaxial dynamical systems we will expect more orbit classes \citep[e.g.][]{Zee85}. 
In particular {\it box orbits} (most frequent
in the centre) without net angular momentum, {\it boxlets} (resonant boxes found 
at larger radii) and {\it inner} and {\it outer major axis tubes} (also {\it
X-tubes} in the following). Major-axis tubes have significant 
angular momentum around the long axis. As in axisymmetric potentials, also triaxial
force fields support minor axis tubes, which have a 
non-zero angular momentum with respect to the short axis. The abundances of different
orbit classes will depend on the 
exact shape of the merger remnant. 

The shape is determined by the ratio of the three principal axes
of the inertial tensor calculated from the particle positions in the merger remnants.
The main axes are denoted: X (long), Y (intermediate) and Z (short) respectively. The
corresponding values of the inertial tensor are $a$, $b$ and $c$, respectively.

%%%%%%%%%%%%%%%%%%%%%%%%%%%%%%%%%%%%%%%%%%
% shape of mergers
%%%%%%%%%%%%%%%%%%%%%%%%%%%%%%%%%%%%%%%%%%
\begin{figure}
\includegraphics[width=84mm,angle=0]{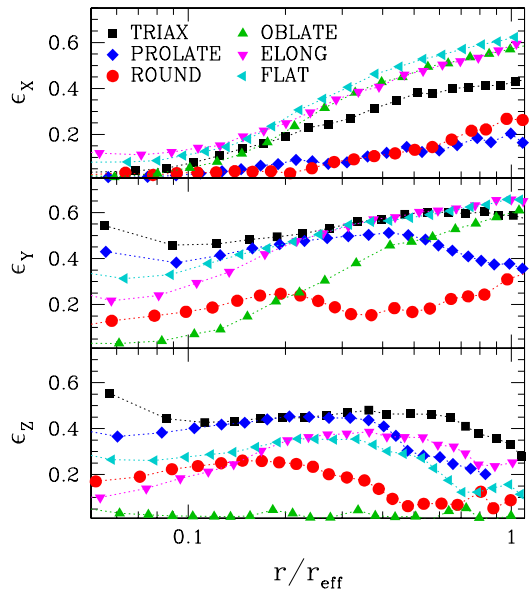}
\caption{Projected ellipticities of merger remnants. From top to bottom:
X-projection, Y-projection and Z-projection. Radii are scaled by the half-light
radius $\reff$ of the respective projection.}
\label{merger:shape}
\end{figure}

\subsection{Sample selection}
\label{sub:samplesel}
We choose six remnants that representatively sample the range of
shapes realised by the collisionless mergers of \citet{Nab03}, including the most
extreme cases: (1) a box-orbit dominated remnant (TRIAX);
(2) one with a high X-tube fraction (PROLATE); (3) a nearly
round object (ROUND); (4, 5) two very flattened remnants
with different inner shape profiles (FLAT and ELONG); (6)
one oblate remnant, dominated by Z-tubes (OBLATE). 
Modeling of further remnants from this sample would bring little additional information.

Table \ref{tab:abund} summarises orbital abundances. Respective
ellipticity profiles 
$\epsx$, $\epsy$ and $\epsz$ that result from projecting the
remnants along the three principal axes $X$, $Y$ and $Z$ are 
shown in Fig.~\ref{merger:shape}. 

\begin{table}
\begin{center}
\begin{tabular}{lccccc}
\multicolumn{3}{c}{merger remnant} &  box \&     & Z-tube & inner \& outer\\
 & &   & boxlet           &    & X-tube \\
\hline
(1) & (2) & (3)    & (4)          &  (5)  & (6) \\
\hline 
TRIAX   & 1:1 & 5  &  0.57   & 0.24    &  0.06 \\
PROLATE & 1:1 & 7  &  0.40   & 0.21    &  0.29 \\
ROUND   & 2:1 & 12  &  0.23   & 0.40    &  0.25 \\
FLAT    & 2:1 & 17  &  0.47   & 0.35    &  0.04 \\
ELONG   & 3:1 & 29  &  0.53   & 0.29    &  0.05 \\
OBLATE  & 4:1 & 11  &  0.12   & 0.76    &  0.03 \\
\hline 
\end{tabular}
\caption{Selected merger remnants. (1) Merger remnant; (2) progenitor mass 
ratio; (3) merging geometry according to Tab. 1 of \citet{Nab03}; (4-6) abundances of major orbit 
classes among the 40\% most bound particles 
(from \citealt{Jes05}). Irregular orbits and orbits without classification are not included in the table. 
\label{tab:abund}}
\end{center}
\end{table}

The Schwarzschild
models considered in this work always assume oblate axial symmetry.
Concerning projected ellipticities, axial symmetry implies either
\begin{equation}
\epsz \equiv 0, \, \epsx \equiv \epsy \,\, \mathrm{(oblate)}
\end{equation}
or 
\begin{equation}
\epsx \equiv 0, \, \epsy \equiv \epsz \,\, \mathrm{(prolate)}.
\end{equation}
Fig.~\ref{merger:shape} reveals that one of the six modelled merger remnants is
consistent with oblate axial symmetry (OBLATE), while two others are marginally
consistent with prolate axial symmetry (ROUND, PROLATE).

\begin{figure}
\includegraphics[width=84mm,angle=0]{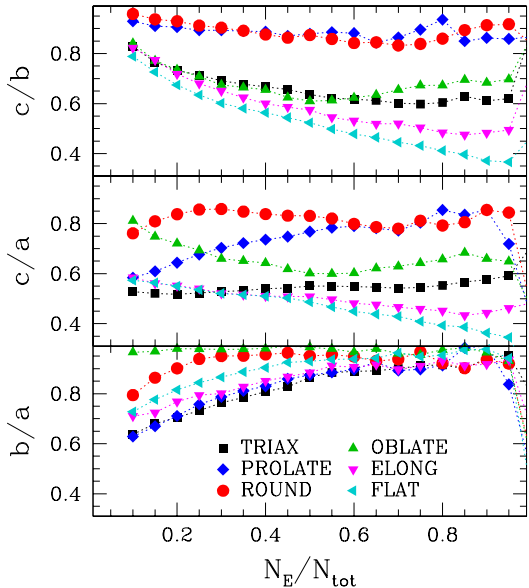}
\caption{Internal axis ratios of the merger remnants shown in Fig.~\ref{merger:shape}
as function of the fraction $N_\mathrm{E}/N_\mathrm{tot}$ 
of particles with the highest binding energy. From top to bottom:
$c/b$, $c/a$ and $b/a$.}
\label{merger:intshape}
\end{figure}

Generally, the remnant sample of \citet{Nab03} is deviant from oblate rotational 
symmetry in the inner regions. This can be inferred from Fig.~\ref{merger:intshape}, 
which shows profiles of internal axis ratios, calculated from the spatial distribution of the
$N_\mathrm{E}$ most bound (luminous) particles ($N_\mathrm{tot}$ is the total number of luminous 
particles in the remnant; cf. \citealt{Jes05}). 
In terms of intrinsic axis ratios, oblate axial symmetry implies
$c < a$ and $a \equiv b$.

%%%%%%%%%%%%%%%%%%%%%%%%%%%%%%%%%%%%%%%%%%
% shape of mergers
%%%%%%%%%%%%%%%%%%%%%%%%%%%%%%%%%%%%%%%%%%

\subsection{Mock Observations}
We will 
model the 'observational' data of the projections along the three principal axes for 
each remnant. Almost all observational properties, photometric or kinematic, will reach 
their maximum or minimum values at these projections. This is so because the principal 
axes are also the symmetry axes of the various orbit classes which means that particles will 
move perpendicular or parallel to one of the axes, depending on orbit class and 
projection. Consequently, the moments of the line-of-sight velocity 
distribution (LOSVD) will reach extreme values for the 
respective projections. Similar reasoning can be applied to the photometric
properties, e.g. the isophotal shape parameter $a_4$ \citep{Ben87}. Concerning Z-tubes
in triaxial potentials, for example, $a_4$ reaches its extreme values in the long-axis
projection (most boxy) and intermediate-axis projection (most disky), respectively 
(cf. \citealt{Jes05} for detailed discussion). 
In summary: by modeling the principal projections, we are testing the extreme 
cases, where the influence of certain orbit classes on the observables is visible 
(or not). Influence of the viewing angle on our results will be discussed in detail 
in the following sections.

As we want to make meaningful statements about the recovery of real
galaxy structure, we try to imitate observational conditions of a comparison
sample of Coma ellipticals modelled by \citet{Tho07} with
the same Schwarzschild code. This
comparison sample consists of long-slit major and minor axis spectra
\citep{Meh00,Weg02}. The photometry for the Coma galaxy models 
is constructed as a composite of HST (centre) and ground based imaging
(outer parts; cf. \citealt{Tho05}).

The photometry of the $N$-body mergers is obtained for two different resolutions as well.
The coarse one is mimicking ground based observations and extends to 
large radii. The one with higher resolution simulates HST 
data at small radii. We include seeing by smoothing the particle distribution 
%in the 'slit' 
with a Gaussian of width $1/20 \, \reff$ (roughly the size of the 
numerical softening parameter) in the high resolution 
case. In the coarse resolution case the seeing amounts to three quarters of the 
effective radius. Low and high resolution photometry are combined into one continuous 
profile, as described in \citet{Tho05}.

To match the observational conditions of the Coma comparison sample
the kinematic information is extracted at an intermediate resolution 
(smoothing width $1\farcs2 \approx 1/6 \, \reff$) along the apparent photometric 
major and minor axes (cf. Sec.~\ref{chidef}). We have kinematic data 
out to about two effective 
radii (see \citealt{Nab03} for extensive 
discussion on how artificial observations are performed on
N-body remnants). 

\section{Schwarzschild modeling}
\label{section:schwarzschild}
Our Schwarzschild models are described
in \citet{Tho04} where a thorough discussion of the modeling implementation is given.

\subsection{Model setup}
\label{sub:basic}
In the following, we briefly recall the basic steps of the Schwarzschild technique:

\begin{enumerate}
\item
The surface brightness of each remnant-projection is
deprojected at three inclination angles: the edge-on deprojection probes the inclination
where the Schwarzschild model is least flattened. 
The other extreme, the most flattened Schwarzschild model, is constructed at an 
inclination angle, for which
the deprojection appears as an E7 galaxy (when seen from the side). Inclination angles
resulting in intrinsically even flatter models are unreasonable because (1) ellipticals
flatter than E7 are not observed and (2) remnants flatter than E7 are not in our
merger remnant sample. As an intermediate case we also probe an inclination angle that
leads to a Schwarzschild model resembling an E5 galaxy (when seen from the side).

The luminosity distribution
$\nu$ is assumed to be axisymmetric and we include surface-brightness,
ellipticity and $a_4$-profiles in the deprojection, which is performed
with the non-parametric program of \citet{mag99}.

\item
Based on the deprojected luminosity-profile $\nu$ a mass distribution is constructed via
\begin{equation}
\label{eq:mass}
\rho = \Upsilon \, \nu + \rho_\mathrm{DM},
\end{equation}
where $\Upsilon$ determines the amount of mass that follows the light 
(we will denote $\Upsilon$ the stellar mass-to-light ratio in the following).
For the additional dark matter density $\rho_\mathrm{DM}$ we adopt a Navarro-Frenk-White 
(NFW) profile and the relation between concentration and mass given in \citet{nfw96}. 
The dark halos of the merger remnants do not follow these profiles exactly. 
The progenitor galaxies are embedded in pseudo-isothermal
halos with a flat central density core. After the merging the central dark matter slope
steepens, but is still shallower than in NFW-profiles (cf. Sec.~\ref{subsec:mdens}).

Our choice for NFW-halos is motivated by results of 
Monte-Carlo simulations showing that one can always find a NFW-halo among the above
introduced family which mimics an (non-singular) isothermal distribution sufficiently 
well over the radial region considered here \citep{Tho05}. Moreover, for a few remnant 
projections we have
also calculated cored logarithmic halos, and the results do not change
significantly (cf. Sec.~\ref{subsec:mdens}). Then, since we do not loose generality, it
is convenient to use the one-parameter family of NFW-halos. To
explore possible effects of halo shapes we model each halo once with a
spherical mass distribution and once with a flattening of the density distribution of
$c/b \equiv c/a=0.7$, where $a$, $b$ and $c$ are the long, intermediate and short-axis of the 
halo mass distribution, respectively. The halos of the merger remnants are close to
oblate-axial symmetry, with $b/a>0.9$ and $0.7 \le c/a \le 1.0$.

With the mass density fixed, the gravitational potential $\Phi$ follows by solving Poisson's 
equation.

\item
In the gravitational potential $\Phi$ a representative set of orbits is
calculated. The orbit sampling is described in detail in \citet{Tho04}.

\item
In the final step the orbits are superposed to fit the
photometric and kinematical constraints. The maximum entropy-technique of
\citet{maxspaper} is applied and the kinematic data is fitted by solving 
for the maximum of
\begin{equation}
\label{maxs}
S - \alpha \chi^2 \rightarrow \mathrm{max}.
\end{equation}
$S$ is the entropy of the model and
\begin{equation}
\label{chilosvd}
\chi^2 \equiv \sum_{j=1}^{N_{\cal L}} \, \sum_{k=1}^{N_\mathrm{vel}} 
\left(
\frac{{\cal L}^{jk}_\mathrm{mod}-{\cal L}^{jk}_\mathrm{in}}
{\Delta {\cal L}^{jk}_\mathrm{in}}
\right)^2
\end{equation}
measures the difference between input LOSVDs $\cal{L}_\mathrm{in}$ and
model LOSVDs $\cal{L}_\mathrm{mod}$ (see \citealt{Tho04} for more details 
about the calculation of $S$ and $\cal{L}$ in this context). 
Each LOSVD is binned into $N_\mathrm{vel}$
velocity-bins and the input data consists of $N_{\cal L}$ LOSVDs in total (cf. 
Sec.~\ref{chidef} for further details). 
The luminosity density is treated as a boundary condition
to equation (\ref{maxs}). 
The regularisation parameter $\alpha$ in equation (\ref{maxs}) allows to
control the relative importance of $\chi^2$-minimisation (fit to data) and 
entropy-maximisation (smoothness of the distribution function). 

In the following we will consider two cases for $\alpha$. Firstly, models obtained 
with $\alpha=0$ will be called $\ms$-models, because for $\alpha=0$ the $\chi^2$-term 
vanishes and the orbital weights are entirely determined by the maximisation 
of $S$ (under the boundary condition related to $\nu$). 
In order to fit an orbit library to a given set of kinematical data, $\alpha$ has to
be positive. The larger $\alpha$, the better the fit will be.
In case of real observations, very large $\alpha$ can result in models that
fit the noise in the data. Concerning our merger
remnant fits, we assume that the input data is not affected significantly by
noise (cf. next Sec.~\ref{chidef}). Therefore, as the second case for $\alpha$, 
we consider a value large enough such that the minimum of $\chi^2$
is reached ($\mc$-models). This usually occurs around $\alpha \approx 1$\footnote{The 
exact value we will use is $\alpha = 0.9143$ and arises from the 
iterative solution of equation (\ref{maxs}); see, for example, \citet{Tho05}.}. 
Larger $\alpha$ do not change $\chi^2$ or other model 
properties significantly.
\end{enumerate}

\subsection{Definition of $\chi^2$} 
\label{chidef}
To solve equation (\ref{maxs}) one needs to 
evaluate the $\chi^2$-term and, thus, to specify the $\Delta {\cal L}^{jk}_\mathrm{in}$
of equation (\ref{chilosvd}).
Insofar as the $N$-body simulations are viewed as a discrete $N$-particle 
realisation of an underlying 
continuous phase-space distribution function, the mock observations should be
interpreted to have some intrinsic Poisson scatter that decreases with increasing 
the number of particles. This describes the case of Sec.~\ref{section:hernquist},
where we test our modeling machinery with an $N$-body representation of a 
Hernquist sphere. It is also valid for the setup of the progenitor systems. In both cases, the
$N$-body system is an imperfect representation of an underlying continuous 
phase-space distribution function. The merger remnants, however, 
are not such an $N$-body sampling of some unknown distribution function. Instead, they just
reflect the dynamical evolution of $N$ particles from their particular initial conditions --
irrespective of how these have been constructed. In this sense, after the relaxation induced
by the merging, we treat the mock observations as 'ideal' observations of a discrete
($N \approx 10^5$ particle) dynamical system, that we try to represent by 
Schwarzschild models. 
There is no obvious way to define $\Delta {\cal L}^{jk}_\mathrm{in}$ in this case, however.

For a statistical analysis the proper
way to proceed is to add random fluctuations
to the raw observations. The resulting noisy 'data' together with the 'error bars'
from which the noise has been constructed provide a statistically consistent input to the
models. However, our merger sample is small and it would be
necessary to model several random realisations of the original raw data in order
to avoid any influence of a particular noise pattern on the results. This is computationally
too expansive as it means to model effectively dozens of data sets. Moreover, it is not the
goal of this study to quantify uncertainties that originate from
observational errors (which has been done elsewhere, e.g. \citealt{Tho05}).
Instead, our aim is to explore possible {\it systematic} biases arising when treating 
non-axisymmetric objects with axisymmetric models.
Therefore, we setup our model input as follows.

First, Gauss-Hermite moments $v$, $\sigma$, $H_3$ and $H_4$ \citep{Ger93,vdMF93}
of the merger remnants are calculated as in \citet{Nab01}. The Gauss-Hermite
moments are then used to calculate the LOSVDs $\cal{L}_\mathrm{in}$ at a
set of radii typical for our comparison sample of Coma ellipticals. Corresponding
observational errors of Coma galaxies at these radii are scaled to the mock
data\footnote{We use fractional errors
in $v$ and $\sigma$ but absolute errors for $H_3$ and $H_4$. As template
to create the error-bars we use the observations of NGC4807, 
which are proto-typical for the Coma sample in terms of 
radial coverage and signal-to-noise.}. The Gauss-Hermite 'error-bars' 
are propagated into
$\Delta {\cal L}^{jk}_\mathrm{in}$ by means of Monte-Carlo simulations. The resulting
LOSVDs $\mathcal{L}_\mathrm{in} \pm \Delta {\mathcal L}^{jk}_\mathrm{in}$ 
are used as input for the Schwarzschild models without adding noise explicitly. 
Neglecting the noise makes uncertainties of derived model quantities 
(masses, internal velocity moments) unreliable. 
But for our purpose of identifying systematic trends it is only important to
flag a bestfit model in a similar way as a bestfit model is determined for a real galaxy.
The role of $\Delta {\cal L}^{jk}_\mathrm{in}$ is
to specify the relative weight of different data points. The usage of error bars from real
observations ensures that in our models data from different spatial regions 
are weighted similar as in models of real galaxies.

\subsection{The bestfit model} 
\label{subsec:bf}
To obtain the bestfit dynamical model we calculate Schwarzschild models on a grid
in the two-dimensional parameter space $(\Upsilon,c)$. Thereby we probe 
$0.3 \le \Upsilon \le 1.3$ with $\Delta \Upsilon = 0.1$ and 
$2.5 \le c \le 30.0$ with $\Delta c = 2.5$. For each pair $(\Upsilon,c)$ on the grid
one model is calculated with a spherical halo and another one with a halo flattening
of $q=0.7$. The procedure is repeated for up to three inclinations (cf. Sec.~\ref{sub:basic}).

For this first set of models we use a coarse library setup with 
$2 \times 3500$ orbits, roughly half the number used to model Coma ellipticals
\citep{Tho07} and roughly twice the number that has been
used by the Nuker team for models of galaxy centres \citep{Geb00}.
Among the low resolution models one, say with parameters
$(\Upsilon_f,c_f,q_f)$, yields the lowest $\chi^2$. Around these parameter values
we recalculate models with a larger number of orbits ($2 \times 9000$ orbits as used
for models of Coma galaxies by \citealt{Tho07}). The overall bestfit model
is chosen among these high resolution fits according to the minimum of 
$\chi^2$.  

For the high resolution models we adjust the
modeling strategy as follows. (1) As it will be discussed in Sec.~\ref{sub:incnote}, the 
best-fit (low resolution) model is always at an inclination $i=90\degr$. 
For the high resolution case we therefore only
consider edge-on geometries. (2) We examine the same grid
for ($\Upsilon,q$) as in the low-resolution case, but restrict concentrations around
$c_f$, usually probing the region between $c_f-2\,\Delta c$ and $c_f+2\,\Delta c$. If
necessary we extent the concentration interval such that the bestfit high resolution
model never occurs at the boundary of the sampled parameter space. When resampling with
a larger number of orbits, we vary the halo-concentrations in smaller steps of $\Delta c = 1.0$.

We do not find systematic differences between the models with $2 \times 3500$ orbits and
those with $2 \times 9000$ orbits, respectively. For example, twelve out of eighteen 
best-fit luminous mass-to-light ratios $\mlfit$ are the same in low-resolution and 
high-resolution models. In the remaining 
cases they change by $\Delta \mlfit = 0.1$ (four models) and $\Delta \mlfit = 0.2$ 
(two models), respectively. There is no preferred direction for the change $\Delta \mlfit$.

\section{Validation: a Hernquist sphere}
\label{section:hernquist}

To check all conversions from $N$-body systems to Schwarzschild models and back
we first model a self-consistent Hernquist sphere \citep{Her90}:
we sample the isotropic Hernquist distribution function
with $N=1.6 \times 10^{5}$ particles and 'observe' the resulting $N$-body
realisation in exactly the same way as 
the merger remnants. The number of particles $N=1.6 \times 10^{5}$ resembles the number of luminous particles
in the 1:1-merger remnants, analysed later on. 
For the goal of verifying our machinery by reconstructing the Hernquist sphere
we modify the modeling procedure as follows: 
(1) We only consider an inclination of $i=90\degr$, such that the
deprojection is unique. (2) We only fit
self-consistent Schwarzschild models,
because the Hernquist sphere is setup self-consistently (without dark matter). 
Finally, in order to evaluate the influence of noise in the
$N$-body representation, we combine Schwarzschild fits to ten different
Monte-Carlo realisations of the Hernquist sphere. 
The only free mass parameter in this test run is the stellar mass-to-light ratio $\Upsilon$. 

Application of our Schwarzschild models to
the Hernquist sphere yields
$\mlrat = 0.993 \pm 0.037$, where $\mlfit$ and $\mlin$ are
the mass-to-light ratio of the Schwarzschild models (averaged over fits to ten
realisations of the Hernquist sphere) and the $N$-body input,
respectively. The quoted uncertainties reflect the variance about the mean. 

%%%%%%%%%%%%%%%%%%%%%%%%%%%%%%%%%%%%%%%%%%
% hernquist
%%%%%%%%%%%%%%%%%%%%%%%%%%%%%%%%%%%%%%%%%%
\begin{figure}
\includegraphics[width=84mm,angle=0]{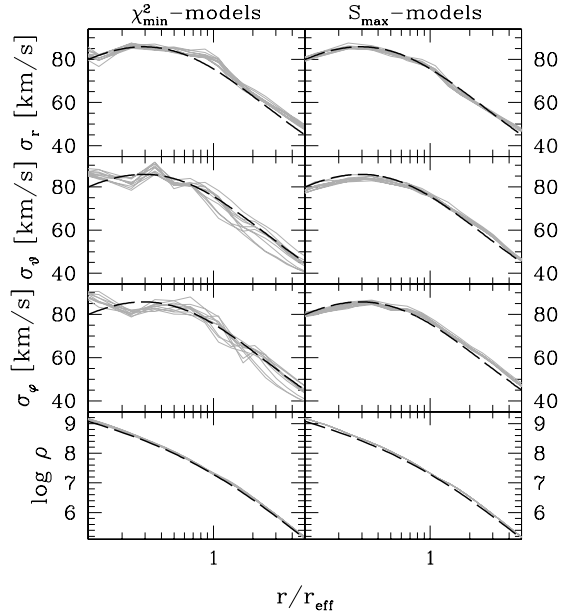}
\caption{Schwarzschild models for a Hernquist sphere. In each panel ten solid grey lines 
represent models for ten Monte-Carlo samplings of a Hernquist sphere. The analytically 
derived reference profiles of the Hernquist sphere are shown by
black, dashed lines. Three upper rows: radial ($\sigr$, top), meridional ($\sigt$, middle) and 
azimuthal ($\sigp$, lower) velocity dispersions (luminosity-weighted, spherical averages); 
bottom row: density. Left column: $\mc$-fits; right column: 
$\ms$-models (details in the text).}
\label{hernquist}
\end{figure}

Internal velocity moments of Schwarzschild fits are shown in Fig.~\ref{hernquist} 
together with the analytic profiles for the isotropic Hernquist
sphere \citep{Her90}. Results from Schwarzschild modeling are spherically averaged.
For the $\mc$-models in the left panels of Fig.~\ref{hernquist} a regularisation 
parameter $\alpha \approx 1$ has been used (cf. Sec.~\ref{sub:basic}; the same value
is used for the merger remnant fits). Apart from some noise
in the Schwarzschild models, the overall agreement between
analytic calculations and Schwarzschild fits is very good.

To understand whether the scatter in $\mlrat$ and
the internal moments originates from uncertainties in the Schwarzschild code or 
whether it comes from noise in the $N$-body realisation, we also tried 
to reconstruct the Hernquist sphere by a method that is independent from
noise in the $N$-body kinematics: the solution of equation (\ref{maxs}) for $\alpha = 0$. 
As stated in Sec.~\ref{sub:basic}, with $\alpha = 0$ the $\chi^2$-term vanishes 
and the orbit distribution is determined
entirely by maximising its entropy ($\ms$-model).
The idea behind considering $\ms$-models here is the following: the
maximisation of $S$ yields, in a sense, the smoothest distribution function (DF) 
for the given density profile.
Assuming that this smoothing isotropises stellar velocities then the
$\ms$-model would be identical to the (unique) isotropic DF, which is connected to 
any self-consistent spherical density profile. Maximising $S$ would therefore 
determine the orbital weights (and internal moments etc.) of our orbit representation 
of the Hernquist sphere without any fit to the kinematics.

Since the orbital weights in the $\ms$-models are fixed, the only degree of freedom
is the velocity scale $\Upsilon$. Results of the corresponding fits are 
shown in the right panels of Fig.~\ref{hernquist}. As can be seen, the internal moments of the 
$\ms$-models follow closely the analytic profiles, confirming the above speculations
about the connection between entropy and isotropy in spherical systems.
That the $\ms$-models in fact match better with the
analytical Hernquist profiles than the fits on the left
implies that the scatter in the fits is mainly caused by noise in the $N$-body LOSVDs.
Uncertainties in the
Schwarzschild code (finite number of orbits and finite numerical resolution) are
instead negligible, as otherwise deviations between reference moments and orbit
representation would be larger.
Likewise, since the $\ms$-models in the right panels of 
Fig.~\ref{hernquist} are based upon the 
deprojected $N$-body light-profiles, noise in the $N$-body light-profiles 
is also not the dominant driver for scatter in the left panels.

Concerning mass-to-light ratios we find $\mlrat = 1.007 \pm 0.016$ in the mean
over all ten $\ms$-models. 
As stated above, the remaining scatter of about $1.5$ percent is due to noise in the
$N$-body kinematics. We do not expect this scatter to have a significant influence on
our results of fits to the merger remnants.

\section{Schwarzschild fits of merger remnants: general notes}
\label{section:schwarznotes}
Now to the models of simulated merger remnants. This section
contains notes
on general properties of the Schwarzschild fits and the deprojections.

\subsection{Luminosity densities}
\label{subsec:depro}
\begin{figure}
\includegraphics[width=84mm,angle=0]{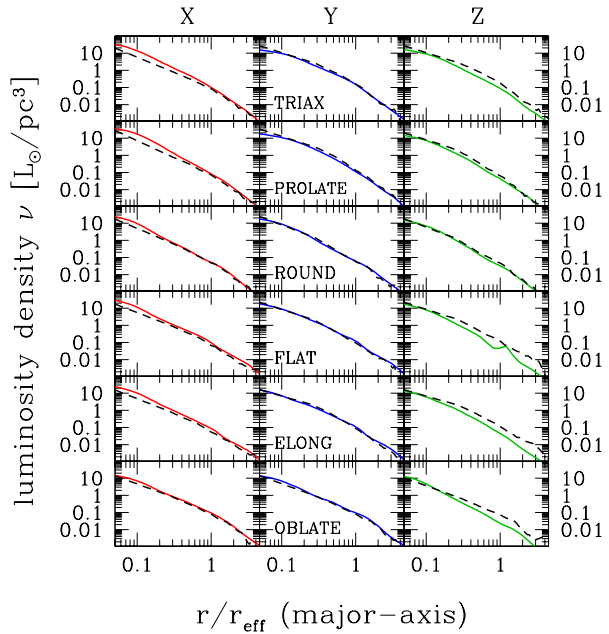}
\caption{Luminosity density of merger remnants (black, dashed) and merger models
(coloured, solid). From left to right: models of X-projections, Y-projections and Z-projections.
Densities are evaluated along the projected major-axis.}
\label{depro}
\end{figure}

Fig.~\ref{depro} compares the axisymmetric deprojections with the internal
luminosity density profiles of the merger remnants. The figure only compares densities
along the projected major-axis. Results along other position angles are similar. For the
merger remnants, the density is averaged over a plan-parallel wedge of size 
$\Delta r \approx 0.05 \, \reff$ along the major-axis, $\Delta z \approx 0.2 \, \reff$ 
perpendicular to this axis (in the plane of the sky) and
$\Delta \phi = 45\degr$ in the plane defined by the line-of-sight and the projected
major-axis.

If a remnant is seen along its long-axis (left panels), then
the axisymmetric deprojection overestimates the density -- especially
near the centre. The opposite occurs if a merger is seen along the
intermediate axis (middle panels): the axisymmetric deprojection 
of the Y-projections underestimates
the remnant density. Note that for the remnant in the bottom row 
(OBLATE) X and
Y-deprojections are almost equal, consistent with its oblate shape ($b \approx a$).

\begin{figure}
\includegraphics[width=64mm,angle=0]{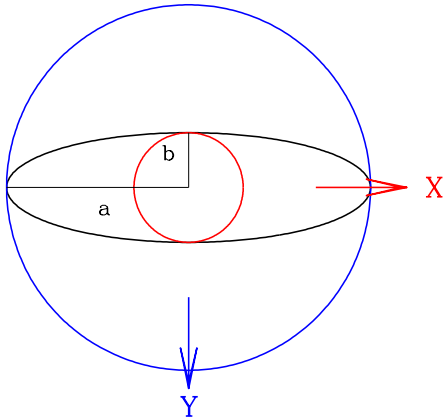}
\caption{Schematic view onto the $(X,Y)$-plane of a prolate body ($b/a < 1$).
If the body is seen along the X-axis, then a deprojection
assuming axial symmetry with the symmetry axis being perpendicular to the
$(X,Y)$-plane, overestimates the density 
inside $r<b$ (small circle, red). Correspondingly, if the body is seen from the
$Y$-axis, an axisymmetric deprojection underestimates the density inside $r<a$ 
(large circle, blue).}
\label{skizze}
\end{figure}

Fig.~\ref{skizze} illustrates that the viewing-angle dependency of
the deprojections reflects the intrinsic non-oblateness $b/a \ne 1$ 
of most of our merger remnants\footnote{We restrict the discussion to the edge-on case,
since all our bestfit Schwarzschild models have $i=90\degr$ 
(cf. Sec.~\ref{sub:incnote}).}. The
light inside an ellipse with $b < a$, if seen along the long-axis, is quenched 
into the region $r<b$ in the axisymmetric deprojection. Accordingly, the mean density
of the deprojection inside $b$ must be larger than the original density inside the
same spatial region. Conversely, if the ellipse is viewed side
on, the axisymmetric deprojection stretches the light into the larger region $r<a$ and, hence,
underestimates the true density. 

Concerning our merger remnants, deviations between deprojection and
intrinsic light profile are largest where $b/a$ is smallest 
(cf. Fig.~\ref{merger:intshape}) -- in accordance with the above reasoning. 
At large radii, the intermediate-to-long axis ratio becomes $b/a \approx 1$ 
and the deprojections
of X and Y-projections approach the luminosity profiles of the remnants.

%%%%%%%%%%%%%%%%%%%%%%%%%%%%%%%%%%%%%%%%%%
% deprojection
%%%%%%%%%%%%%%%%%%%%%%%%%%%%%%%%%%%%%%%%%%

Concerning the short axis, $c/a$ quantifies the quenching of light along the line-of-sight as
much as $b/a$ quantifies it along the intermediate axis. Insofar, the Z-projection is 
similar to the Y-projection, which explains why Z-deprojections underestimate luminosity
densities of the mergers as well. A difference arises at large radii because
$b/a \rightarrow 1$, whereas
$c/a$ stays roughly constant (e.g. FLAT, ELONG, OBLATE).
Consequently, Z-deprojections deviate over the whole radial range plotted in Fig.~\ref{depro}
and have a steeper slope than the luminosity profiles of the mergers. 

%%%%%%%%%%%%%%%%%%%%%%%%%%%%%%%%%%%%%%%%%%
% total mass: bestfit
%%%%%%%%%%%%%%%%%%%%%%%%%%%%%%%%%%%%%%%%%%

\subsection{Kinematic fits}
\label{section:fits}
\begin{figure}
\includegraphics[width=84mm,angle=0]{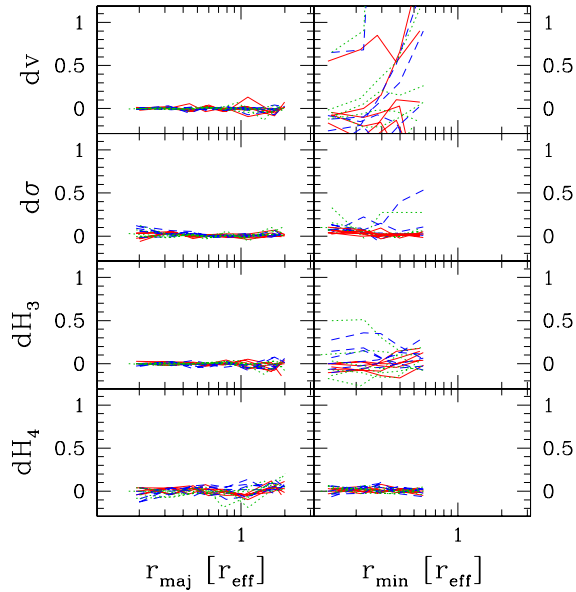}
\caption{Residuals between Schwarzschild fits and remnant LOSVDs normalised to the
assigned error-bars: $d v \equiv (v_\mathrm{fit}-v_\mathrm{in})/\Delta v_\mathrm{in}$
(and analogously for $\sigma$, $H_3$ and $H_4$). Left: major-axis; right: minor-axis; 
red/solid: X-projections; blue/dashed: Y-projections; green/dotted: Z-projections.}
\label{losvd-match:bestfit}
\end{figure}

Because the merger remnants do not obey oblate axial symmetry  
it is not clear whether
their kinematics can be fit by our models -- which respect this
symmetry -- at all. Residuals in the kinematic fits are shown in
Fig.~\ref{losvd-match:bestfit}. Except from minor-axis rotation $v$ and asymmetry 
of the LOSVD $H_3$ 
Schwarzschild models reproduce the data very well, to an accuracy of about a tenth of the 
assigned 'error bars'. Since these 'error bars' are taken from observations,
a comparable degree of triaxiality in real galaxies would be hardly 
recognisable in terms of a systematic offset between models and data.

Discrepancies between merger remnants and Schwarzschild models 
in minor-axis profiles of $v$ and $H_3$ are the result of oblate axial 
symmetry enforcing $v \equiv H_3 \equiv 0$ in the models. 
Hence, the upper-right panel of 
Fig.~\ref{losvd-match:bestfit} in fact shows the amount of 
minor-axis rotation in the remnants. Neglecting the latter
in our fits implies that part of the
kinetic energy of the merger remnants is missing in the Schwarzschild models. 
This could lead to an underestimation 
of the mass. However, the minor-axis rotation
in Fig.~\ref{losvd-match:bestfit} is of the order of the
assigned error bars ($d v \la 1$), e.g. below 10 percent of the 
kinetic energy in the dispersion (cf. radial profiles of 
$v$, $\sigma$, $H_3$ and $H_4$ and their assigned errors in App.~\ref{app:data}). 
We therefore do not expect that neglecting minor-axis rotation of the merger 
remnants has a dominant effect on our results.

%%%%%%%%%%%%%%%%%%%%%%%%%%%%%%%%%%%%%%%%%%
% losvd-match: best fits
%%%%%%%%%%%%%%%%%%%%%%%%%%%%%%%%%%%%%%%%%%

\subsection{Inclinations of bestfit models}
\label{sub:incnote}
Although we probe models at three different inclinations for each remnant
projection, the bestfit model (with $2 \times 3500$ orbits; cf. Sec.~\ref{subsec:bf}) 
always occurs at an inclination
of $i=90\degr$ (edge-on). This is not surprising for X and Y-projections.
However, Z-projections could have been
expected to be better represented by nearly face-on models, e.g. with $i \approx 0\degr$. 

However, according to the
lower panel of Fig.~\ref{merger:shape} all remnants except the OBLATE one appear
flattened when projected along the Z-axis ($\epsz >0$). 
Axisymmetric models, on the other hand, are
necessarily round when seen along the axis of symmetry. Thus, an 
axisymmetric $i=0\degr$ model {\it cannot} fit the Z-projection of most remnants.

Only one remnant (OBLATE) is close enough to axial symmetry that its Z-projection is
almost round. Why is the bestfit model for this remnant again achieved
for $i=90\degr$? The main reason is probably the small rotation signal
$v \ne 0$ and $H_3 \ne 0$ along the apparent major axis of its Z-projection 
(the face-on view is not exactly round, cf.
Fig.~\ref{merger:shape}). At a viewing angle of $i=90\degr$ the model can adjust
the balance between prograde and retrograde orbits to 
fit $v \ne 0$ and $H_3 \ne 0$. Instead, any rotation and asymmetric deviation
from a Gaussian LOSVD disappear when looking at an axisymmetric system face-on: 
$v \equiv H_3 \equiv 0$ (for all position angles). Thus, everything else being equal, 
a face-on model will necessarily have a larger $\chi^2$ than an edge-on model. 
In fact $83 \, \%$ of the $\Delta \chi^2$ between 
the bestfit edge-on and the
bestfit face-on model\footnote{The deprojection of axisymmetric bodies at $i=0\degr$ is
infinitely uncertain \citep{Ryb87,Ger96}. For the above comparison a face-on model has 
been constructed using the ($i=90\degr$) deprojection of the X-projection. If the remnant
would be exactly axisymmetric, then this deprojection would uniquely recover its intrinsic
luminosity distribution.} of the OBLATE remnant, respectively, is due to differences
in the fit to $v$ and $H_3$. This is not a proof, but a strong indication
that the residual rotation in the Z-projection of the OBLATE remnant is the main driver 
for the bestfit model to occur at an inclination of $i=90\degr$.

A triaxial dynamical system can exhibit various degrees of rotation in the Z-projection.
If this indeed causes the corresponding axisymmetric fit to prefer an inclination
of $i=90\degr$, then the inclination mismatch is an unavoidable consequence of the 
false symmetry assumption. Concerning models of real galaxies, an additional complication
enters by measurement errors: even for an exactly axisymmetric face-on galaxy one would
determine $v \ne 0$ and $H_3 \ne 0$ due to measurement uncertainties. In such a case,
a best-fit axisymmetric inclination of $90\degr$ would be an artifact related to the 
ability of the modeling machinery to fit the noise in the data.
Proper regularisation could provide a way out of the inclination mismatch
then. For the Z-models of the OBLATE remnant we find indeed 
a best-fit inclination $i=0\degr$ for $\alpha<0.005$ (strong regularisation).

A systematic investigation of the question whether noise in real data 
can bias axisymmetric models towards $i=90\degr$ and whether this possible
bias can be reduced by using proper regularisation is out of the scope of this paper.
For simplicity, we adopt the same regularisation scheme to all merger remnants
in the following. We expect this to significantly
affect only the fits to the face-on projection of the OBLATE remnant. 
Specifically, assuming the wrong inclination makes our Z-model worse than 
it could possibly be with optimised regularisation. 
In all other remnants the inclination mismatch is due to intrinsic non-axisymmetry.

\section{Mass distribution in remnants and models}
\label{section:mass}
Having discussed general features of the Schwarzschild models we now
turn to the comparison of the mass distribution in models
and the corresponding merger remnants.

%%%%%%%%%%%%%%%%%%%%%%%%%%%%%%%%%%%%%%%%%%
% best-fit histograms
%%%%%%%%%%%%%%%%%%%%%%%%%%%%%%%%%%%%%%%%%%
\begin{figure}
\includegraphics[width=84mm,angle=0]{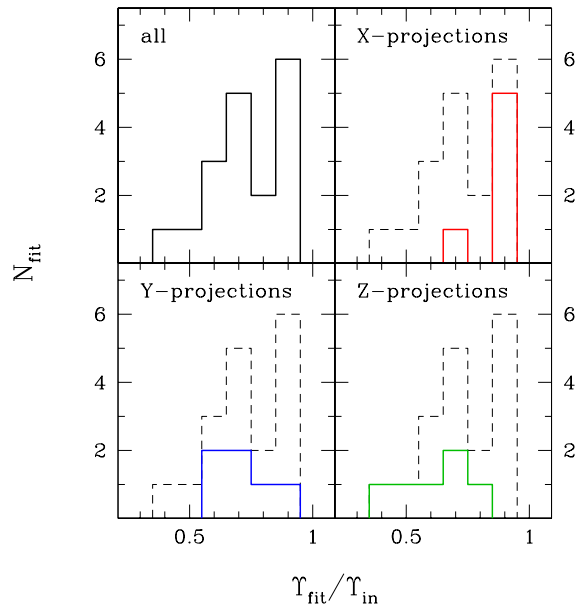}
\caption{Distribution of bestfit mass-to-light ratios $\mlfit$ (scaled by
the input value $\mlin$). Top left panel: whole sample; other panels: results for different
projections (indicated in each panel; dashed lines: total distribution for comparison).}
\label{hist:bestfit}
\end{figure}

\subsection{Stellar mass-to-light ratio}
\label{subsec:stemass}
Fig.~\ref{hist:bestfit} shows the distribution of (scaled) mass-to-light
ratios $\mlrat$ obtained from our bestfit Schwarzschild models.
The reconstructed $\mlfit$ systematically underestimate $\mlin$.
Seventeen out of the eighteen models have mass-to-light ratios in the range
$0.5 \le \mlrat \le 0.9$, one model has $\mlrat = 0.4$.
Separating the results according to the viewing angle yields that models of 
X-projections (shortly X-models below) recover the true mass-to-light ratio
very well ($\mlfit = 0.9 \, \mlin$ in all but one case; 
cf. upper-right panel of Fig.~\ref{hist:bestfit}). 
In contrast, Schwarzschild models of Y and Z-projections 
have mass-to-light ratios distributed almost homogeneously in the range
$\mlrat = 0.5 - 0.9$ (bottom panels
of Fig.~\ref{hist:bestfit}).
{\it Although} the luminosity density of the deprojection predicts less light
in the corresponding models than there is in the merger. The reason for this behaviour 
will be further discussed in Secs.~\ref{sec:totm} and \ref{sec:lumml}.

It should be restated that our mock observations are not drawn from random projections. Therefore,
Fig.~\ref{hist:bestfit} does not equal the distribution of mass-to-light
ratios that would result from modeling real galaxies (even 
if they would be structurally similar
to the merger remnants). The most significant result here
is that axisymmetric models tend to {\it underestimate} 
the mass fraction that follows the light. We have no proof for the generality of this 
result, but since we have modelled all three principal projections for each remnant 
we do not expect models from other viewing angles to 
deliver $\mlfit > \mlin$.

\subsection{Mass densities}
\label{subsec:mdens}
\begin{figure}
\includegraphics[width=84mm,angle=0]{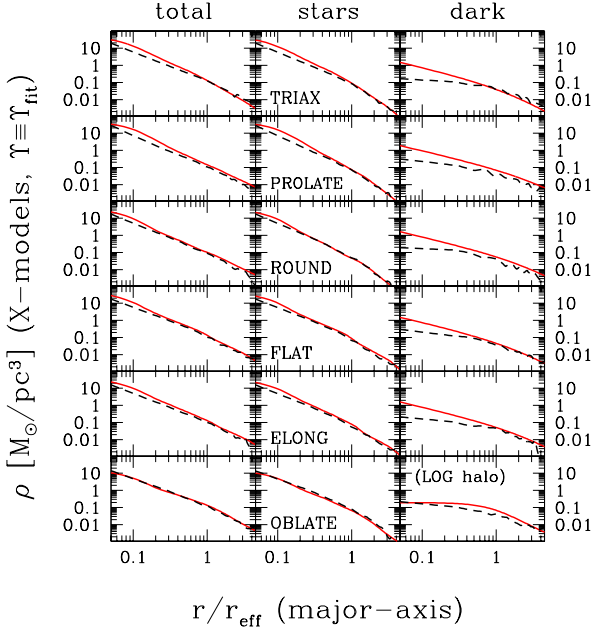}
\caption{Comparison of remnant (black/dashed) and Schwarzschild model
(coloured/solid) mass-density profiles (along projected major-axis). Left/middle/right: 
total/luminous/dark mass. The figure surveys results from modeling X-projections.}
\label{rho:x}
\end{figure}

\begin{figure}
\includegraphics[width=84mm,angle=0]{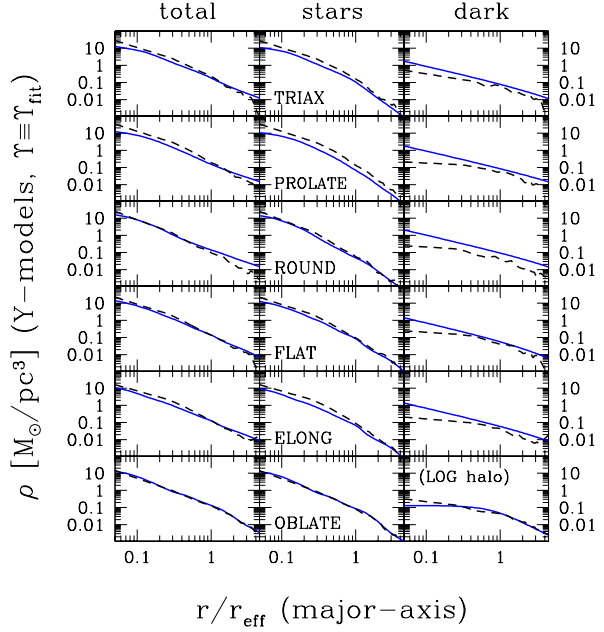}
\caption{As Fig.~\ref{rho:x}, but for Y-projections.}
\label{rho:y}
\end{figure}

\begin{figure}
\includegraphics[width=84mm,angle=0]{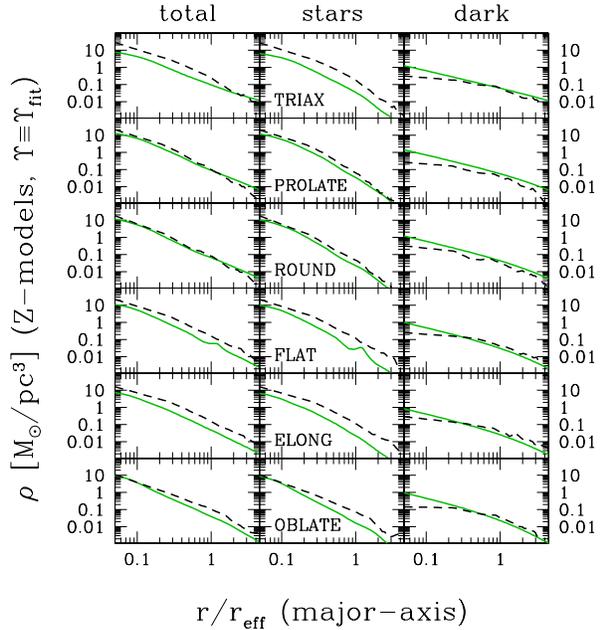}
\caption{As Fig.~\ref{rho:x}, but for Z-projections.}
\label{rho:z}
\end{figure}

Our Schwarzschild models (and the merger remnants as well) contain both luminous as well as dark
mass and $\Upsilon$ only represents a fraction of the total mass. The next
question is how well total and dark matter density profiles are represented
in the Schwarzschild models. To explore this, Figs.~\ref{rho:x} -
\ref{rho:z} survey radial density profiles of models and remnants separately
for the three principal projections. The figures show intrinsic densities along 
the projected major
axis. The middle panels (luminous mass density) differ from the $\nu$-profiles 
of Fig.~\ref{depro} only in the scaling (the stellar mass density 
equals $\mlfit \times \nu$; cf. equation \ref{eq:mass}).

Evidently, in X-models not only the luminosity density, but also the total mass in the
inner regions is overestimated. Exceptional is the X-model of the OBLATE remnant: because
the remnant is close to axial symmetry, no overestimation of the central density occurs.
In Y and Z-models -- parallel to the underestimation of the light -- also the total mass 
density is underestimated. Again exceptional is the Y-model of the OBLATE remnant: the
total mass is well recovered. This reflects again the axial symmetry of the remnant, according
to which X and Y-projections are equivalent and both should allow a good reconstruction
with our models. 

The case of the OBLATE remnant also reveals a slight degeneracy in the
mass recovery. The best-fit X-model has $\mlfit = 0.7$, while the best-fit 
Y-model is obtained with $\mlfit =0.9$. Despite these different $\mlfit$, the
total mass inside $\reff$ is
recovered with high accuracy in both models: $2.8$ percent fractional accuracy in the 
X-model and $0.4$ percent in the Y-model, respectively. 
Thus, the total mass can be recovered with about the same accuracy, even if luminous masses
differ by about 20 percent.

%%%%%%%%%%%%%%%%%%%%%%%%%%%%%%%%%%%%%%%%%%
% total mass: bestfit
%%%%%%%%%%%%%%%%%%%%%%%%%%%%%%%%%%%%%%%%%%

Independent of projection, central dark matter densities are overestimated in
all Schwarzschild models. Most likely, this reflects our choice
of NFW-profiles for the halos of the models (cf. Sec.~\ref{sub:basic}).
In principle, an overestimation of the central dark matter density could
cause an underestimation of the luminous mass for compensation.
Near the centre, where the dark matter excess is most prominent, the luminous matter
is, however, still a factor of ten larger than the dark matter density 
(at $0.1 \, \reff$, for example). We therefore do not expect the central 
over-prediction of dark matter to be important for the recovery
of $\Upsilon$. Moreover, while the dark matter excess is 
projection-{\it independent}, the underestimation of 
$\mlrat$ is projection-{\it dependent}.

Nevertheless we have additionally calculated a set of logarithmic (LOG) halos for 
one merger remnant (OBLATE; the grid used to sample the halos is 
described in \citealt{Tho07}). In case of the X and Y-projection LOG-halos allow
a slightly better fit than NFW-halos (cf. bottom-right panels of Figs.~\ref{rho:x} and
\ref{rho:y}). As will become clear from the discussions in the next sections, these
models are in no respect systematically different from the
models of other remnants which are calculated with NFW-halos. 

We have also calculated logarithmic halos for the X-models of the TRIAX, PROLATE and ELONG 
remnants. In these cases as well as concerning the Z-model of the OBLATE remnant, LOG halos do not 
provide better fits. As a consequence, considering LOG-halos does not change $\mlfit$ of
this models. It follows that the particular choice 
of the halo profile (between NFW and LOG) has little effect on 
our results. It merely influences the match to the dark matter component 
in a spatial region, where dark matter is a minor contributor to the total mass.

Towards the outer edge of the kinematical data ($\reff \la r \la 2 \, \reff$), mass densities 
of Schwarzschild models and merger remnants agree reasonably well. 
This holds for the total mass, as well as for
luminous and dark components, separately. 
Around $1 - 2 \, \reff$, integrated total masses of Schwarzschild models are accurate
to about 20 percent. The Z-models of the most flattened remnants (FLAT, ELONG, OBLATE) 
are deviant by up to 40-50 percent.

%%%%%%%%%%%%%%%%%%%%%%%%%%%%%%%%%%%%%%%%%%
% total mass: bestfit
%%%%%%%%%%%%%%%%%%%%%%%%%%%%%%%%%%%%%%%%%%

%%%%%%%%%%%%%%%%%%%%%%%%%%%%%%%%%%%%%%%%%%
% total mass: bestfit
%%%%%%%%%%%%%%%%%%%%%%%%%%%%%%%%%%%%%%%%%%

\section{Velocity anisotropies in remnants and models}
\label{section:intmom}
We now consider in more detail 
the internal dynamical structure of the merger remnants 
and how it is represented by our Schwarzschild fits. 

\subsection{Anisotropy profiles}
\label{sub:aniso}
\begin{figure}
\includegraphics[width=84mm,angle=0]{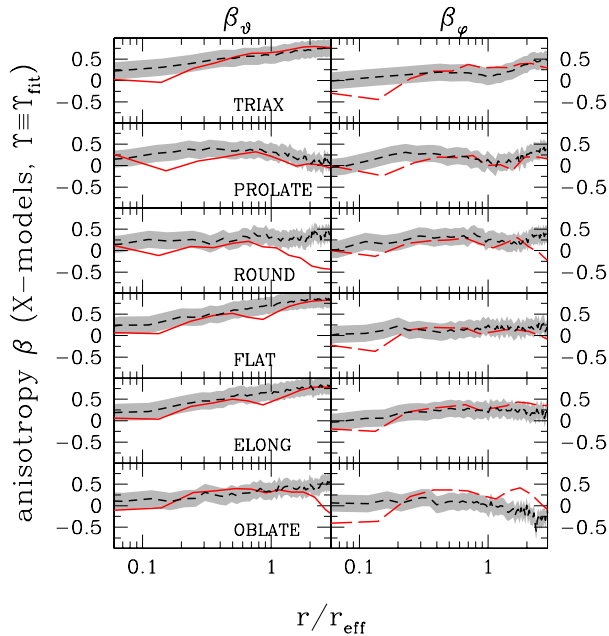}
\caption{Comparison of velocity anisotropies. Left: meridional $\betatheta$ of
Schwarzschild models (solid/coloured) and of merger remnants (short-dashed/black);
right: azimuthal $\betaphi$ of Schwarzschild models (long-dashed/coloured) and
of merger remnants (short-dashed/black); grey: $\pm 0.2$ absolute deviations 
from merger values for comparison. The figure shows results of X-models.}
\label{beta:x}
\end{figure}

\begin{figure}
\includegraphics[width=84mm,angle=0]{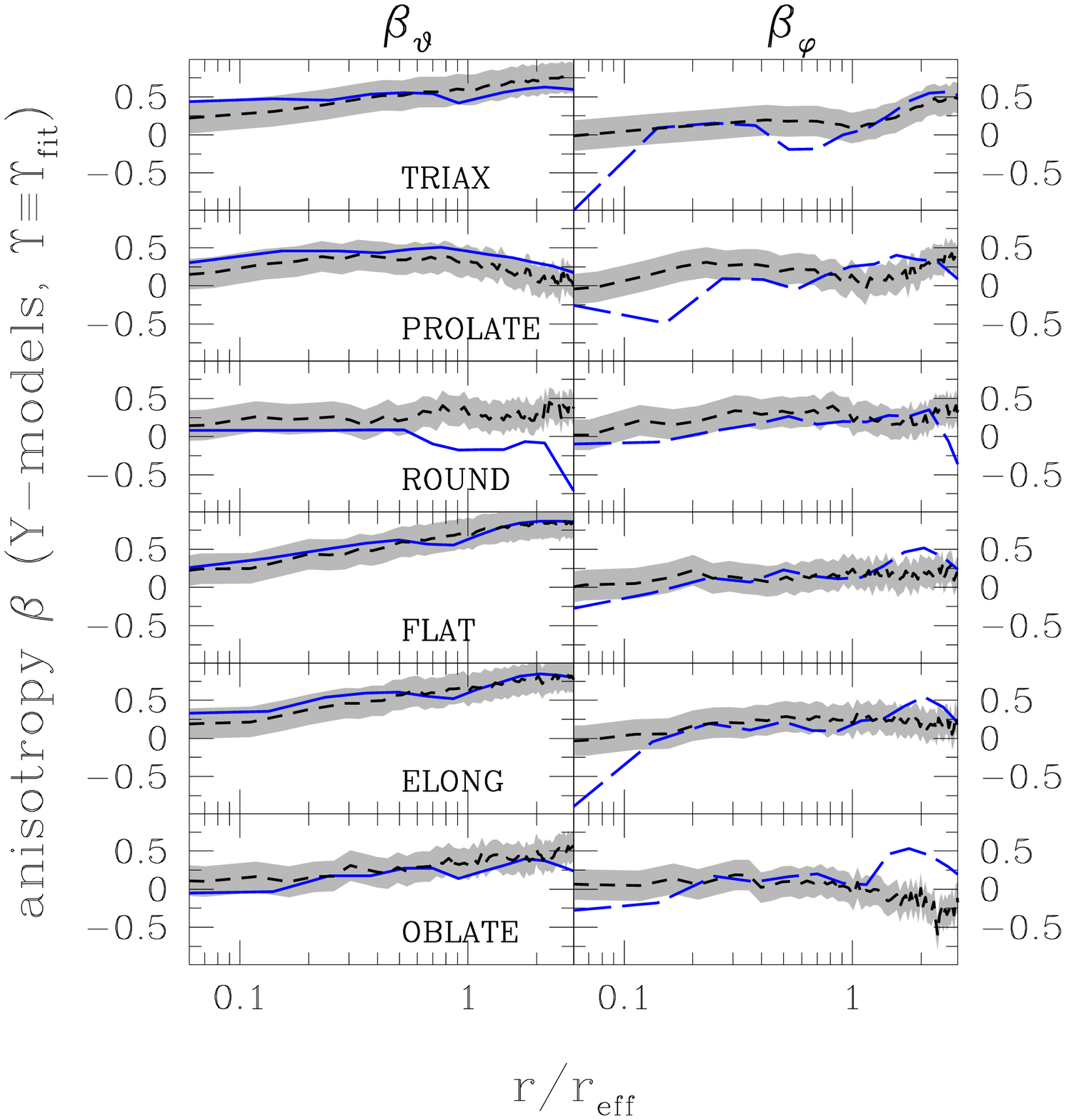}
\caption{As Fig.~\ref{beta:x}, but for Y-models.}
\label{beta:y}
\end{figure}

\begin{figure}
\includegraphics[width=84mm,angle=0]{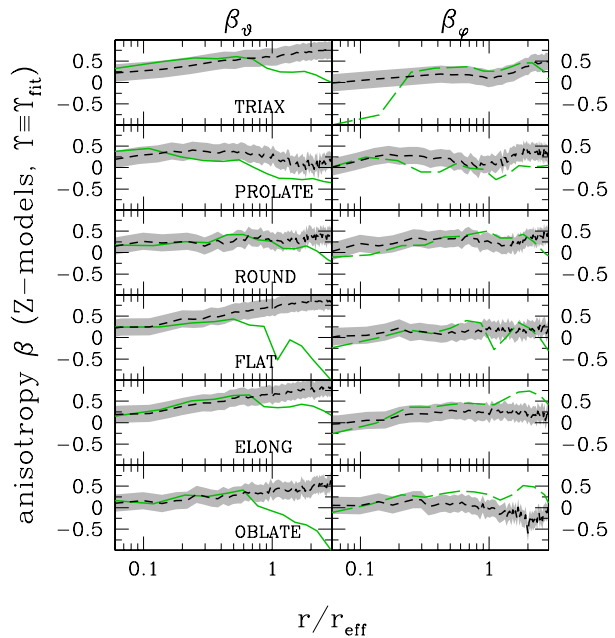}
\caption{As Fig.~\ref{beta:x}, but for Z-models.}
\label{beta:z}
\end{figure}

Figs.~\ref{beta:x} - \ref{beta:z} compare profiles of meridional
anisotropy
\begin{equation}
\betatheta \equiv 1 - \frac{\sigma_\vartheta^2}{\sigma_r^2}
\end{equation}
and azimuthal anisotropy
\begin{equation}
\betaphi \equiv 1 - \frac{\sigma_\varphi^2}{\sigma_r^2}
\end{equation}
of Schwarzschild models and merger remnants. 
We use spherical coordinates
$r$, $\vartheta$ and $\varphi$, oriented along the principal axes such that 
$\varphi$ is the azimuth in the $(X,Y)$-plane and 
$\vartheta$ is the latitude. 
The velocity dispersions are luminosity weighted spherical averages.

%%%%%%%%%%%%%%%%%%%%%%%%%%%%%%%%%%%%%%%%%%
% beta: X
%%%%%%%%%%%%%%%%%%%%%%%%%%%%%%%%%%%%%%%%%%

%%%%%%%%%%%%%%%%%%%%%%%%%%%%%%%%%%%%%%%%%%
% beta: Y
%%%%%%%%%%%%%%%%%%%%%%%%%%%%%%%%%%%%%%%%%%

%%%%%%%%%%%%%%%%%%%%%%%%%%%%%%%%%%%%%%%%%%
% beta: Z
%%%%%%%%%%%%%%%%%%%%%%%%%%%%%%%%%%%%%%%%%%

In Figs.~\ref{beta:x} - \ref{beta:z}  $\Delta \beta = \pm 0.2$ margins are highlighted. The
choice of these margins is arbitrary, and is only to guide a quantification of
deviations between mergers and models. In most cases these are 
smaller than $\Delta \beta < 0.2$. But there are some outliers (mostly
among Z-models).
As a general rule, X and Y models fit better with the
intrinsic properties of the merger remnants than Z-models.

The mismatch of the Z-models is partly due to the fact that the bestfit 
Schwarzschild models are always achieved for an inclination $i=90\degr$ (cf.
Sec.~\ref{sub:incnote}). Apart from the related mismatch in
the deprojection it raises a complication concerning the comparison of the 
internal moments:
in the Z-projection of a merger remnant, according to the above 
definitions, the azimuth $\varphi$ appears as the angle in 
the plane of the sky. In the Schwarzschild models, however, 
$\vartheta$ as defined above is the angle in the plane of the sky, 
as long as $i=90\degr$.

%%%%%%%%%%%%%%%%%%%%%%%%%%%%%%%%%%%%%%%%%%
% beta: Z - exchanged
%%%%%%%%%%%%%%%%%%%%%%%%%%%%%%%%%%%%%%%%%%
\begin{figure}
\includegraphics[width=84mm,angle=0]{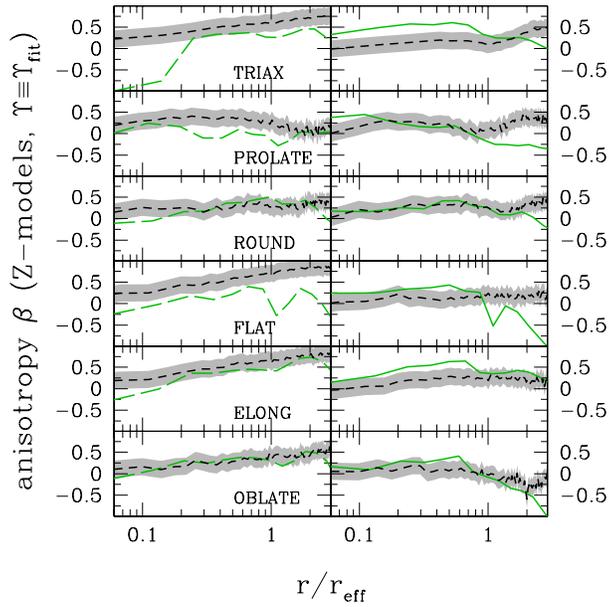}
\caption{As Fig.~\ref{beta:z}, but Schwarzschild $\betaphi$ (long-dashed) are compared to
merger remnant $\betatheta$ (left panels) and Schwarzschild $\betatheta$ (solid) 
are compared with $\betaphi$ of the remnants (right panels).}
\label{beta:z2}
\end{figure}

Much of the discrepancies between Schwarzschild models and merger
remnants can be attributed to these different coordinate definitions. To show
this Fig.~\ref{beta:z2}
replicates the same profiles as Fig.~\ref{beta:z}, but 
$\betaphi$ of the Schwarzschild models is now compared to $\betatheta$ of the merger
remnants and vice versa. The differences between the
mergers and the models are significantly smaller in Fig.~\ref{beta:z2} than in 
Fig.~\ref{beta:z}, especially among the most strongly flattened remnants.

\subsection{Interpretation in terms of orbits}
\label{sub:axialorbits}
The remaining deviations between the anisotropy profiles of 
merger remnants and their corresponding Schwarzschild fits are most likely
related to the different orbit families supported by $N$-body potentials
on the one side and axisymmetric potentials on the other.

\begin{figure}
\includegraphics[width=84mm,angle=0]{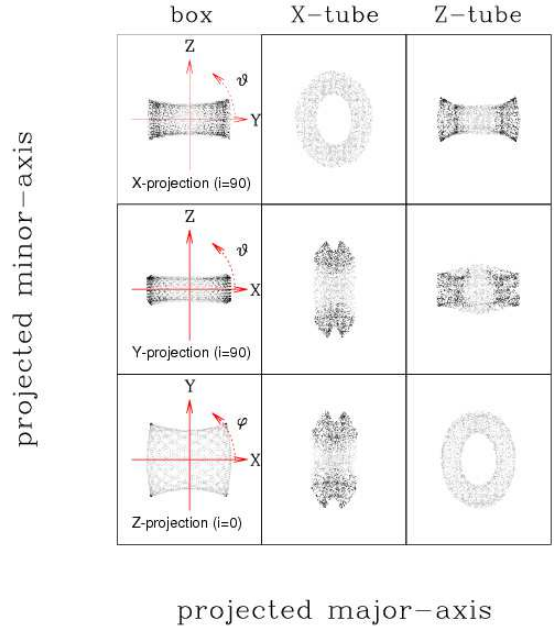}
\caption{Numerically integrated orbits in an $N$-body potential. From left
to right: box orbit, X-tube and Z-tube. From top to bottom: X, Y and Z-projection. 
Coordinate definitions are illustrated on the left hand side. 
Black: $\vlos>\vperp$; light grey: $\vlos<\vperp$ (details in the
text).}
\label{orbits:triax}
\end{figure}

\begin{figure}
\includegraphics[width=84mm,angle=0]{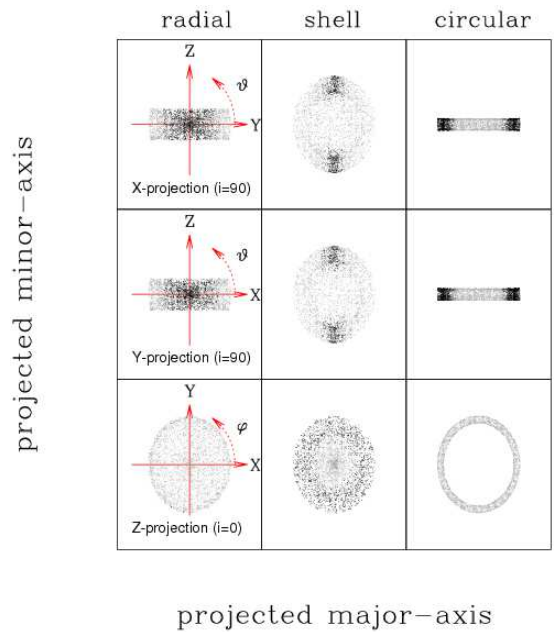}
\caption{As Fig.~\ref{orbits:triax}, but for the case of an axisymmetric 
potential. From left to right: radial, shell-like and nearly circular orbit.}
\label{orbits:axial}
\end{figure}

Figs.~\ref{orbits:triax} and \ref{orbits:axial} review 
principal projections of orbits numerically
integrated in an $N$-body potential (Fig.~\ref{orbits:triax}) 
and in an axisymmetric potential
(Fig.~\ref{orbits:axial}). Regions with $\vlos>\vperp$ are plotted dark and regions
with $\vlos<\vperp$ are plotted grey. Thereby $\vlos$ is the absolute line-of-sight velocity in
the given projection and $\vperp$ is the absolute magnitude of the velocity
perpendicular to the line-of-sight. In dark areas most of the kinetic energy 
of an orbit is directed towards the observer, whereas in
grey areas most of the kinetic energy is in motion perpendicular to the line-of-sight.

The tangential anisotropy of the X-model for the PROLATE remnant can be explained
by the dominance of X-tubes in this remnant. According to Fig.~\ref{orbits:triax}
their round appearance in the X-projection makes them most similar to
the edge-on projection of axisymmetric shell orbits (cf. Fig.~\ref{orbits:axial}). 
The latter, in turn, have
large $\sigt$ and low $\sigr$ and cause the tangential anisotropy in the Schwarzschild
model. 

Likewise, the similarity of $\betatheta$ in Z-models with $\betaphi$ of the merger
remnants discussed at the end of Sec.~\ref{sub:aniso} can be explained by the
fact that the dominant orbits in the outer parts of merger remnants, Z-tubes, appear nearly 
round when seen face-on. Again, they are likely mapped onto axisymmetric shell
orbits, with the same consequence for the model's anisotropy as discussed for the X-model
of the PROLATE remnant.

%%%%%%%%%%%%%%%%%%%%%%%%%%%%%%%%%%%%%%%%%%
% orbits: triaxial
%%%%%%%%%%%%%%%%%%%%%%%%%%%%%%%%%%%%%%%%%%

%%%%%%%%%%%%%%%%%%%%%%%%%%%%%%%%%%%%%%%%%%
% orbits: axial
%%%%%%%%%%%%%%%%%%%%%%%%%%%%%%%%%%%%%%%%%%

\section{Modeling uncertainties}
\label{section:virial}
Up to now we have presented the viewing-angle dependency of the masses and anisotropies
which we reconstructed with our axisymmetric orbit models. 
The behaviour of the anisotropy could be
explained by the way in which projected properties of major remnant orbit families 
match with different axisymmetric orbits. The recovered masses
are less easy to understand, in particular the low $\mlfit$. This section and the
following two Secs.~\ref{sec:totm} and \ref{sec:lumml} are aimed to discuss the
mass recovery in more detail. We start this discussion here by investigating
whether the projection-dependency of the mass-recovery in our
axisymmetric dynamical models is an artifact of the modeling machinery.

\subsection{Stationarity assumption}
As it has been stated in the introduction, the Schwarzschild method is based on 
Jeans' theorem and the assumption that the object to be modelled is stationary.
Non-stationarity of the merger remnants can have a significant influence
on the recovered masses. For example, if a remnant contracts because the ratio 
of its kinetic and potential energies is smaller than in virial equilibrium then a
stationary model could deliver a mass smaller than the true one. Likewise,
if a remnant expands then the recovered mass-to-light ratio could be too large.

Stationarity or virial equilibrium, respectively, implies that
\begin{equation}
\label{eq:vireq}
2 T_{ij} = -W_{ij},
\end{equation}
where $T_{ij}$ denotes the kinetic energy tensor and $W_{ij}$ denotes the potential energy 
tensor. Equation (\ref{eq:vireq}) holds for the luminous and the dark components
separately, if both are stationary. In the following we only consider the luminous component.
The calculation of its kinetic and potential energies is straight forward:
\begin{equation}
\label{tij}
T_{ij} = \frac{1}{2} \sum_{\alpha=1}^{N_{l}} m_{\alpha} \dot{x}^{(\alpha)}_{i}
\dot{x}^{(\alpha)}_{j},
\end{equation}
where the sum extends over all the $N_{l}$ luminous particles of the merger remnant
(with mass $m_{\alpha}$ each) and $\dot{x}^{(\alpha)}_{i}$ is the $i$-component of the
velocity of particle $\alpha$.
The potential energy of the luminous component comprises the two contributions
\begin{equation}
\label{wij}
W_{ij} = W_{ij}^{(ll)} + W_{ij}^{(ld)},
\end{equation}
where
\begin{equation}
W_{ij}^{(ll)} = -G \sum_{\alpha=1}^{N_{l}-1} m_{\alpha} 
\sum_{\beta=\alpha+1}^{N_{l}} m_{\beta} 
\frac{\xi^{(\alpha \beta)}_{ij}}{r^3_{\alpha \beta}}
\end{equation}
and
\begin{equation}
\xi^{(\alpha \beta)}_{ij} \equiv 
(x^{(\alpha)}_{i}-x^{(\beta)}_{i})(x^{(\alpha)}_{j}-x^{(\beta)}_{j}).
\end{equation}
The sum extends over luminous particles only and $r_{\alpha \beta}$ is the distance between 
particles $\alpha$ and $\beta$, respectively. The contribution of dark matter comes in through
\begin{equation}
W_{ij}^{(ld)} = -G \sum_{\alpha=1}^{N_{l}} m_{\alpha} 
\sum_{\beta=1}^{{\cal N}_{d}} {\cal M}_{\beta} 
\frac{\xi^{(\alpha \beta)}_{ij}}{r^3_{\alpha \beta}}.
\end{equation}
In the last sum ${\cal N}_{d}$ denotes the total number of dark matter particles
in the remnant (with mass ${\cal M}_{\beta}$ each). In virial equilibrium total
potential energy $W = \sum W_{ii}$ and total kinetic energy $T = \sum T_{ii}$ 
obey $2T/|W| = 1$

The six modelled merger remnants have $2T/|W| \in [0.960,0.981]$ with the lowest value
for the FLAT remnant and the largest value for the OBLATE one. Thus, the
remnants are very close to virial equilibrium and we expect that the assumption of
stationarity in the models should affect the models' masses at most at the
5 percent level. Hence it is not the main driver for the low $\mlrat$ in our
models. In addition $\mlrat$ is projection-dependent whereas $2T/|W|$ is
projection-independent.

\subsection{Phase space sampling}
\label{sub:norb}
Another potential uncertainty in the modeling procedure is the difference in phase-space
structure of merger remnants on the one side and Schwarzschild models on the other: 
while the remnants are composed of a 
relatively large number of particles, each sampling a different orbit at one point, the
Schwarzschild model is composed of a relatively low number of orbits, each sampled
very densely (we use about $10^5$ time-steps for each orbit integration).

Concerning the sampling of the orbit (the time-step and total integration time used),
our implementation of Schwarzschild's method has been successfully tested on
continuous analytical dynamical models, like for example a Hernquist sphere \citep{Tho04}.
To check whether a similarly good agreement
can be achieved when modeling $N$-body systems, we have repeated
the tests with discrete $N$-body realisations as modeling targets (cf.
Sec.~\ref{section:hernquist}). The small uncertainties that we find imply that 
differences in phase-space structure are negligible.

\subsection{Chaotic orbits}
\label{sub:chaotic}
In the implementation
of Schwarzschild's method applied here (as in most others) 
chaotic orbits are treated in the same way as regular orbits. 
This is not necessary in Schwarzschild models, but makes
them computationally more efficient. A chaotic region in phase-space
at fixed $E$ and $L_z$ has to have a constant phase-space density according to Jeans'
theorem. If such a region is represented by one (chaotic) orbit in the library, then the
method works fine. However, it may happen that the (finite) integration time 
of the first orbit 
that is launched in the chaotic region is insufficient to cover the accessible phase-space 
volume entirely. Then the program will launch one or more other orbits to 
fill up the rest of the
chaotic region. It is then likely (although not necessary) that these fractional
orbits will have different phase-space densities in the final model. 
As a consequence, the model no longer satisfies Jeans' theorem. Several suggestions have
been made to overcome this problem \citep[e.g.][]{Mer96F,Haf00}.

Since
the main consequence of chaos in phase-space is to break
the stationarity of the Schwarzschild models, it should manifest itself in deviations
from virial equilibrium and, thus, can be quantified by
evaluating the virial equations of the Schwarzschild models. To calculate the kinetic 
energy tensors $T_{ij}$ and $W_{ij}$ defined in equations (\ref{tij}) and (\ref{wij}) 
we have constructed $N$-body realisations of each bestfit Schwarzschild model 
as described in App.~\ref{app:nbodyreal}.

For the obtained virial ratios we find $2T/|W| \approx 1$ to within 15 percent. 
This limits the amount of chaos in our orbit libraries. 
Deviations from $2T/|W| = 1$ are not correlated with viewing-angle but with 
halo-concentration, which is an artifact of the $N$-body realisation and 
further discussed in App.~\ref{app:nbodyreal}. 
Thus, the margins for intrinsic non-stationarity are even smaller than the above
quoted 15 percent. The related uncertainties
are not sufficient to explain the trends in the mass recovery.

\section{The viewing angle dependency of the total mass recovery}
\label{sec:totm}
The last section has ruled out modeling uncertainties as the main source for the magnitude
and projection-dependency of the mass recovery. We now investigate whether the different 3-dimensional shapes of models and merger remnants
are the main driver of this dependency. 

Globally, mass and kinetic energy are linked by the 
virial theorem, $2T = |W| = \kappa \, M$, where $\kappa$ depends on the density profile.
Accordingly, reconstructed masses $M^{(\mathrm{fit})}$ and input masses $M^{(\mathrm{in})}$ 
are related via 
\begin{equation}
\label{virrat:princ}
\frac{M^{(\mathrm{fit})}}{M^{(\mathrm{in})}} = 
\frac{\kappa^{(\mathrm{in})}}{\kappa^{(\mathrm{fit})}}
\times
\frac{T^{(\mathrm{fit})} + T^{(\mathrm{fit})}_\mathrm{DM}}{T^{(\mathrm{in})} + T^{(\mathrm{in})}_\mathrm{DM}},
\end{equation}
with $T\fit$ and $T\fit_\mathrm{DM}$ denoting the kinetic energy of luminous and dark matter 
in the Schwarzschild fit and $T\inp$ and $T\inp_\mathrm{DM}$ being
the analogue quantities of the merger
remnant, respectively. Since it is basically $T^{(\mathrm{fit})}$ that is
constrained by the LOSVD-fits, it is instructive to study first the energy budget
of the Schwarzschild models in comparison to the merger remnants. Then, 
equation (\ref{virrat:princ}) can be used to evaluate the implications on the
reconstructed masses. 

Because the virial theorem relates energies to {\it total} masses this
section deals with the viewing-angle dependency of the total mass recovery. The
mass-to-light ratio of the stellar component will be discussed separately in the next
Sec.~\ref{sec:lumml}.

\subsection{Energy budget of the  Schwarzschild models}
\label{subsec:energybudget}
In the merger remnants, by definition of the axes (cf. Sec.~\ref{section:sample})
$T_{xx} \ge T_{yy} \ge T_{zz}$, whereas oblate axial symmetry implies
$T_{xx} \equiv T_{yy} \ge T_{zz}$ in the Schwarzschild fits\footnote{We will
only consider the diagonal elements $T_{ii}$ in the following, because in the
merger remnants as well as in the $N$-body realisations of the Schwarzschild models
the other components are at least two orders of magnitude lower and, thus,
energetically negligible.}. In the following, it is convenient to switch from
axis-labels referring to the intrinsic shape of the remnant (e.g. X, Y and Z as defined
in Sec.~\ref{section:sample}) to projection-based labels: let us define the kinetic energy 
$T_\mathrm{los}$ as the energy parallel to the
axis that points towards the observer, $T_\mathrm{maj}$ as the energy parallel to the
axis that projects to the apparent major-axis and $T_\mathrm{min}$ as the energy
directed parallel to the apparent minor-axis.

%%%%%%%%%%%%%%%%%%%%%%%%%%%%%%%%%%%%%%%%%%
% tkin: los
%%%%%%%%%%%%%%%%%%%%%%%%%%%%%%%%%%%%%%%%%%
\begin{figure}
\includegraphics[width=84mm,angle=0]{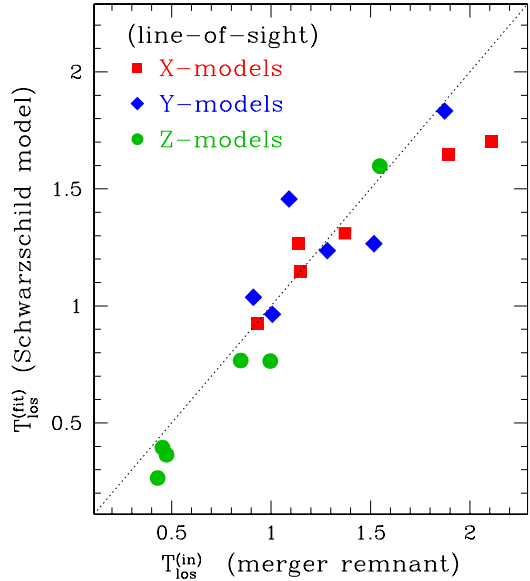}
\caption{Correlation between kinetic energies of merger remnants (in)
and Schwarzschild models (fit; along the line-of-sight). Dotted line: one-to-one
relation for comparison.}
\label{tkin:los}
\end{figure}

%%%%%%%%%%%%%%%%%%%%%%%%%%%%%%%%%%%%%%%%%%
% virial flattening
%%%%%%%%%%%%%%%%%%%%%%%%%%%%%%%%%%%%%%%%%%
\begin{figure}
\includegraphics[width=84mm,angle=0]{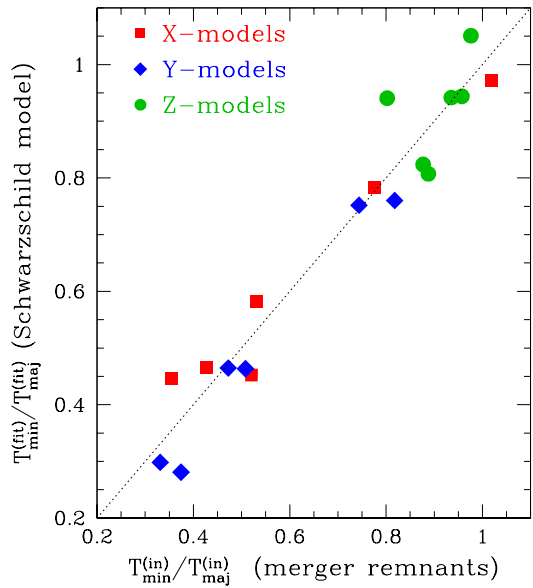}
\caption{Correlation between the ratio $T_\mathrm{min}/T_\mathrm{maj}$ of the two
transversal energies in merger remnants (in) and Schwarzschild models (fit), respectively.
Dotted: one-to-one relation for comparison.}
\label{tkin:flat}
\end{figure}

Figs.~\ref{tkin:los} and \ref{tkin:flat} show that the
line-of-sight energy $T_\mathrm{los}$ and the ratio $T_\mathrm{min}/T_\mathrm{maj}$ 
of the two transversal energies are well recovered by the Schwarzschild models.
This could have been expected since $T_\mathrm{los}$ is the energy mapped by the
projected kinematics. A mismatch in $T_\mathrm{los}$ should manifest itself in the kinematic 
fits. Some scatter remains, however, because we do not assume full sky-coverage with 
kinematic data.
That the Schwarzschild models match also with $T_\mathrm{min}/T_\mathrm{maj}$ of the 
remnants is plausible, because this energy ratio determines the shape (e.g. \citealt{Bin87}).
And the shape is accounted for in the Schwarzschild fits through 
the deprojected luminosity density, which is used as a boundary condition for our models.

The two relations revealed by Figs.~\ref{tkin:los} and \ref{tkin:flat} 
have several important implications for the energy budget and, thus, the recovered 
masses of the Schwarzschild models.

To see this, let's assume for simplicity that
\begin{equation}
\label{rel:los}
T_\mathrm{los}\fit \equiv T_\mathrm{los}\inp
\end{equation}
and
\begin{equation}
\label{rel:shape}
\frac{T_\mathrm{min}\fit}{T_\mathrm{maj}\fit} \equiv \frac{T_\mathrm{min}\inp}{T_\mathrm{maj}\inp}
\end{equation}
(in other words, we replace the two approximate one-to-one correlations of Figs.~\ref{tkin:los} and \ref{tkin:flat}
by identities). By symmetry $T_\mathrm{maj}\fit \equiv T_\mathrm{los}\fit$ in the
Schwarzschild fits and it follows
\begin{equation}
\label{rel:maj}
T_\mathrm{maj}\fit \equiv T_\mathrm{los}\inp.
\end{equation}
Moreover, according to equation (\ref{rel:shape})
\begin{equation}
T_\mathrm{min}\fit \equiv T_\mathrm{maj}\fit \frac{T_\mathrm{min}\inp}{T_\mathrm{maj}\inp},
\end{equation}
and by equation (\ref{rel:maj}):
\begin{equation}
\label{rel:min}
T_\mathrm{min}\fit \equiv T_\mathrm{los}\inp \frac{T_\mathrm{min}\inp}{T_\mathrm{maj}\inp}.
\end{equation}
Hence, the two relations (\ref{rel:los}) and (\ref{rel:shape}) uniquely link the 
three relevant components of the kinetic energy tensor of the Schwarzschild model 
to the energy components of the merger remnant via 
equations (\ref{rel:los}), (\ref{rel:maj}) and (\ref{rel:min}). Note that this holds only
for edge-on models. If a model is calculated at $i=0\degr$, then 
$T_\mathrm{min}\fit \equiv T_\mathrm{maj}\fit$ by symmetry. In this case, even the identity
$T_\mathrm{los}\fit = T_\mathrm{los}\inp$ does not constrain the two ratios 
$T_\mathrm{los}\fit/T_\mathrm{min}\fit$ and $T_\mathrm{los}\fit/T_\mathrm{maj}\fit$, 
respectively. This reflects the uncertainty in the flattening along the line-of-sight
of the model.

To express the expected energy budget for a Schwarzschild fit of a given
merger projection more quantitatively, it is necessary to figure out to which intrinsic axes
$T_\mathrm{los}$, $T_\mathrm{maj}$ and $T_\mathrm{min}$ correspond. Simple algebra leads
to Tab.~\ref{tab:energies}, in which Schwarzschild model energies relative
to remnant energies are given explicitly for each projection.
The table allows to draw some important conclusions.
In fact, for X-models the energy in the
Schwarzschild model has to be larger than in the remnant, $\trat \ge 1$, unless
the remnant is oblate-axisymmetric. Contrary, 
in Y and Z-models $\trat \le 1$ and the energy in the Schwarzschild model has to be
smaller than in the remnant.
\begin{table}
\begin{center}
\begin{tabular}{lcccc}
prj. & $\frac{T_\mathrm{los}\fit}{T_\mathrm{los}\inp}$ & $\frac{T_\mathrm{maj}\fit}{T_\mathrm{maj}\inp}$ & $\frac{T_\mathrm{min}\fit}{T_\mathrm{min}\inp}$
& $\frac{T\fit}{T\inp}$\\
\hline
(1) & (2) & (3)    & (4) & (5) \\
\hline 
X  & $1$ & $\frac{T_{xx}\inp}{T_{yy}\inp}\ge 1$ & $\frac{T_{xx}\inp}{T_{yy}\inp}\ge 1$ & $\ge 1$\\
\\
Y  & $1$ & $\frac{T_{yy}\inp}{T_{xx}\inp}\le 1$ & $\frac{T_{yy}\inp}{T_{xx}\inp}\le 1$ & $\le 1$\\
\\
Z  & $1$ & $\frac{T_{zz}\inp}{T_{xx}\inp}\le 1$ & $\frac{T_{zz}\inp}{T_{xx}\inp}\le 1$ & $\le 1$\\
\hline 
\end{tabular}
\caption{Ratios of kinetic energies of Schwarzschild fits and merger remnants 
according to equations (\ref{rel:los}), (\ref{rel:maj}) and (\ref{rel:min}). (1) projection; (2) line-of-sight
energy; (3) projected long-axis energy; (4) projected short-axis energy; (5) total energy ($T = T_\mathrm{los} + T_\mathrm{maj} + T_\mathrm{min}$).}
\label{tab:energies}
\end{center}
\end{table}

%%%%%%%%%%%%%%%%%%%%%%%%%%%%%%%%%%%%%%%%%%
% total mass versus kinetic energy
%%%%%%%%%%%%%%%%%%%%%%%%%%%%%%%%%%%%%%%%%%
\begin{figure}
\includegraphics[width=84mm,angle=0]{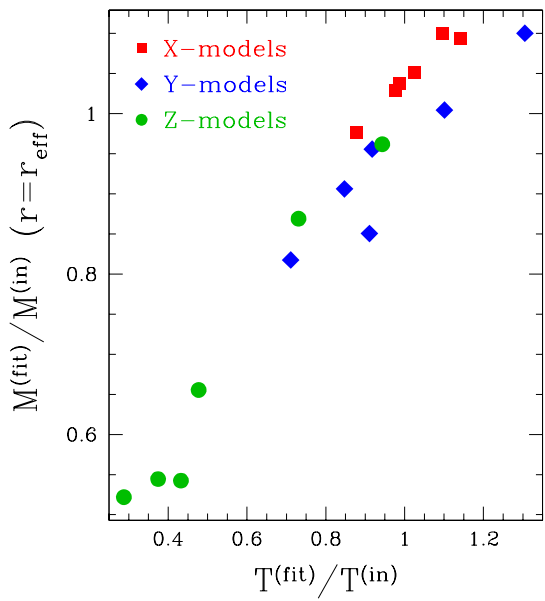}
\caption{Accuracy of reconstructed total (luminous + dark) mass inside
$\reff$ versus ratio $\trat$ of total kinetic energies in Schwarzschild fits
and merger remnants.}
\label{totmekin}
\end{figure}

\subsection{Mass budget of the  Schwarzschild models}
\label{subsec:massbudget}
What are the consequences for the masses of the Schwarzschild fits? 
For simplicity let's first assume that $\kappa\fit=\kappa\inp$.
Then, equation (\ref{virrat:princ}) predicts that the ratio 
$M\fit/M\inp$ of reconstructed and input masses equals the ratio
of the corresponding total kinetic energies (luminous + dark). However,
due to the lack of kinematic information about the constituents of the dark halo, 
the models have no access to the dark matter kinetic energy of the remnants and 
$T\fit_\mathrm{DM}$ is not constrained. As a consequence, the results on the energy budget 
are significant for the comparison of reconstructed and input masses only inside a radius 
where luminous matter dominates, for example
inside $\reff$. There, the contribution of the dark matter kinetic energy is small.
Assuming that its contribution is in fact negligible, then 
equation (\ref{virrat:princ}) and the last column of Tab.~\ref{tab:energies} 
imply $\mrat \ge 1$ inside $\reff$ for X-models and $\mrat \le 1$ for Y and Z-models. The amount by which $\mfit$ exceeds 
$\mmin$ in X-models should be comparable to the amount by which $\mfit$ is reduced relative 
to $\mmin$ in Y-models and masses of Z-models should be smaller than those of Y-models.
Fig.~\ref{totmekin} relates the mass ratio $\mrateff$ (at the effective radius) 
to kinetic energies $\trat$ and globally confirms the just discussed trends between 
reconstructed masses and kinetic energies. 

Hence, at the half-light radius,
masses of our Schwarzschild fits are closely related to the
energy budget of the models. The energy, in turn, derives from 
the fit to the kinematics and the shape of the modeling target via relations
(\ref{rel:los}) and (\ref{rel:shape}). For edge-on systems, the restrictions 
imposed by the assumption of axial symmetry together with these two relations already uniquely
determine the luminous kinetic energy and, hence, the mass budget of the axisymmetric 
fits. Thereby it turns out that X-models have to 
overestimate the true mass, while Y and Z-models have to underestimate it.

We close this section with a few further comments on Fig.~\ref{totmekin}.
According to equation (\ref{virrat:princ}),
the relation between reconstructed masses and reconstructed energies can be tilted 
with respect to a one-to-one relation if the $\kappa$-ratio varies systematically over
the sample. In Sec.~\ref{subsec:depro} we have discussed systematic variations of the
deprojections with viewing-angle which could cause the tilt with respect to the 
one-to-one case revealed by Fig.~\ref{totmekin}. In addition there
is a correlation of the dark halo properties in the models with remnant projection, which 
could also contribute to the tilt in Fig.~\ref{totmekin} (cf. Sec.~\ref{sub:roledm}).
On top of that, the fact that we compare luminous kinetic energies with masses inside
$\reff$ adds to the uncertainties in the step from $\trat$ to $\mrateff$ and can
also tilt the relation or increase its scatter. Other sources of scatter
are scatter in the relations (\ref{rel:los}) and (\ref{rel:shape}). For example, the two 
Y-models with $\mrateff>1$ in Fig.~\ref{totmekin} 
correspond to the two Y-models above the one-to-one relation in Fig.~\ref{tkin:los}.

\section{Central remnant structure and the luminous mass-to-light ratio}
\label{sec:lumml}

In this section we discuss the 
results for the modelled luminous mass-to-light ratios. In contrast to the total masses the 
disagreement between models and mergers cannot be simply traced to the energy budget. 

\begin{figure}
\includegraphics[width=84mm,angle=0]{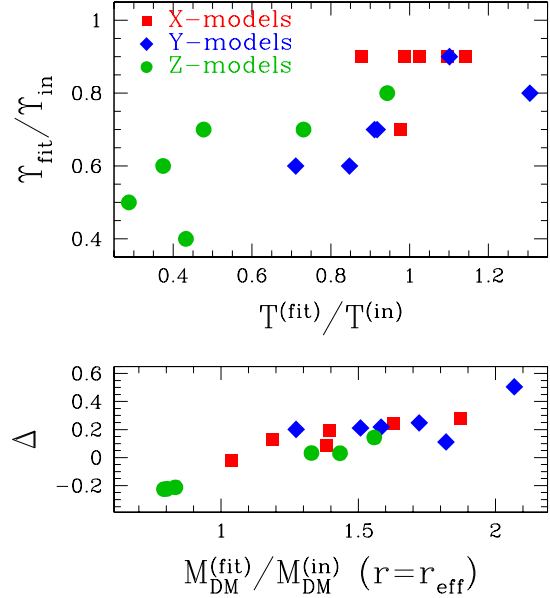}
\caption{Top: Accuracy of reconstructed luminous mass-to-light ratio $\mlrat$
versus ratio $\trat$ of total kinetic energies in Schwarzschild fits
and merger remnants. Bottom: Deviations $\Delta \equiv \trat - \mlrat$ 
from a one-to-one correlation in the
top panel versus accuracy of reconstructed dark mass (inside $\reff$).}
\label{upsekin}
\end{figure}

The upper panel of Fig.~\ref{upsekin} shows $\mlrat$ versus total kinetic energy $\trat$. 
The scatter in Fig.~\ref{upsekin} is much larger than in the corresponding 
Fig.~\ref{totmekin} which deals with total masses. As expected
from the tightness of the correlation in Fig.~\ref{totmekin}, 
deviations from a one-to-one correlation between $\mlrat$ and $\trat$ are 
correlated with the dark matter content of the fits: where $\mlrat$ is too low,
the dark matter in the models overestimates the dark matter in the remnant and vice
versa. A larger scatter in Fig.~\ref{upsekin} than in
Fig.~\ref{totmekin} is not surprising because the reconstruction of the mass decomposition 
is less certain than the reconstruction of the total mass: there is some 
freedom in the modeling to shift mass from the luminous to the dark component 
(and vice versa) without changing the fit significantly. 
There are however two striking trends: (1) $\mlfit$ is generally smaller than $\mlin$ and (2) 
at a given value of $\trat$ Y-models suffer from
a slightly stronger underestimation of $\Upsilon$ than models of other projections.

%%%%%%%%%%%%%%%%%%%%%%%%%%%%%%%%%%%%%%%%%%
% luminous mass versus kinetic energy
%%%%%%%%%%%%%%%%%%%%%%%%%%%%%%%%%%%%%%%%%%

\begin{figure}
\includegraphics[width=84mm,angle=0]{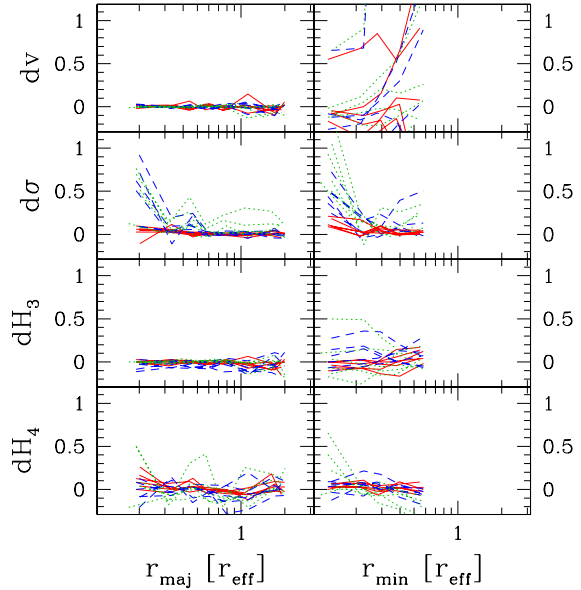}
\caption{As Fig.~\ref{losvd-match:bestfit}, but for Schwarzschild models with
$\Upsilon \equiv \mlin$.}
\label{losvd-match:mlin}
\end{figure}

The systematics in the lower panel of Fig.~\ref{upsekin} 
indicate that $\mlfit$, unlike the total mass, is not 
merely set by the total kinetic energy
but must instead depend on something else. To investigate this
further, we have redetermined bestfit Schwarzschild models (with $2 \times 3500$ orbits;
cf. Sec.~\ref{subsec:bf}) under the condition that the
luminous mass-to-light ratio is fixed to the true value of the mergers, 
$\Upsilon \equiv \mlin$. The corresponding fits
are illustrated in Fig.~\ref{losvd-match:mlin} in the same way as fits with
optimised $\mlfit$ are shown in Fig.~\ref{losvd-match:bestfit}. The most important result
is that Y and Z-models constrained to
have the true mass-to-light ratio fail to fit the central kinematics: they predict
too large central velocity dispersions. X-models in Fig.~\ref{losvd-match:mlin} are only
marginally different from those in Fig.~\ref{losvd-match:bestfit}, because most X-models yield
$\mlfit \approx \mlin$ to within ten percent in any case.

%%%%%%%%%%%%%%%%%%%%%%%%%%%%%%%%%%%%%%%%%%
% losvd-match: Upsilon = 1
%%%%%%%%%%%%%%%%%%%%%%%%%%%%%%%%%%%%%%%%%%

\subsection{The relation between flattening and central kinematics}
\label{sub:generalflat}
That axisymmetric models of Y and Z-projections with $\mlfit = \mlin$ 
overpredict the velocity dispersion somewhere
along the equator, can be qualitatively understood as follows. 
The flattening of an axisymmetric body always comes along
with some excess of energy parallel to the $(X,Y)$-plane. According to rotational
symmetry, this excess energy is equally distributed in X and Y, respectively. 
Consequently, the projected kinetic energy in directions where the object looks most
flattened is always relatively large. Only details of the 
radial distribution of the projected 
energy are not strongly constrained by the flattening alone. For example, if the flattening 
comes predominantly from near-circular orbits, then the central projected energy can be 
relatively low (circular orbits cross the central line-of-sight with zero 
line-of-sight velocity), while most energy resides in the outer regions. If radial 
equatorial orbits are responsible for the flattening, then the central projected energy 
(velocity dispersion) is relatively large, instead \citep[e.g.][]{DG93}.

The situation in triaxial bodies is different: again there is an excess of 
energy in the $(X,Y)$-plane as soon as $c<a$ and $c<b$. But this energy is no longer 
distributed equally between
X and Y. If $b<a$, then there is more energy parallel to X than parallel to Y. It follows
that the Y-projection, in which the object is most flattened (one sees the shortest and
longest axis in projection), has relatively low projected kinetic energy. 
In other words, Y-projections of triaxial systems can be highly flattened in combination
with a low specific projected energy (e.g. line-of-sight energy per mass). Now, since
the flattening in axisymmetric systems is connected to a relatively large specific 
projected energy around the equatorial plane, a mismatch of projected dispersions
somewhere around the equator is plausible (if the masses are equal).

%%%%%%%%%%%%%%%%%%%%%%%%%%%%%%%%%%%%%%%%%%
% H4 - b/a
%%%%%%%%%%%%%%%%%%%%%%%%%%%%%%%%%%%%%%%%%%
\begin{figure}
\includegraphics[width=84mm,angle=0]{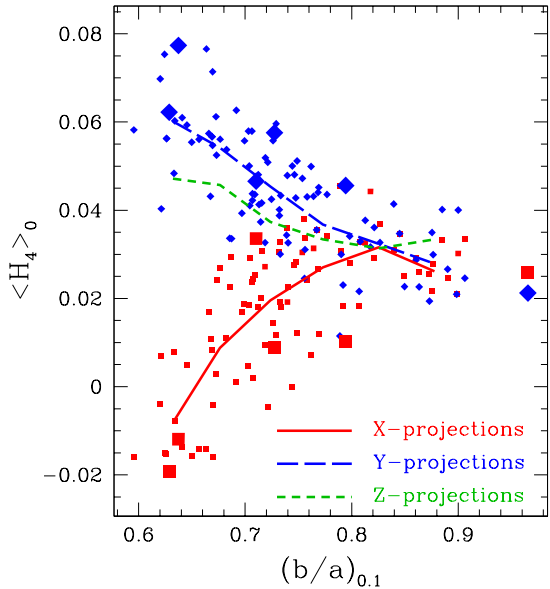}
\caption{Central $\shape$ in X and Y-projections as function of $b/a$
(10 percent most bound particles). Small symbols: merger remnants of \citet{Nab03}; 
large symbols: subset of modelled merger remnants in this work; lines trace the mean
of the distributions. For comparison also the mean of the Z-projections is
indicated.}
\label{h4ba01}
\end{figure} 

\subsection{Flattening by box orbits and $\mlfit$}
\label{sub:cenaniso}

So far the general situation. The fact that the specific dispersion of the axisymmetric
models with $\mlfit = \mlin$ is too large near the centre (cf. Fig.~\ref{losvd-match:mlin})
is most likely connected to the specific structure of the here analysed merger remnants.
They become prolate near the centre (cf. Sec.~\ref{sub:samplesel}) and this inner prolateness 
is connected to particles moving preferentially on box orbits \citep{Jes05}.
Fig.~\ref{h4ba01} shows the interplay between intrinsic central shape and projected
central kinematics. The former is quantified by the
axis-ratio $b/a$ calculated from the spatial distribution of the 10 percent most bound
particles and the latter is expressed in terms of $\shape$ (calculated inside an aperture of 
$2 \, \arcsec$ -- about $\reff/3$ or 1 kpc at the Coma distance of 
$d = 100 \, \mathrm{Mpc}$). The figure shows that high $\shape$ occur in Y-projections (low
line-of-sight velocities), 
while low or negative $\shape$ appear when viewing the prolate centres end-on (large
line-of-sight velocities). 
Differences between X and Y-projections increase with decreasing $b/a$. 
Finally, as expected for nearly prolate systems,
Z and Y-projections are almost equivalent.

\begin{figure}
\includegraphics[width=84mm,angle=0]{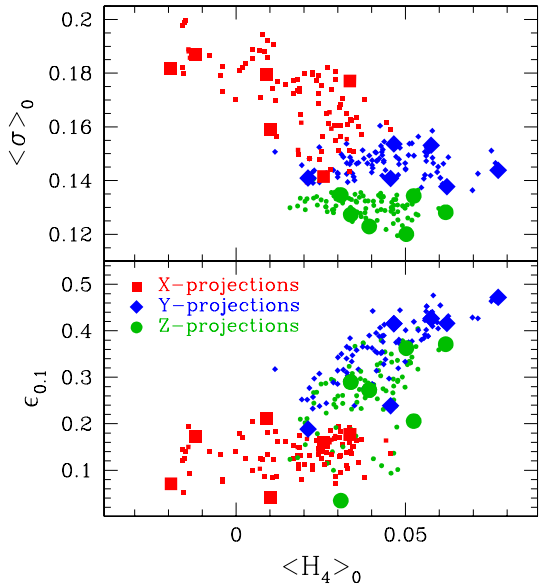}
\caption{Central $\shape$ versus $\sig$ (top) and projected ellipticity $\epsilon_{0.1}$ of
the 10 percent most-bound particles (bottom). Small symbols: all remnants; large symbols:
remnants with Schwarzschild models.}
\label{sigh4}
\end{figure}

In principle, the projected central velocity dispersion should show analogue trends.
However, it also depends on the total mass $M$ and size $R_h$ of a system and
has to be normalised before different objects can be compared.
One option is to use
\begin{equation}
\sig \equiv \frac{\sigma_0}{\sqrt{GM/R_{h}}},
\end{equation}
where $R_h$ is the half-mass radius of the light distribution and
$\sigma_0$ is the central velocity dispersion, measured in the same 
aperture as $\shape$. 
The top panel of Fig.~\ref{sigh4} shows that $\shape$ and
$\sig$ are closely correlated: high $\shape$ come along with low projected dispersions
and vice versa, as expected from our above discussion. In addition, high $\shape$ (and low
$\sig$, respectively) are connected to large projected ellipticities, as illustrated
in the lower panel of the figure. Concluding, 
inner box orbits in the merger remnants indeed cause a situation as described
in Sec.~\ref{sub:generalflat}: 
Y and Z-projections of the remnant centres have high flattening in combination with
low $\sig$ and high $\shape$. 

%%%%%%%%%%%%%%%%%%%%%%%%%%%%%%%%%%%%%%%%%%
% sigma/eps_0.1, H4
%%%%%%%%%%%%%%%%%%%%%%%%%%%%%%%%%%%%%%%%%%

\begin{figure}
\includegraphics[width=84mm,angle=0]{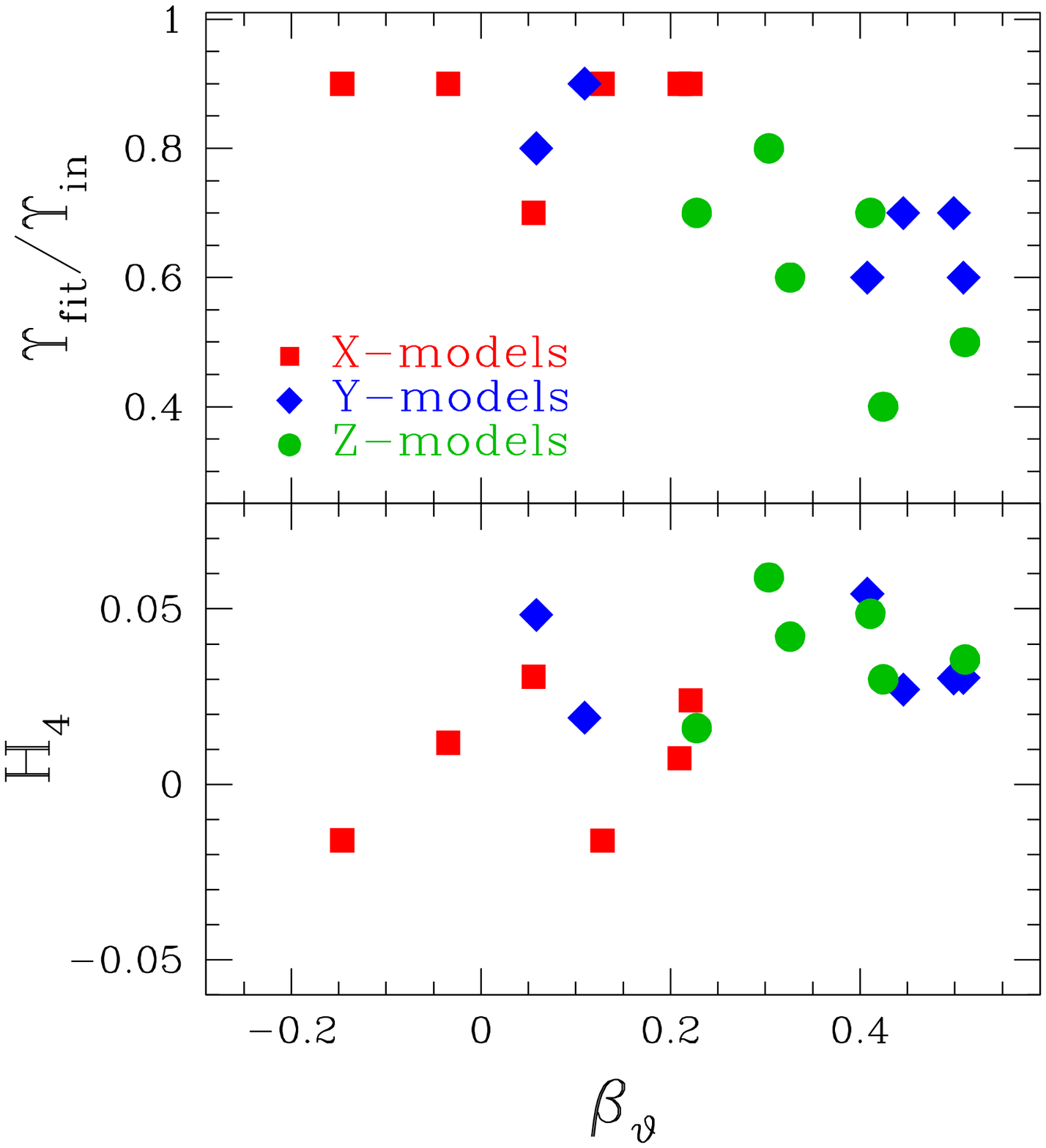}
\caption{Meridional anisotropy $\betatheta$ at $r=0.2 \, \reff$ versus
$\mlrat$ (top) and versus $H_4$ at same radius (bottom).}
\label{aniso:centre}
\end{figure}

How are the remnant centres mapped onto the axisymmetric Schwarzschild models?
Particles on box-orbits stream perpendicular to the line-of-sight of Y and Z-projections.
The respective high $H_4$ values provoke radial anisotropy in the inner 
regions of Y and Z-models (lower panel of Fig.~\ref{aniso:centre}). It has been
mentioned above that axisymmetric systems which are flattened by equatorial radial
orbits have large central velocity dispersions (e.g. \citealt{DG93}; see also the edge-on 
projection of the axisymmetric radial orbit in 
Fig.~\ref{orbits:axial}). However, the Y and Z-projections of the remnants are
characterised by the opposite: low central velocity dispersions. Consequently, Y and Z-models 
constrained to have $\Upsilon \equiv \mlin$ predict a too large central velocity dispersion 
(cf. Fig.~\ref{losvd-match:mlin}).

Assuming that the anisotropy in the models is fixed by the constraints imposed
through $H_4$, then there is only one way
to match the low central dispersion of the remnants: to reduce the inner mass. 
The inner mass, in turn, depends on only one parameter, $\Upsilon$, because both, 
mergers and models, are virtually free of dark matter in their centres.
Accordingly, if it is indeed the flattening by box orbits in the merger remnants,
e.g. the combination of high ellipticity, positive $\shape$ but low $\sig$ that causes
the low $\mlfit$ in Y and Z-models, then we would expect the following behaviour of the
models: the larger
$\shape$, the larger the radial anisotropy in the Schwarzschild fits. The larger
the anisotropy, in turn, the larger $\sig$ for $\Upsilon = \mlin$. Thus, we expect that
models which are forced to be more radially anisotropic will have a lower $\mlfit$ for
compensation. The upper panel of Fig.~\ref{aniso:centre} confirms that indeed the lowest
$\mlrat$ appear in models that are most strongly radially anisotropic near the centre
(at $0.2\,\reff$).

%%%%%%%%%%%%%%%%%%%%%%%%%%%%%%%%%%%%%%%%%%
% luminous m/l and anisotropy
%%%%%%%%%%%%%%%%%%%%%%%%%%%%%%%%%%%%%%%%%%

Hence, the systematically low $\mlfit$ of Y and Z-models in Fig.~\ref{upsekin} most
likely reflect a lack of appropriate counter-parts of box orbits in axisymmetric potentials
that can support a high flattening in combination with a low central velocity dispersion 
(per mass). 

The underestimation of $\Upsilon$ in Y and Z-models is at first glance similar to
the underestimation of the total mass in these models (cf. Sec.~\ref{subsec:massbudget}).
However, models of X-projections also deliver $\mlfit < \mlin$. 
This is different from the reconstruction of total masses,
which are instead overestimated in X-models. On the one side, it should be noted that 
-- irrespective of $\mlfit < \mlin$ -- the total mass of X-models, even close to the centre, 
is often larger than in
the remnants because of the overestimation of light in the deprojection 
(cf. Sec.~\ref{subsec:depro}). On the other hand, the discussion of the Y and Z-models has 
revealed that $\mlfit$ depends primarily on the central orbital structure and not, as the
total mass, on the global energy budget. Therefore the reconstruction of $\Upsilon$
is different from the reconstruction of the total mass. In particular, it 
most likely depends on the specific structure of the here considered merger
remnants. We cannot rule out that systems exist in which, say, $\shape$ and the inner flattening
combine in a way such that X-models would be forced to have $\mlfit>\mlin$. This needs further 
exploration of a broader sample of modeling targets. The X-projections of the here
analysed merger remnants apparently do not require $\mlfit>\mlin$ to be modelled adequately.

\subsection{The role played by dark matter}
\label{sub:roledm}
In Fig.~\ref{aniso:centre} Y and Z-models behave similarly (low $\mlfit$, large radial
anisotropy) as expected from the central box orbits in the merger remnants. In 
Fig.~\ref{upsekin}, however, Y-models are different from Z-models in their dark halos:
the underestimation of luminous mass in Y-models is compensated for by relatively
massive halos, while this is apparently not the case in Z-models. According to 
Sec.~\ref{sec:totm} higher projected kinetic energies in Y-projections are responsible
for their higher total masses (compared to
Z-models). The difference between Y and Z-models could be related to the different views
on minor-axis tubes, that dominate the outer parts of the remnants \citep{Jes05}. 
Y-projections map them edge-on such that they contribute significantly to the total
line-of-sight energy. In Z-projection they appear face-on and
most of their kinetic energy is perpendicular to the line-of-sight. The transition from
side-on inner box-orbits to edge-on outer Z-tubes may be the origin for the local maximum
in some dispersion profiles of Y-projections (TRIAX, FLAT; cf. App.~\ref{app:data}).
Z-projections lack of a similar maximum, as expected if it is caused by the
edge-on view on Z-tubes. The increase of $\sigma$ at the transition from box orbits
to Z-tubes could explain why the low $\mlfit$ of Y-models (set by the fit to the
central remnant kinematics) have 
to be compensated for with massive halos at larger radii. Z-models do not need such 
compensating halo components, because $\sigma$ drops smoothly. 

It appears that for our modeling the inclusion of dark matter in the fits has two
main effects: (1) it allows to trace the true mass structure of the modeling
targets in the outer parts (because the remnants contain dark matter); (2) in some cases
it can improve the fit to the {\it central} kinematics by allowing for 
an artificial $(M/L)$-gradient over the spatial region dominated by luminous mass.
In particular it offers to combine low central $\Upsilon$ with larger outer
$M/L$. This second issue related to dark matter in the models raises the question
whether fitting self-consistent triaxial systems with self-consistent axisymmetric
models might be different: the mass-to-light ratio $\Upsilon$ of the fits is then 
constrained by the inner as well as the outer kinematics. Accordingly, the quality of the 
overall fit could be less good. For example, in the self-consistent case $\Upsilon$ cannot
be reduced arbitrarily to match low central $\sig$ such as those arising in specific 
Y-projections, as otherwise the mass in the outer parts would be insufficient to fit the
corresponding kinematics there.
Consequently, the scatter in the relation shown in Fig.~\ref{tkin:los} could increase.
This, in turn, would affect the conclusions drawn about the projection-dependency of the mass 
budget of the axisymmetric fits, because they are partly based on this relation.

In summary, the luminous mass-to-light ratio, because it is the only parameter that
controls the central mass, depends not primarily on the total projected energy (like 
the total mass). Instead it is more sensitive to the central orbital
structure of the merger remnants.
In particular the low $\mlfit$ of Y and Z-models result from the lack of orbits that
resemble side-on views of box orbits, e.g. can support high flattening, large
transversal motions and low central line-of-sight dispersions. The radial anisotropy
induced in the axisymmetric models by the transversal streaming of box orbits in the
remnants requires a lowering of $\mlfit$ to keep the central dispersion low. In Y-models
massive halos partly compensate for the low outer luminous masses that result from
the low $\mlfit$ required for the central fit. These massive halos are needed to fit
side-on views of outer Z-tubes and the related relatively large velocity dispersions.
Z-models do not require massive halos, because when viewed face-on the outer Z-tubes
produce relatively low dispersions that do not need particularly large masses to be fit.

\section{Implications for models of real galaxies}
\label{sec:implications}
In the last two Secs.~\ref{sec:totm} and \ref{sec:lumml} we have followed 
the projection-trends in the recovery of total and luminous masses back to the
restrictions imposed by axial symmetry. We now discuss very briefly some 
implications for models of real galaxies.

\subsection{A possible bias in the reconstruction of central masses}
The dependency of $\mlfit$ on the central kinematics of the merger remnants discussed
in Sec.~\ref{sec:lumml} implies that $\mlrat$ is
connected to $\shape$ and $\sig$. Insofar as models and remnants are dominated by
luminous mass near their centres, one would expect that $\mlrat$ resembles the
ratio $\mrat$ between reconstructed and input mass, evaluated at some radius
near the centre. Consequently, $\mrat$ should be connected to
$\shape$ and $\sig$ as well.

%%%%%%%%%%%%%%%%%%%%%%%%%%%%%%%%%%%%%%%%%%
% Upsilon, sigma & H4
%%%%%%%%%%%%%%%%%%%%%%%%%%%%%%%%%%%%%%%%%%
\begin{figure}
\includegraphics[width=84mm,angle=0]{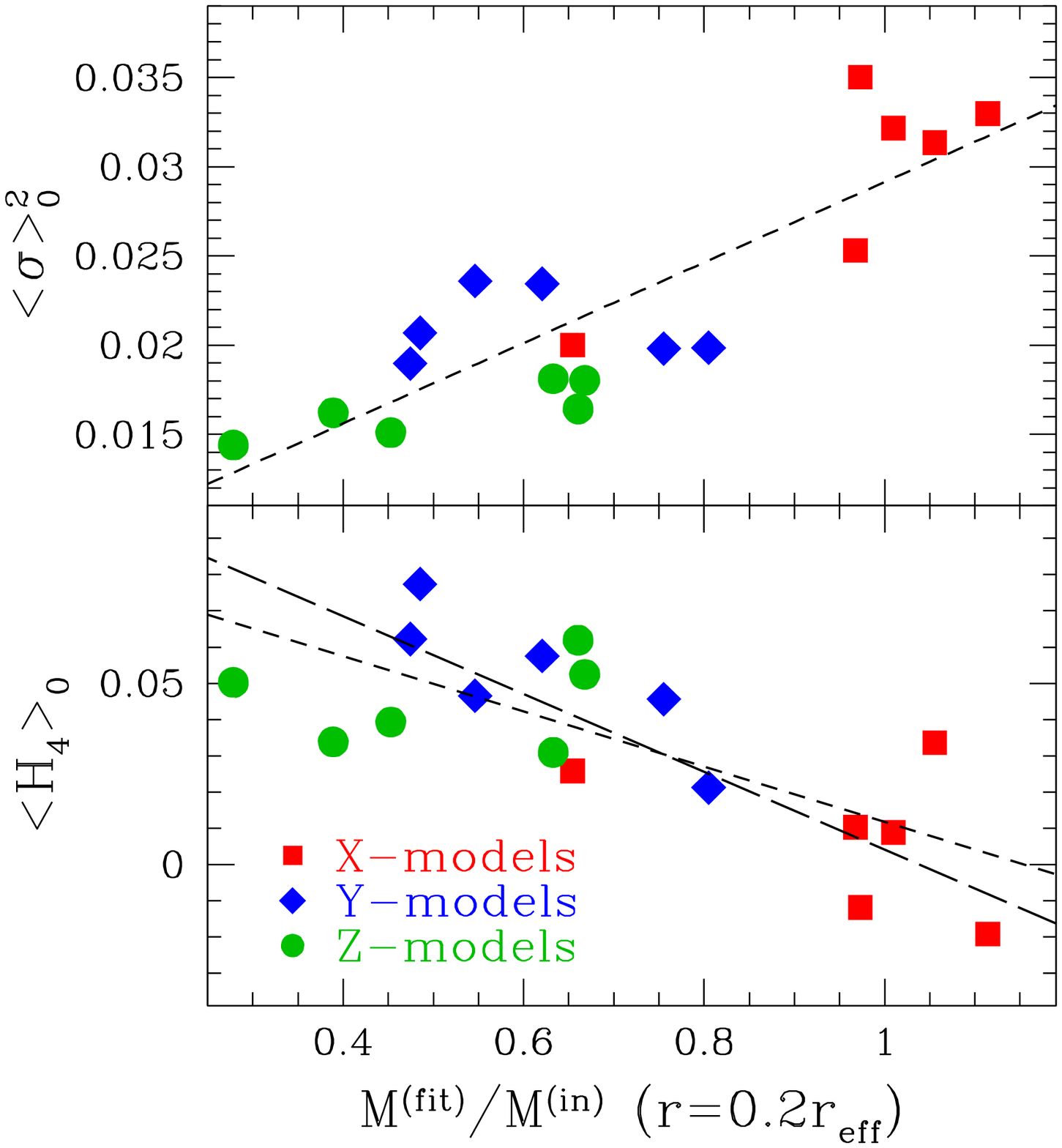}
\caption{Central velocity dispersion $\sig^2$ (top) and $\shape$ (bottom) 
versus $\mrat$ (evaluated at $0.2 \, \reff$). 
Short-dashed lines: linear fits. The long-dashed line in the bottom
panel shows the relation expected according to
the fit from the top-panel and the correlation between $\sig$ and $\shape$
shown in the top panel of Fig.~\ref{sigh4}. Symbols as labelled in the bottom panel.}
\label{mlsigh4}
\end{figure}

The relationship between $\mratcen$ (evaluated at $0.2 \, \reff$) 
and the central kinematical parameters $\shape$ and $\sig$
is shown in Fig.~\ref{mlsigh4}. Because masses globally scale
with velocities squared, we have chosen $\sig^2$ as the ordinate in the top panel.
Indeed the inner $\mratcen$ is closely correlated to $\shape$ and $\sig^2$.
Linear fits yield
\begin{equation}
\label{relation:sig0}
\left. \frac{\mfit}{\mmin} \right|_{0.2} = 44.33 \times \sig^2 - 0.29
\end{equation}
and
\begin{equation}
\label{relation:mlh4}
\left. \frac{\mfit}{\mmin} \right|_{0.2} = 1.2 - 13.1 \times \shape,
\end{equation}
respectively (short-dashed lines in Fig.~\ref{mlsigh4}). The two relations should be
almost equivalent, because according to Fig.~\ref{sigh4} 
$\shape$ and $\sig$ are closely correlated in the merger remnants. In fact, the predicted 
relation between $\mratcen$ and $\shape$ that derives from a linear fit to the top panel 
of Fig.~\ref{sigh4} in combination with equation (\ref{relation:sig0}) is consistent
with the actual relation between the mass ratio and $\shape$ that is shown in the 
lower panel of Fig.~\ref{mlsigh4}. For comparison, this relation is indicated by the 
long-dashed line.

\begin{figure}
\includegraphics[width=84mm,angle=0]{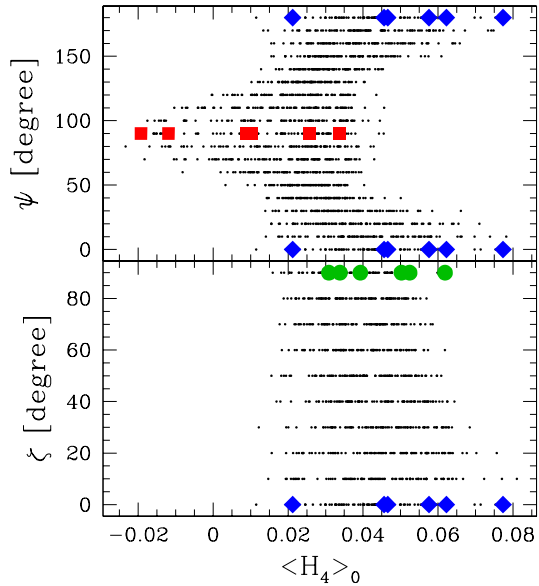}
\caption{Central $\shape$ versus viewing angle. Top: azimuth $\Psi$ in
(X,Y)-plane ($\zeta \equiv 0$); bottom: latitude $\zeta$ in (Y,Z)-plane 
($\Psi \equiv 0$); large symbols (as in 
Fig.~\ref{sigh4}): modelled merger remnants.}
\label{h4angle}
\end{figure}

Both quantities, $\sig$ as well as $\shape$ are viewing-angle dependent. The case of
the principal axes has already been discussed in Sec.~\ref{sub:cenaniso}. The full
viewing-angle dependency of $\shape$ for all merger remnants of \citet{Nab03} 
is shown in Fig.~\ref{h4angle}. 
Thereby $\Psi$ is defined as the azimuth in the $(X,Y)$-plane and $\zeta$ is the latitude. 
In accordance with our previous discussion $\shape$ peaks in Y and Z-projections
and varies little with $\zeta$ between these projections (prolateness). In the $(X,Y)$-plane
$\shape$ decreases smoothly when approaching the long-axis projection.
A similar behaviour of $H_4$ with viewing-angle has been observed in $N$-body binary
mergers of discs with massive bulges, but without dark matter \citep{Hey95}.

%%%%%%%%%%%%%%%%%%%%%%%%%%%%%%%%%%%%%%%%%%
% H4 - angle
%%%%%%%%%%%%%%%%%%%%%%%%%%%%%%%%%%%%%%%%%%

\begin{figure}
\includegraphics[width=84mm,angle=0]{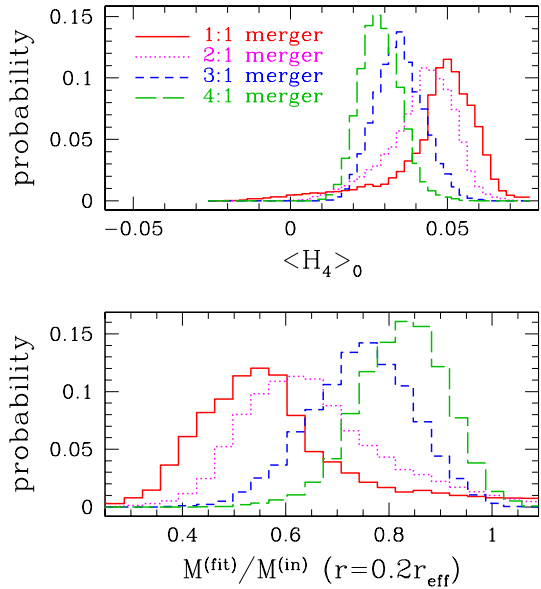}
\caption{Top: central $\shape$ from random projections of the merger remnant sample of 
\citet{Nab03} (results are separated for different progenitor mass ratios); bottom: 
resulting central $\mratcen$ according to relation (\ref{relation:mlh4}).}
\label{h4mlhist}
\end{figure}

If the here studied merger remnants would be seen at 
random projections on the sky, then the viewing-angle dependency of $\shape$ would give rise
to the frequency distribution shown in the top panel of Fig.~\ref{h4mlhist}. 
Equal mass mergers would have on average the highest $\shape$, because they are
nearly prolate and the $\shape$-distribution is dominated by the positive values around
Y and Z-projections, respectively. Towards 4:1 mergers the average $\shape$ decreases slightly. 

Assuming that relation (\ref{relation:mlh4}) holds for all viewing-angles and merger remnants,
then the frequency distribution of $\shape$ can be used to predict the distribution of 
reconstructed central $\mratcen$. The latter is plotted in the lower panel of Fig.~\ref{h4mlhist}: 
axisymmetric models of these merger remnants would be always biased towards too low central 
masses. The bias would be strongest for equal mass mergers, while masses of the more 
axisymmetric 4:1 mergers would be recovered better.

%%%%%%%%%%%%%%%%%%%%%%%%%%%%%%%%%%%%%%%%%%
% H4,M/L random projections
%%%%%%%%%%%%%%%%%%%%%%%%%%%%%%%%%%%%%%%%%%

Note, however, that for the construction of the lower panel of Fig.~\ref{h4mlhist} we have 
assumed that the relation (\ref{relation:mlh4}) holds for models of all merger remnants and 
at all viewing angles. This needs to be verified on the basis of a broader sample of models. 
Likewise it is not clear whether the connection between $\shape$ and the central 
$\mratcen$ also holds for non-axisymmetric targets of more general shapes.

\subsection{A possible bias in the reconstruction of luminous mass-to-light ratios}
\begin{figure}
\includegraphics[width=84mm,angle=0]{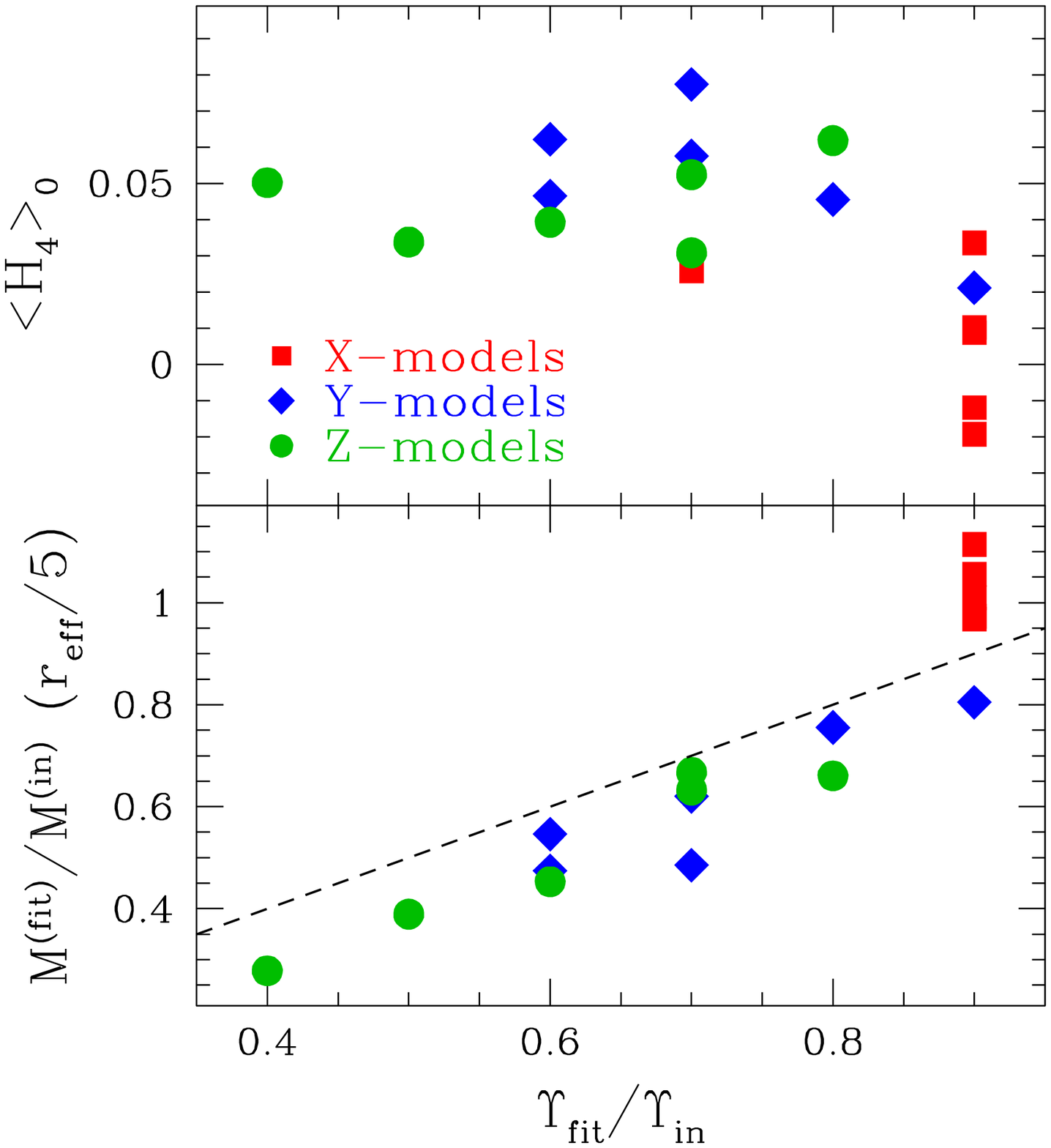}
\caption{Top: $\shape$ versus $\mlrat$; bottom: connection between the parameter ratio $\mlrat$ and the actual
ratio $\mratcen$ of integrated masses (luminous + dark) inside $0.2 \, \reff$. The
dashed line marks the identity relation.}
\label{massml}
\end{figure}

As stated above, if models and merger remnants are void of dark matter in their
central regions, then $\mratcen \approx \mlrat$. Accordingly, if the here analysed merger remnants
would be seen at random projections on the sky, then $\mlrat$ would be subject to a 
similar bias as the central $\mratcen$. However, the relation between
$\mlrat$ and $\shape$, which determines the bias, is more scatterish than the
one with the inner $\mratcen$ and it cannot be described by a straight line (upper
panel of Fig.~\ref{massml}). The reason 
is that the total mass near the centre, say inside $0.2 \, \reff$, is
determined by the product of the central luminosity with $\mlfit$ (and a
negligible amount of central dark matter). The central luminosity however, is overestimated
in X-models, but underestimated in Y and Z-models (cf. Sec.~\ref{subsec:depro}). Consequently,
X-models with a specific $\Upsilon$ have larger $\mratcen$ than Y or Z-models
with the same $\Upsilon$. This is illustrated in the lower panel of Fig.~\ref{massml}.

%%%%%%%%%%%%%%%%%%%%%%%%%%%%%%%%%%%%%%%%%%
% Ypsilon ratio versus central mass ratio
%%%%%%%%%%%%%%%%%%%%%%%%%%%%%%%%%%%%%%%%%%

Regardless of the complicated shape of the relation between $\shape$ and $\mlrat$ it
is clear that reconstructed luminous mass-to-light ratios of the here 
analysed merger remnants would be always underestimated.

\subsection{The mass recovery in models with $\Upsilon \equiv \mlin$}
\label{subsec:massmlin}
If one is interested in the recovery of luminous mass-to-light ratios, one could
use the relationship between $\shape$ and $\mlrat$ (upper panel of Fig.~\ref{massml}) 
to correct dynamically derived mass-to-light ratios (roughly) for the effects of
intrinsic non-axisymmetry. However, the accuracy of this step is limited by the
systematics of the deprojections (lower panel of Fig.~\ref{massml}).
Another way to obtain unbiased luminous mass-to-light ratios would be to apply
stellar population models to line indices, although this requires knowledge 
of the initial-stellar-mass function. In any case it is interesting to ask, whether
knowledge of the true $\Upsilon$ could help to improve other aspects of the dynamical
modelling, for example the recovery of the intrinsic anisotropy. To investigate this,
we now compare Schwarzschild fits obtained
under the condition that the luminous mass-to-light ratio is fixed to its true value, 
$\Upsilon \equiv \mlin$, with the merger remnants.
Since most best-fit X-models with variable $\Upsilon$ already have
$\mlfit \approx \mlin$ to within ten percent (cf. Sec.~\ref{subsec:stemass}),
fixing $\Upsilon \equiv \mlin$ does not change X-models significantly. We therefore
skip X-models in the remainder of this section.

Fig.~\ref{rho:y,mlin} compares the intrinsic mass densities of Y-models with
$\Upsilon \equiv \mlin$ to the merger remnants in a similar fashion as models with 
variable $\Upsilon$ were compared in Fig.~\ref{rho:y}. The figure clearly shows
an improvement in the mass recovery when $\Upsilon$ is known. This holds for both,
total as well as dark mass densities. Fig.~\ref{rho:z,mlin} covers the case of 
Z-models. Here, although again the mass recovery improves, the discrepancy between
the density profiles of models and remnants in the outer parts remains. As it has already been
discussed in Sec.~\ref{subsec:depro}, the light-profile of Z-models differs from the
remnants mainly in having the
wrong {\it slope}. Knowing just the true scaling $\mlin$ cannot remove this mismatch. 

%%%%%%%%%%%%%%%%%%%%%%%%%%%%%%%%%%%%%%%%%%
% rho: Y
%%%%%%%%%%%%%%%%%%%%%%%%%%%%%%%%%%%%%%%%%%
\begin{figure}
\includegraphics[width=84mm,angle=0]{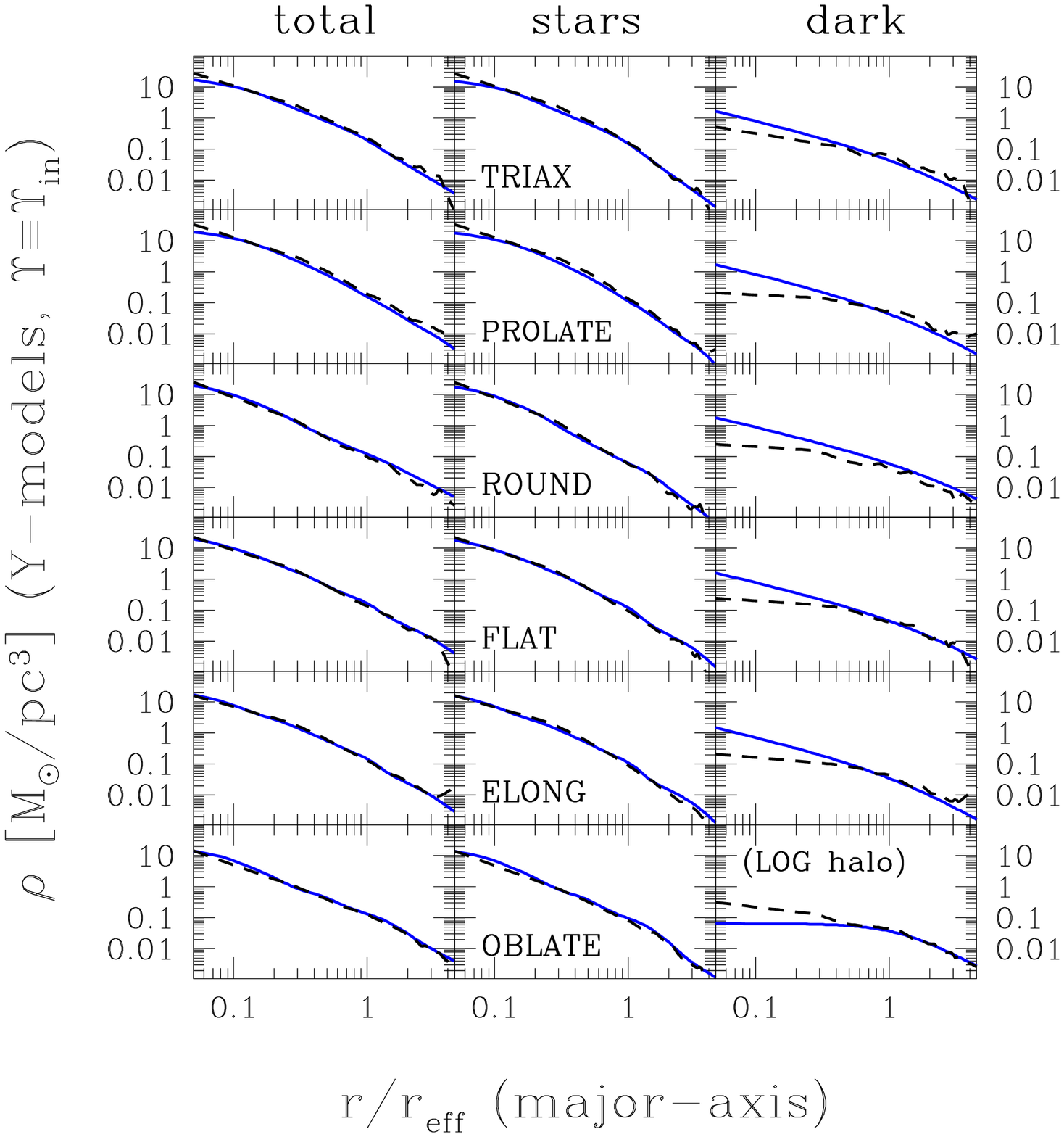}
\caption{As Fig.~\ref{rho:y}, but for models with $\Upsilon \equiv \mlin$.}
\label{rho:y,mlin}
\end{figure}

%%%%%%%%%%%%%%%%%%%%%%%%%%%%%%%%%%%%%%%%%%
% rho: Z
%%%%%%%%%%%%%%%%%%%%%%%%%%%%%%%%%%%%%%%%%%
\begin{figure}
\includegraphics[width=84mm,angle=0]{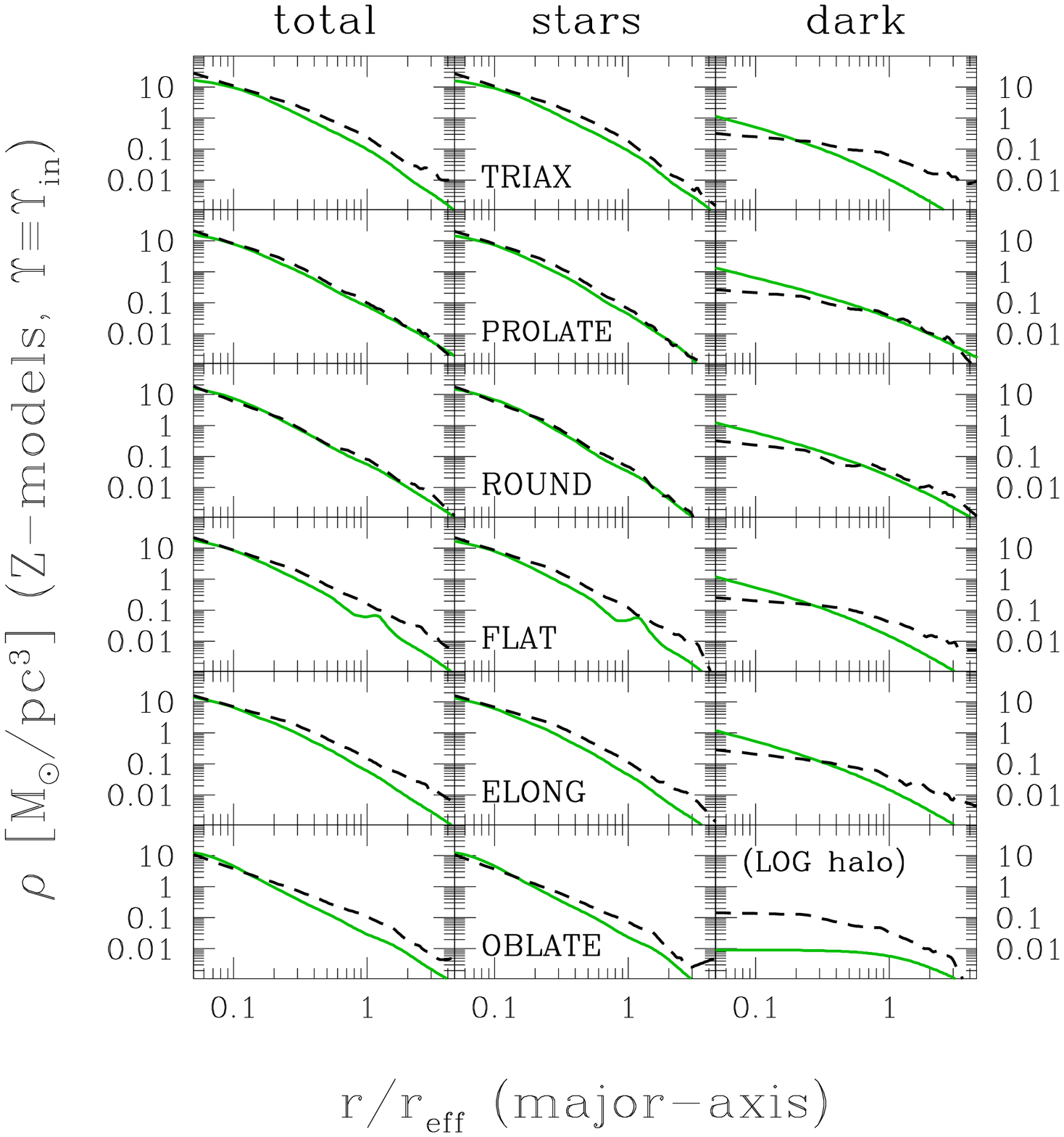}
\caption{As Fig.~\ref{rho:z}, but for models with $\Upsilon \equiv \mlin$.}
\label{rho:z,mlin}
\end{figure}

\subsection{Anisotropy in models with $\Upsilon \equiv \mlin$}
\label{subsec:intmommlin}
Figs.~\ref{beta:y,mlin} and \ref{beta:z,mlin} compare anisotropy profiles of
Y and Z-models with $\Upsilon \equiv \mlin$ to the merger remnants. The general trend
is that models with $\Upsilon \equiv \mlin$ become strongly tangentially anisotropic 
($\betatheta < 0$, $\betaphi < 0$) in the outer regions ($r \ga 0.5 \, \reff$), 
especially Z-models. Towards the centres most of the models shown
in Figs.~\ref{beta:y,mlin} and \ref{beta:z,mlin} become radially anisotropic
($\betaphi > 0$), with a local peak around $0.1 - 0.3 \, \reff$. 

All in all then, fixing $\Upsilon \equiv \mlin$ improves the reconstruction of the
intrinsic mass structure, but deviations in internal velocity moments increase.

%%%%%%%%%%%%%%%%%%%%%%%%%%%%%%%%%%%%%%%%%%
% beta: Y
%%%%%%%%%%%%%%%%%%%%%%%%%%%%%%%%%%%%%%%%%%
\begin{figure}
\includegraphics[width=84mm,angle=0]{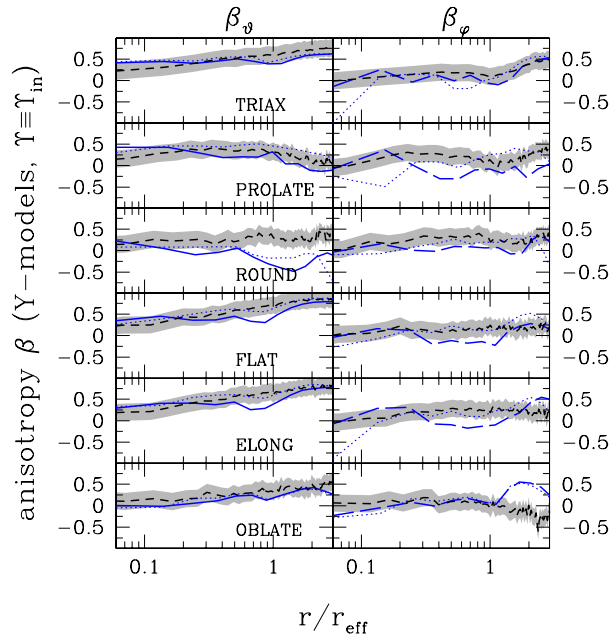}
\caption{As Fig.~\ref{beta:y}, but for models with $\Upsilon \equiv \mlin$. Profiles of the
bestfit models shown in Fig.~\ref{beta:y} are repeated for comparison (dotted).}
\label{beta:y,mlin}
\end{figure}

%%%%%%%%%%%%%%%%%%%%%%%%%%%%%%%%%%%%%%%%%%
% beta: Z
%%%%%%%%%%%%%%%%%%%%%%%%%%%%%%%%%%%%%%%%%%
\begin{figure}
\includegraphics[width=84mm,angle=0]{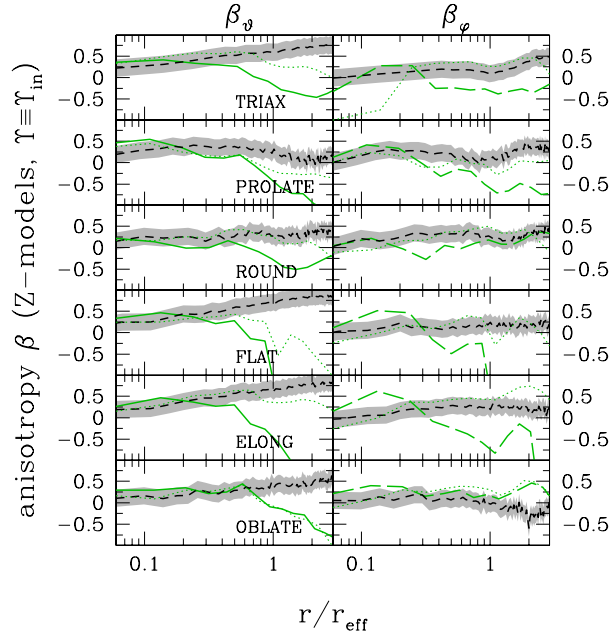}
\caption{As Fig.~\ref{beta:y,mlin}, but for Z-models.}
\label{beta:z,mlin}
\end{figure}

\section{Summary}
\label{section:conclusions}
We have modelled a set of collisionless disc-disc mergers with exactly the
same axisymmetric orbit superposition program that has been used to model
a sample of medium bright giant Coma ellipticals. The models assume a constant mass-to-light
ratio for the luminous matter and a dark halo of the NFW type \citep{nfw96}.
The remnants we model result from the collisionless 
merger of progenitor systems composed of a bulge, a disc
(both with the same mass-to-light ratio) and a dark halo. They are chosen to cover 
representatively 
the range of intrinsic shapes and dynamical structures of the \citet{Nab03} merger sample,
including the most extreme cases.

Intrinsic triaxiality causes a strong viewing-angle dependency
of projected properties of the remnants. As in axisymmetric models many viewing
angles are equivalent this must result in a corresponding dependency of the fits
on the viewing angle of the remnant. The goal of this study is to investigate this
dependency.

Some projected properties of triaxial systems, for example minor-axis rotation or 
isophotal twists, are obviously incompatible with oblate axial symmetry. Apart from 
these, we do not find any obvious mismatch between our Schwarzschild models and the 
merger remnants: residuals in the kinematic fits are smaller than 
typical observational errors.

\subsection{Remnant shapes and the mass recovery}
We find that the reconstruction of the total (luminous + dark) mass at the 
effective radius depends primarily on viewing-angle and not on the orbital structure
of the merger remnants. This is so, because the global mass budget of the axisymmetric
models is fixed by two constraints. Firstly, the match of the total line-of-sight energies 
of the luminous components of models and mergers, which is a consequence of the fit
to the projected kinematics. Secondly, although the two transversal kinetic energies 
are not constrained in the models, the ratio of both has to be the same in models and mergers,
respectively: the ratio determines the flattening and is constrained
by the fit to the luminosity density. For edge-on models the restrictions imposed through axial symmetry
then already fix the total mass budget of the Schwarzschild models. Thereby 
models of X-projections overestimate the true mass, models of Y-projections
underestimate it on about the same level and models of Z-projections have masses
even lower than those of Y-models. The exact amount of over or underestimation depends
on the intrinsic shape of the merger remnants.

In the here analysed merger remnants deviations of cumulative masses inside
$1-2 \, \reff$ are mostly below 20 percent. Extreme values of underestimations are 
larger than those of overestimations. The strongest underestimations occur among 
intrinsically very flattened, face-on systems, where the mass can be underestimated
by up to 50 percent. The underestimation is due to a wrong
inclination of the corresponding models, which arises
from the fact that most remnants appear either flattened
or show some residual rotation along the minor-axis, respectively,
when viewed along their short axis. 
Both phenomena exclude face-on ($i=0\degr$) oblate axisymmetric 
models, cause an underestimation of the luminous kinetic energy and, hence,
an underestimation of the mass inside the effective radius.

The luminous mass-to-light ratio is always underestimated, $\mlfit < \mlin$. 
Unlike the total mass, it does not derive primarily from 
the total kinetic energy of the Schwarzschild model. Although $\mlfit$
varies similarly with viewing-angle as the total mass, this variation is mediated by the central kinematics
of the merger remnants. Box orbits cause a combination of high
projected flattening and low line-of-sight motions (high central $H_4$, low central $\sigma$) 
in Y and Z-projections. Box orbits are mapped onto radial orbits in the Schwarzschild fits. 
The corresponding increase of the central velocity dispersion in the model then requires a lowering of 
the central mass to achieve a good match to both, 
the high central $H_4$ and the low $\sigma$. As the luminous
mass-to-light ratio $\mlfit$ is the only parameter that controls
the central mass, $\mlfit < \mlin$ in Y and Z-models (up to a factor of $2.5$).
Models of long-axis projections yield the best approximations to the mergers, 
in all but one case $\mlfit \approx \mlin$ to within 10 percent.

The asymmetric motion of particles, especially those on inner box orbits, 
correlates central kinematical
parameters like $\sig$ and $\shape$ with viewing-angle, a result that has already been 
found in $N$-body
simulations without dark matter \citep{Hey95}. The link between box orbit kinematics
and $\mlfit$ gives also rise to a correlation of the reconstructed central mass 
(relative to the merger mass) $\mratcen$ with $\shape$ -- and
viewing-angle.

The deficit of luminous mass in models of projections along the intermediate axis
is compensated
for by dark halos that are more massive than in the remnants. Such massive halos
are necessary to fit the relatively large outer
dispersions of these projections. Models of short-axis projections
do not have massive dark halos: their outer dispersions are low. The different
outer dispersions of projections along the intermediate and short axis,
respectively, arise due to different views on minor-axis tubes dominating
the outskirts of the remnants.

If the luminous mass-to-light ratio is fixed to its true value, then the mass recovery of
Y-models improves (on average 6 percent accuracy at $2 \, \reff$). Models of
Z-projections suffer from a wrong slope in the deprojection arising from the
mismatch in the inclination already discussed above. As the inclination is wrong, 
mere knowledge of the true $\Upsilon$ does not improve the mass reconstruction 
of Z-models much.

In general, improvements of the mass recovery are to
the expense of strong outer tangential anisotropy in the Schwarzschild models, which
weakens the match with the remnants' dynamics.

\subsection{Anisotropies in Schwarzschild models of mergers}
The viewing-angle dependency of merger projections induces a 
viewing-angle dependency of anisotropies in corresponding Schwarzschild models. For example,
X-tubes, when seen end-on, are nearly round and are represented by shell orbits in
axisymmetric models, resulting in strong tangential anisotropy in the
Schwarzschild fit. The same X-tubes
seen side-on appear radially extended and are represented by radially extended orbits
in the models, increasing their radial anisotropy. Because different orbit families give rise
to different viewing-angle dependencies (e.g. \citealt{Jes05}) the anisotropy of
Schwarzschild fits depends on projection as well as on the orbital make-up. 

Z-models of intrinsically flat, disc-like merger remnants are dominated by meridional motions
($\sigt>\sigr$). This can be explained as
an inclination effect: as stated above, the flattening of most remnants requires an
inclination $i>0\degr$ in the axisymmetric models. In this case, Z-tubes dominating the
outskirts of the remnant project to nearly round shapes and are mapped onto axisymmetric
shell orbits of the same shape -- causing a meridional anisotropy.

As found for the mass reconstruction, modeling the long-axis projection of a merger remnant
yields a better match to the intrinsic structure, while the largest deviations between
remnants and models appear among the short-axis models. In any case, deviations 
$\Delta \beta$ between Schwarzschild fits and merger remnants are 
below $\Delta \beta < 0.2$ at most radii. Towards the centre and/or towards the outer regions
deviations can be larger, however.

\subsection{Real galaxies}
We have tested our axisymmetric orbit superposition code on a sample of rather extreme
merger remnants, covering a wide range of non-axisymmetric as well as axisymmetric, but
highly flattened, dynamical systems. The aim was to probe the limits of the method.
If real ellipticals would resemble the here considered merger remnants in terms of their
orbital structure, then random viewing angles would provoke scatter in anisotropies of
axisymmetric dynamical models. Furthermore, dynamically derived stellar mass-to-light ratios
would be on average underestimated and the amount of underestimation would be
correlated with the central value of $H_4$. We plan a detailed comparison of the
here described models of merger remnants with models of Coma ellipticals in a future
publication. If models of real ellipticals show
less scatter in anisotropies or if there is no sign for a systematic underestimation
of stellar mass-to-light ratios, then this would be an indication for their intrinsic
shapes to be closer to axial symmetry than in our $N$-body merger sample.

This would have two consequences. Firstly, axisymmetric galaxies would be recovered with 
higher accuracy, even better than the here analysed merger remnants whose masses and anisotropies
at the effective radius are mostly well matched. This holds especially for flattened and rotating
systems, which are known to be close to edge-on. The case of round and non-rotating galaxies 
(even if they are axisymmetric) is more ambigous as the inclination of these systems is 
apriori unknown.

Secondly, knowing that galaxies are close to axisymmetry
provides clues about their formation: dissipation during the
formation can change elliptical galaxy properties significantly \citep[e.g.][]{Cox06,Rob06}.
Particularly, it can drive the final object towards axial symmetry 
\citep[e.g.][]{Bar96,Nab06b}. We therefore also plan to extend our
study to binary disc mergers including gas physics, star formation and feedback from
a central black hole as well as on simulations from cosmological initial conditions.

\section*{Acknowledgements}
We thank Scott Tremaine, Ortwin Gerhard and the anonymous referee for comments 
and suggestions that helped to improve the manuscript. This work was supported by DFG
Sonderforschungsbereich 375 ``Astro-Teilchenphysik''
and DFG priority program 1177.

\appendix

\section{$N$-body realisations of Schwarzschild models}
\label{app:nbodyreal}
In the following we describe the construction of $N$-body realisations of Schwarzschild 
models. Thereby we assume that all particles have the same mass, e.g. that there is
a global constant of proportionality linking the number 
density of particles in phase-space to the phase-space mass density.
To simplify the notation, we also assume $\Upsilon \equiv 1$.
The generalisation to particle masses varying from orbit to orbit or to systems with 
$\Upsilon \ne 1$ is straight forward.

The orbit library of any Schwarzschild model provides
a natural partition of phase-space into cells $\der V(i,j)$, where $\der V(i,j)$ is the phase-space
volume of the cell that orbit $i$ covers during time-step $\der t^{(j)}_i$. The probability
$p(i,j)$ to find a particle in cell $\der V(i,j)$ is proportional to
\begin{equation}
p(i,j) \propto f_i \, \der V(i,j),
\end{equation}
where $f_i$ is the phase-space density along orbit $i$. 
The latter is related to the weight $w_i$ of the corresponding
orbit and its total phase-space volume $V_i$ via $f_i=w_i/V_i$ \citep[e.g.][]{Tho04}. According
to the time-averages theorem 
\begin{equation}
\frac{\der V(i,j)}{V_i} \propto \frac{\der t_i^{(j)}}{\tau_i},
\end{equation}
where $\tau_i$ is the total integration time of orbit $i$. Hence, the probability to find a 
particle in phase-space cell $\der V(i,j)$ is 
\begin{equation}
\label{probij}
p(i,j) = w_i \frac{\der t_i^{(j)}}{\tau_i}.
\end{equation}
If the luminosity is normalised such that $\sum w_i \equiv 1$ then the
$p(i,j)$ of equation (\ref{probij}) provide a complete partition of the
interval $[0,1]$:
\begin{equation}
\label{eq:partition}
[0,1] = \bigcup_{i=1}^{N_\mathrm{orb}} \, \bigcup_{j=1}^{N_{t}(i)} {\cal P}(i,j),
\end{equation}
where $N_\mathrm{orb}$ is the number of orbits in the library, $N_{t}(i)$
is the number of time-steps in the integration of orbit $i$ and
\begin{equation}
{\cal P}(i,j) =
\left\{ 
\begin{array}{lcl}
\left[a_{ij},b_{ij}\right.)&:&  1 \le i \le N_\mathrm{orb}, \, 1 \le j \le N_{t}(i), \\
& & (i,j) \ne (N_\mathrm{orb},N_{t}(N_\mathrm{orb}))\\
\left[a_{ij},b_{ij}\right]&:&  i = N_\mathrm{orb}, \, \, j = N_{t}(N_\mathrm{orb})
\end{array}
\right.
\end{equation}
%${\cal P}(i,j) = [a_{ij},b_{ij})$ 
with
\begin{equation}
a_{ij} \equiv \sum_{k=0}^{i-1} w_k + \sum_{m=0}^{j-1} w_i \frac{\der t_i^{(m)}}{\tau_i}
\end{equation}
and
\begin{equation}
b_{ij} \equiv \sum_{k=0}^{i-1} w_k + \sum_{m=0}^{j} w_i \frac{\der t_i^{(m)}}{\tau_i}
\end{equation}
($w_0 \equiv 0$ and $\der t_i^{(0)} \equiv 0, \, \, 1 \le i \le N_\mathrm{orb}$).
The length of each subinterval equals the probability to find a particle in phase-space
cell $\der V(i,j)$:
\begin{equation}
b_{ij}-a_{ij} = w_i \frac{\der t_i^{(j)}}{\tau_i} = p(i,j).
\end{equation}

An $N$-body realisation with $N_l$ particles of, say, 
the luminous component can now be constructed by choosing 
$N_l$ random numbers $k \in [0,1]$. Each $k$ falls into one subinterval
${\cal P}(i_k,j_k)$. Accordingly, particle $k$ has to be dropped on
orbit $i_k$ during time-step $j_k$. The $N$-body realisation of the dark halo can be
constructed in a similar way, provided corresponding orbital weights are given. For the discussion
in this paper we only need an $N$-body representation of the dark halo density
profile, not of the halo kinematics. Thus, we can choose {\it any} 
distribution function for the halo that supports its density profile. The $N$-body
realisations of the halos in this work have been calculated from orbit weights
that maximise the entropy of the dark matter distribution function. Their
calculation is described in \citet{Tho07}. We use 
$N_l={\cal N}_d = 50000$ particles to sample the Schwarzschild models up to $10 \, \reff$. 

%%%%%%%%%%%%%%%%%%%%%%%%%%%%%%%%%%%%%%%%%%
% virial and halo-concentration
%%%%%%%%%%%%%%%%%%%%%%%%%%%%%%%%%%%%%%%%%%
\begin{figure}
\includegraphics[width=84mm,angle=0]{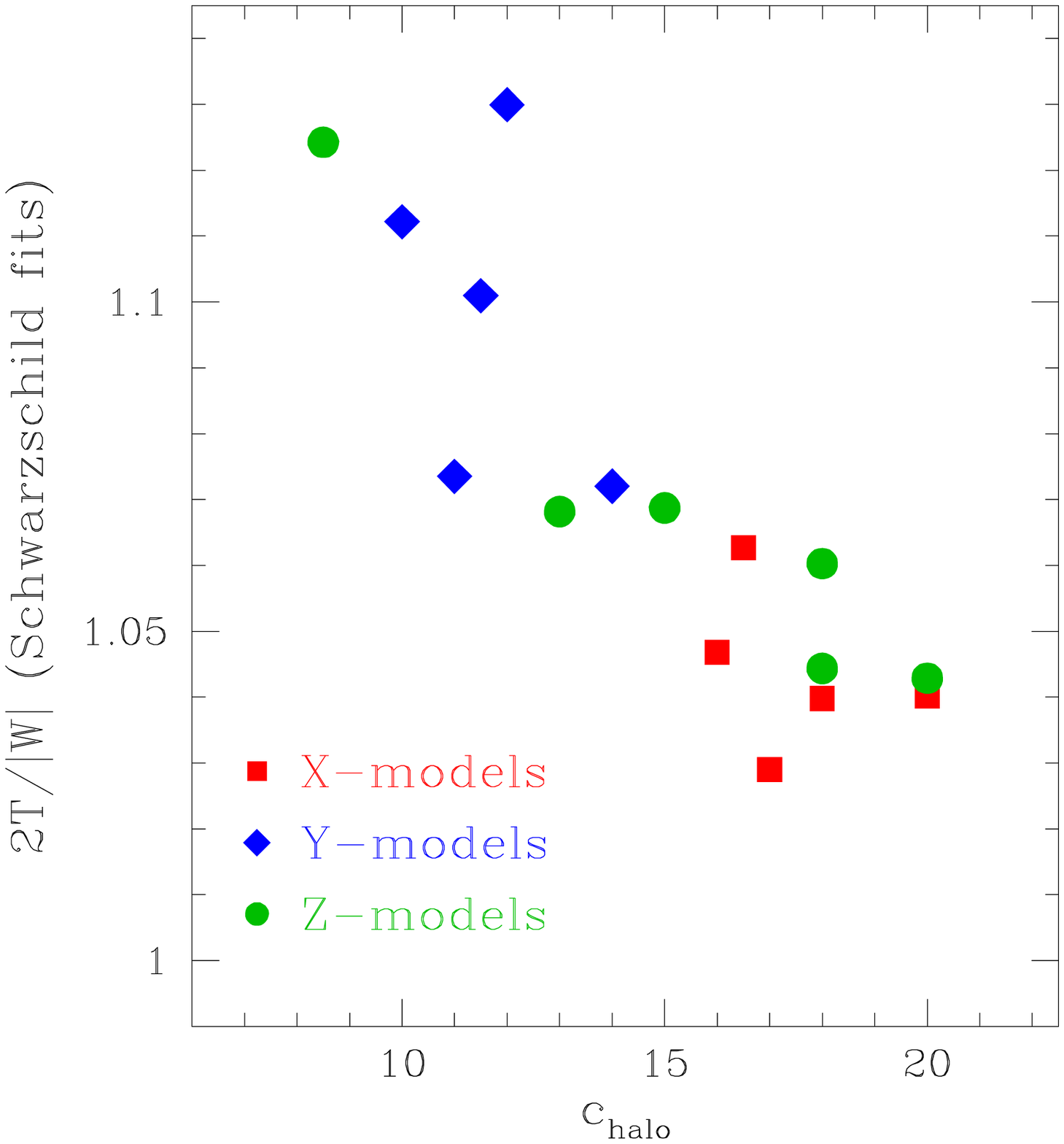}
\caption{Correlation between the virial ratio $2T/|W|$ and the halo-concentration in the
Schwarzschild models.}
\label{chalocor}
\end{figure}

Cutting the $N$-body realisation beyond $10 \, \reff$ introduces a spurious
correlation between the virial ratio $2T/|W|$ and the halo-concentration $c_\mathrm{halo}$
in the Schwarzschild fits. This is shown in Fig.~\ref{chalocor}:
the less concentrated the halo, the larger $2T/|W|$. The systematic offset $2T/|W|>1$
of the $N$-body realisation is thereby most likely caused by the fact that the
halo is undersampled if the concentratation is low ($c_\mathrm{halo} \la 13$) 
and the halo-scaling radius is correspondingly large. Then, the potential 
well is too shallow and $|W|$ is underestimated. For the deviations from virial
equilibrium discussed in Sec.~\ref{sub:chaotic} it follows that they are mostly
an artifact of the $N$-body realisation and
not due to intrinsic non-stationarity of the models.

\section{Data and models}
\label{app:data}
\begin{figure}
\includegraphics[width=84mm,angle=0]{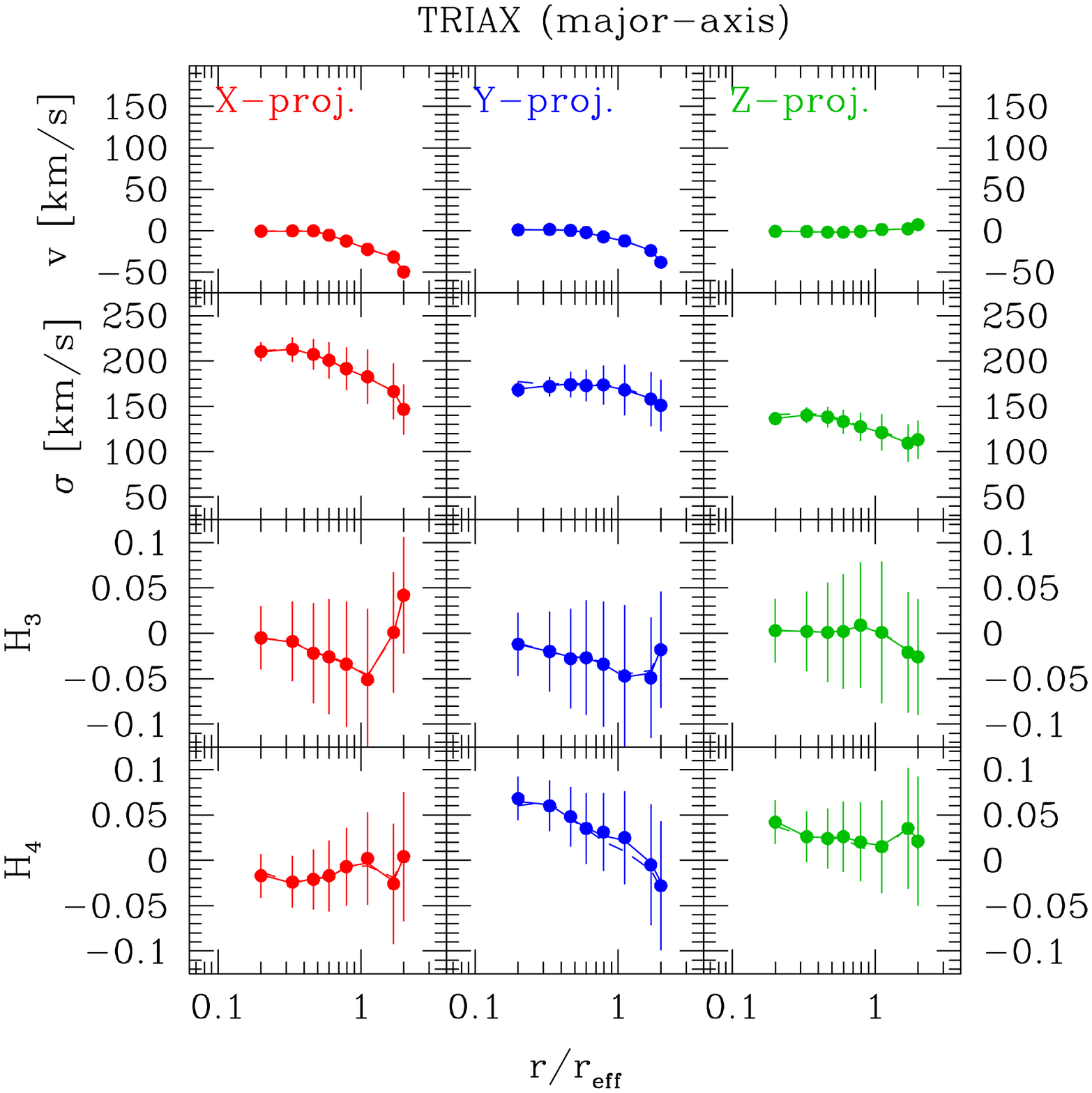}
\includegraphics[width=84mm,angle=0]{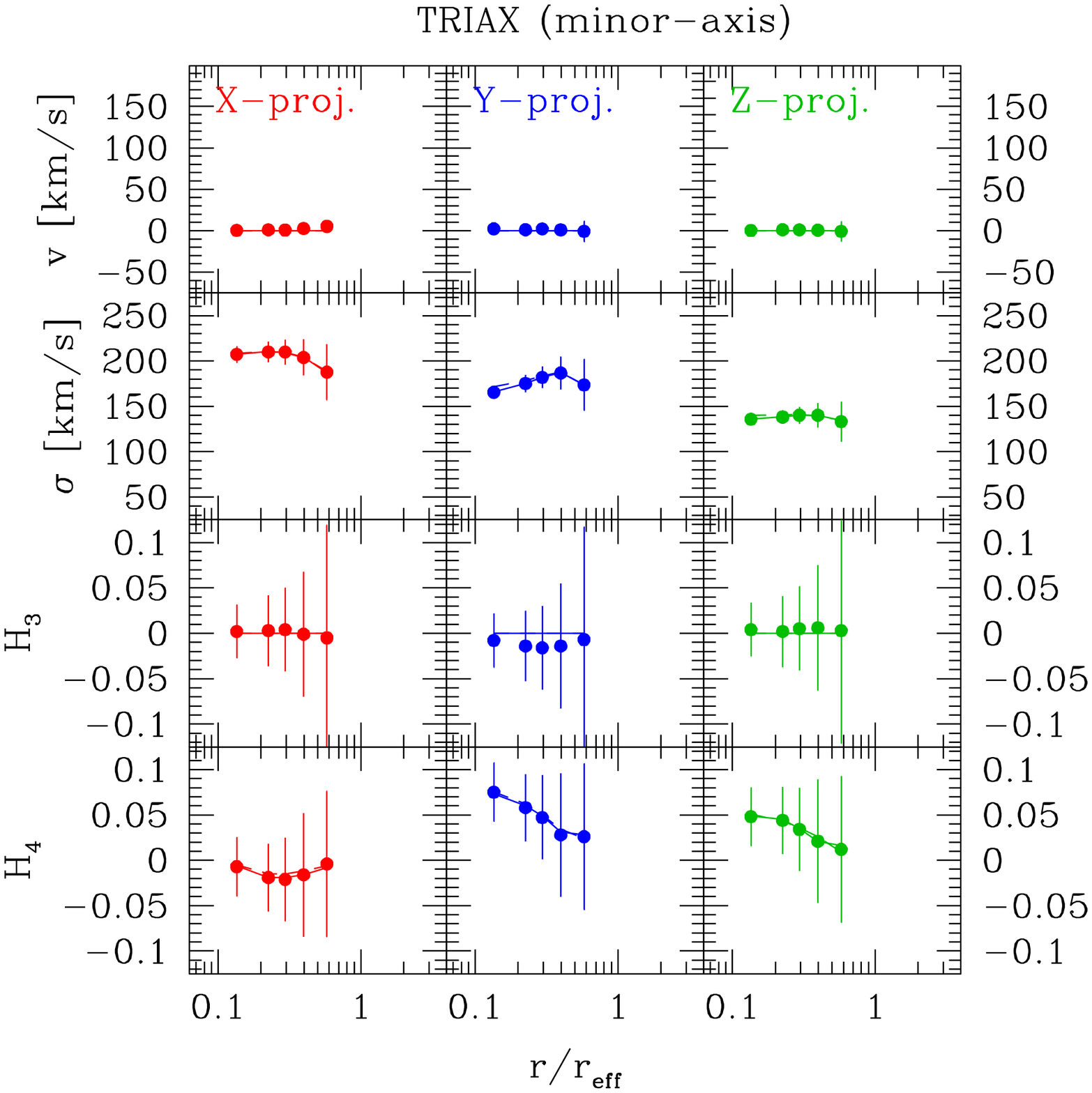}
\caption{Comparison of model kinematics (solid lines: bestfit; dashed 
lines $\mlfit \equiv \mlin$) and input data (dots with error bars) for the
TRIAX target. Top panel: major-axis; bottom panel: minor-axis. In each
panel, from top to bottom $v$, $\sigma$, $H_3$ and $H_4$; 
left: X-projection; middle: Y-projection;
right: Z-projection. The luminous mass-to-light ratios of the Schwarzschild fits
are $\mlrat=0.9,\,0.7,\,0.4$ (X,Y and Z-model, respectively).}
\label{11mcs05:kinematics}
\end{figure}

\begin{figure}
\includegraphics[width=84mm,angle=0]{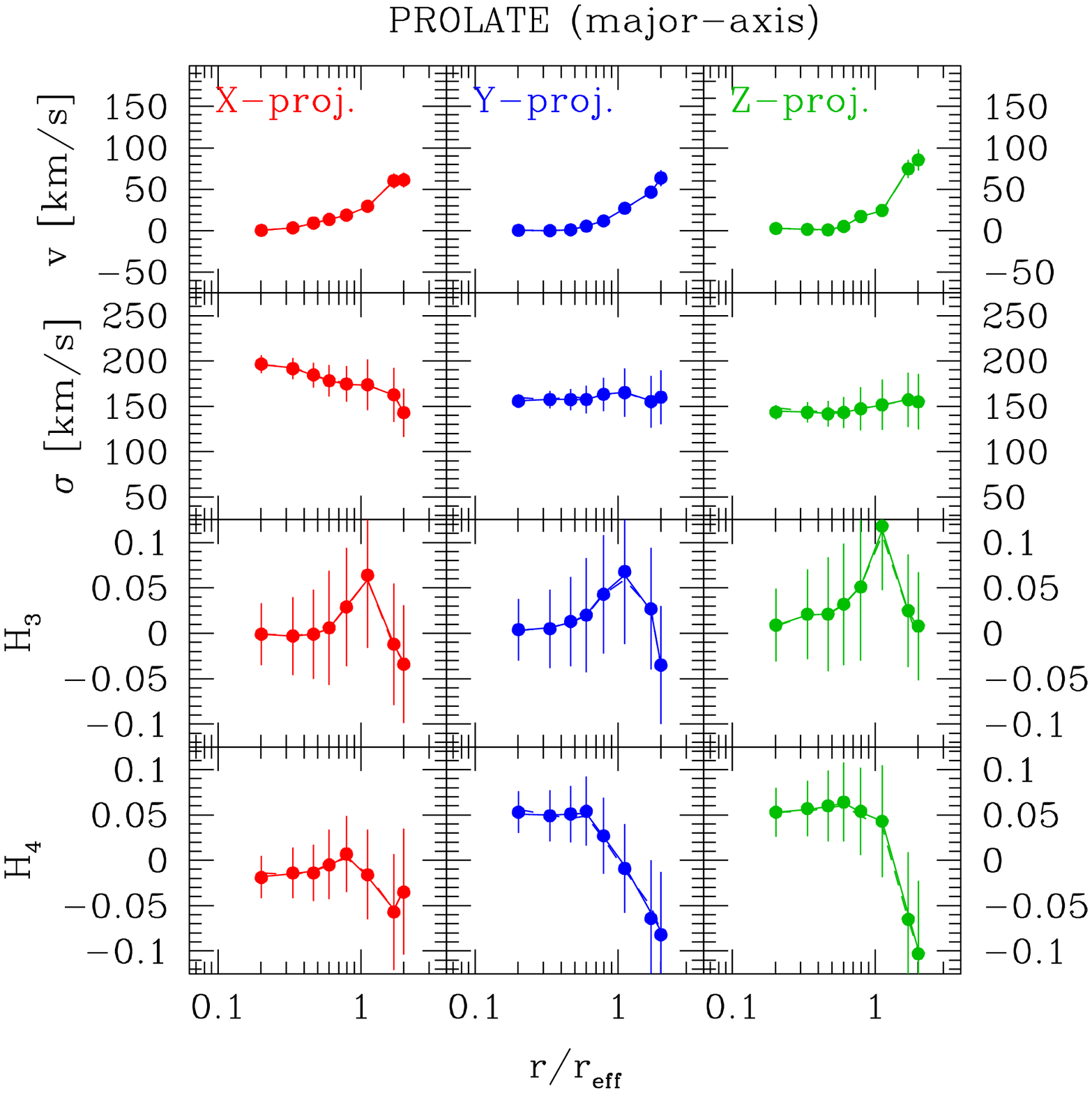}
\includegraphics[width=84mm,angle=0]{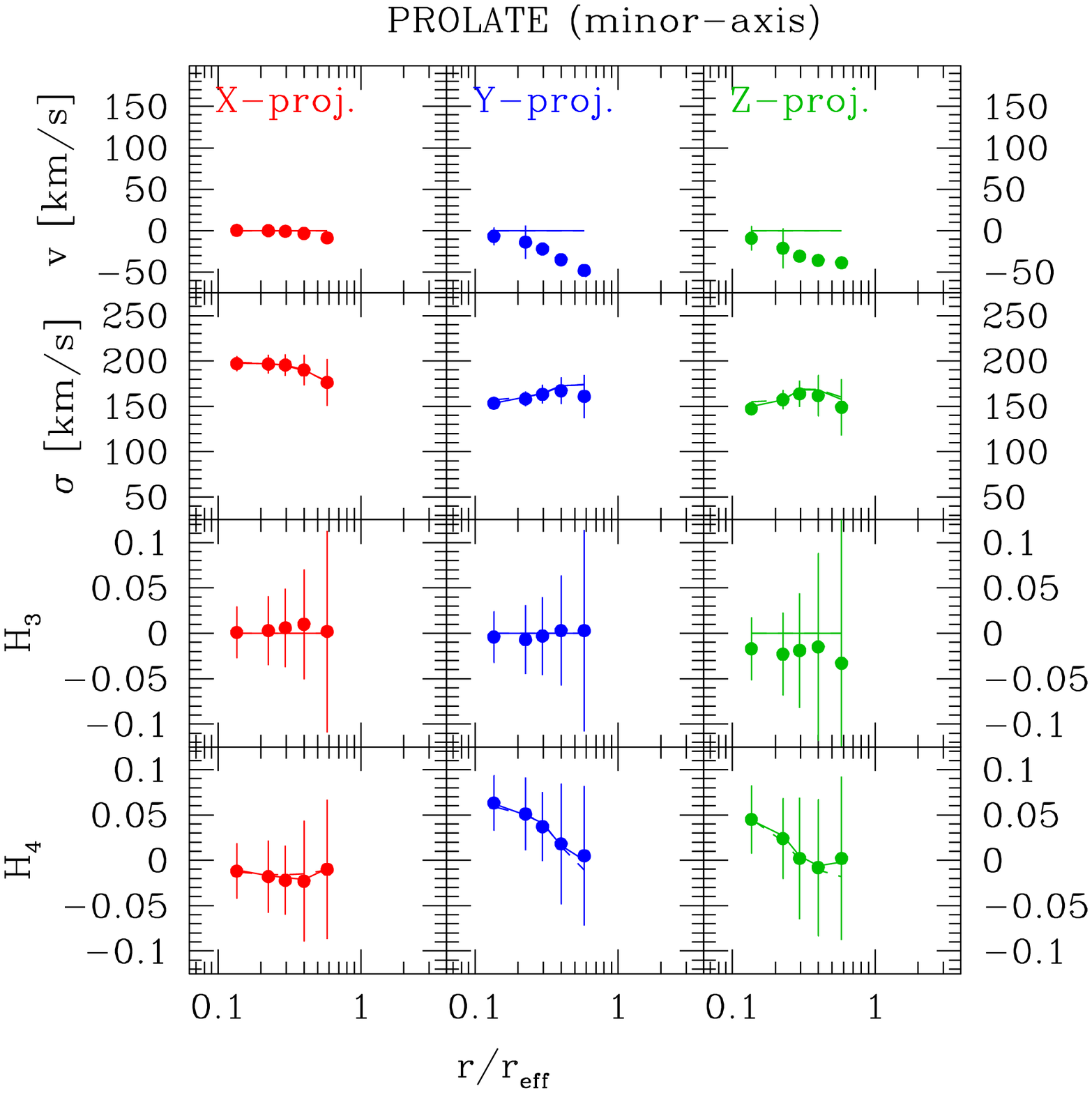}
\caption{As Fig.~\ref{11mcs05:kinematics}, but for the PROLATE target.
$\mlrat=0.9,\,0.6,\,0.8$ (X,Y and Z-model, respectively).}
\label{11mcs07:kinematics}
\end{figure}

\begin{figure}
\includegraphics[width=84mm,angle=0]{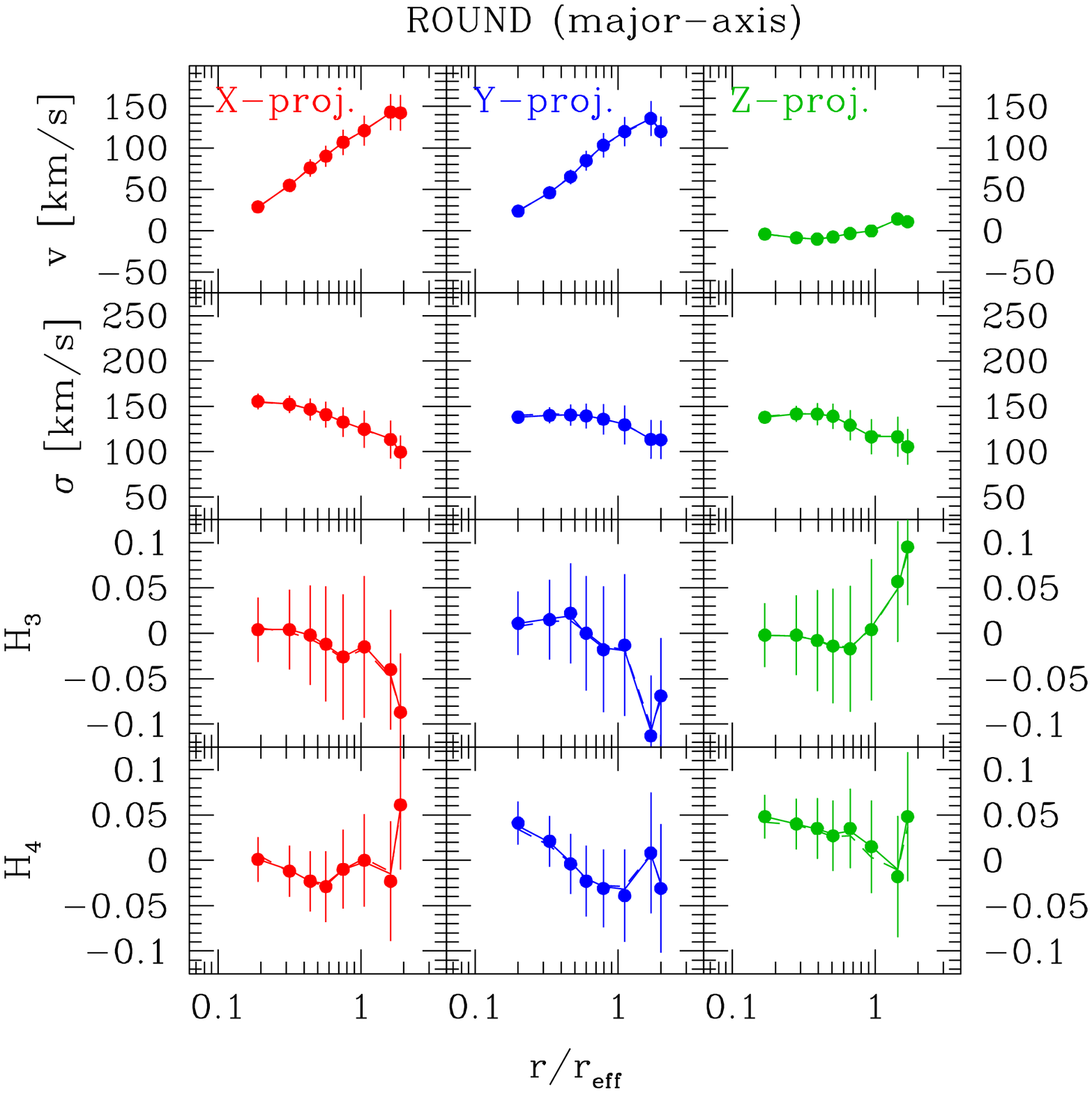}
\includegraphics[width=84mm,angle=0]{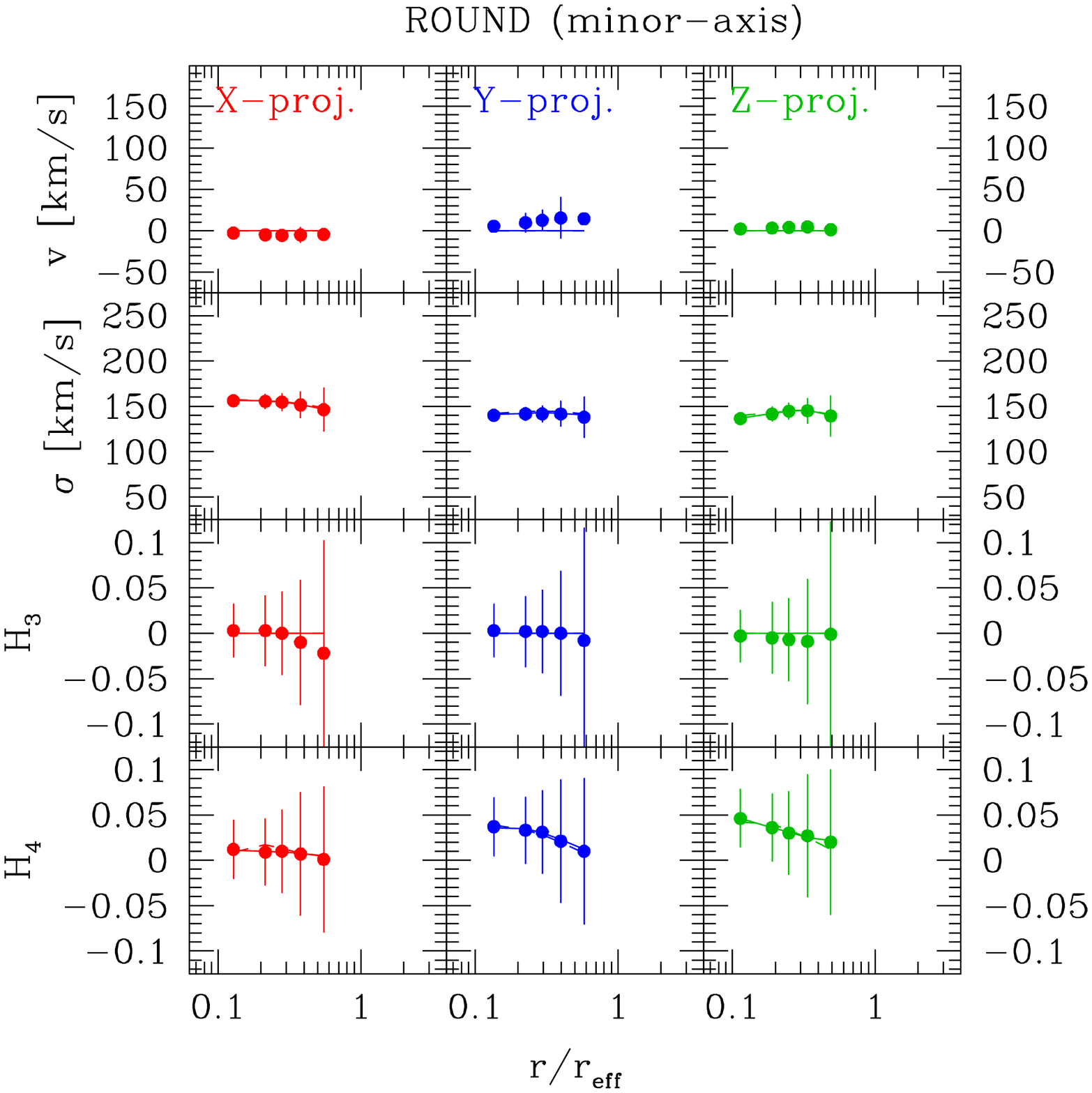}
\caption{As Fig.~\ref{11mcs05:kinematics}, but for the ROUND target.
$\mlrat=0.9,\,0.8,\,0.7$ (X,Y and Z-model, respectively).}
\label{21mcs12:kinematics}
\end{figure}

\begin{figure}
\includegraphics[width=84mm,angle=0]{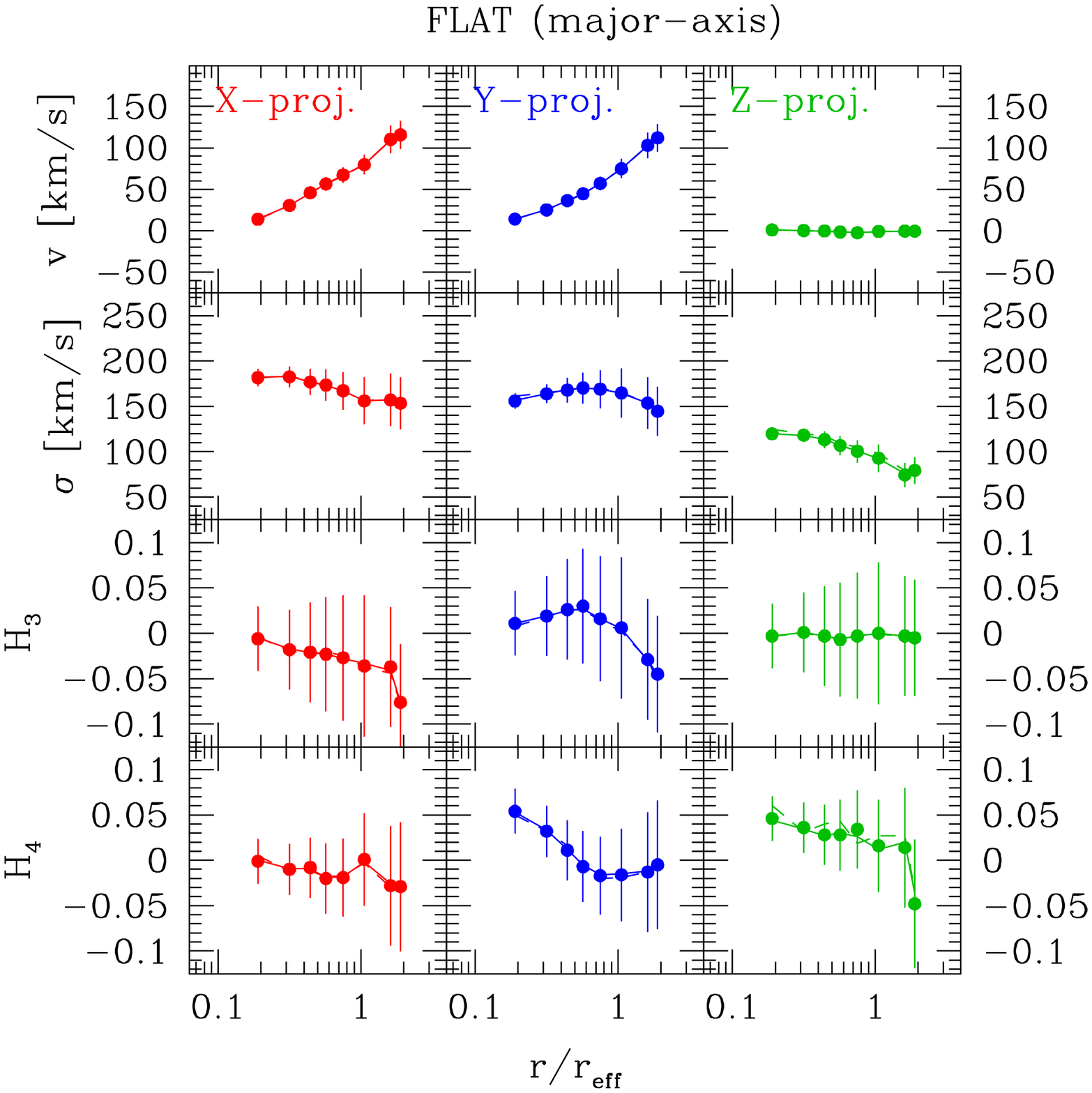}
\includegraphics[width=84mm,angle=0]{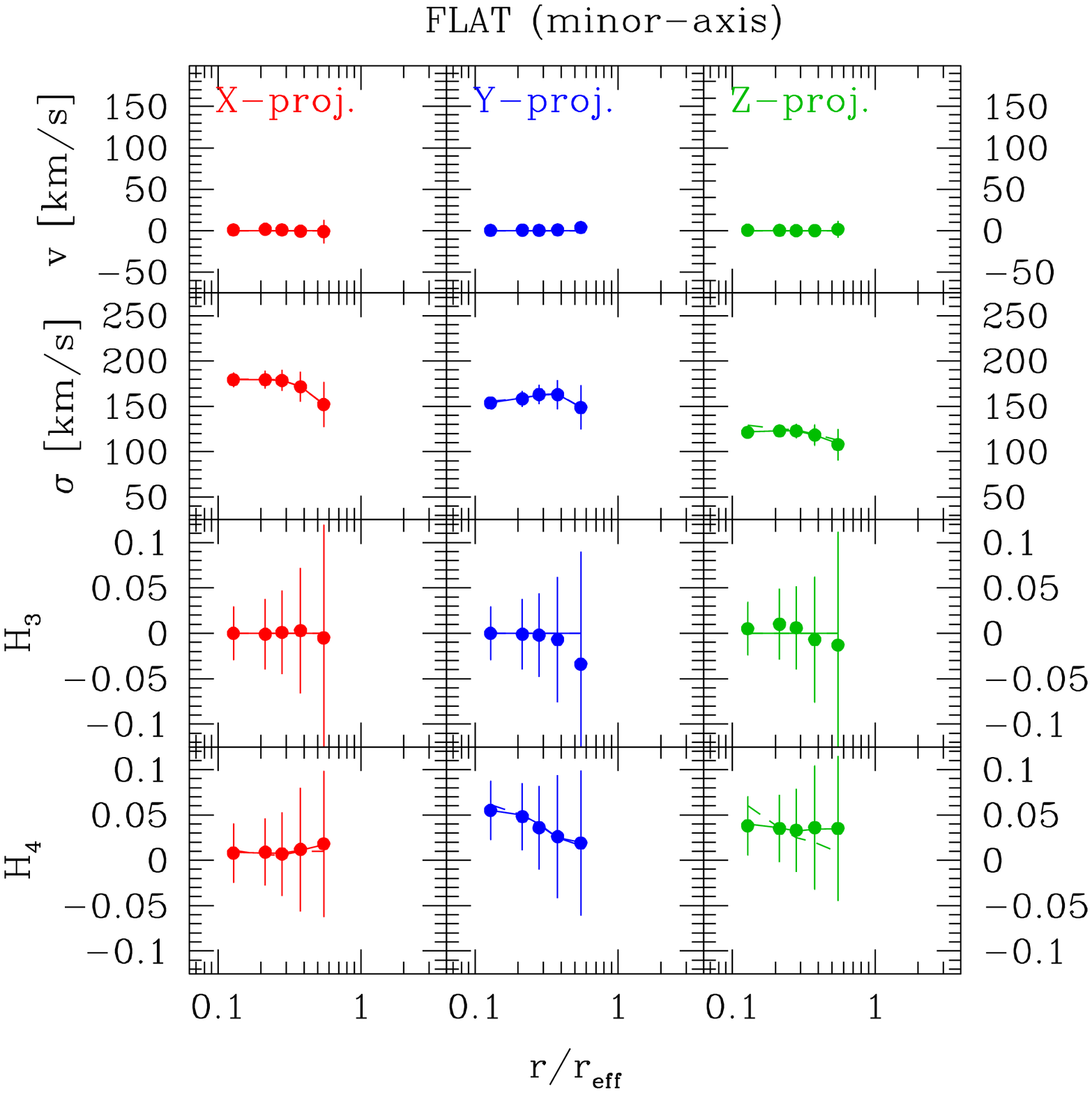}
\caption{As Fig.~\ref{11mcs05:kinematics}, but for the FLAT target.
$\mlrat=0.9,\,0.7,\,0.6$ (X,Y and Z-model, respectively).}
\label{21mcs17:kinematics}
\end{figure}

\begin{figure}
\includegraphics[width=84mm,angle=0]{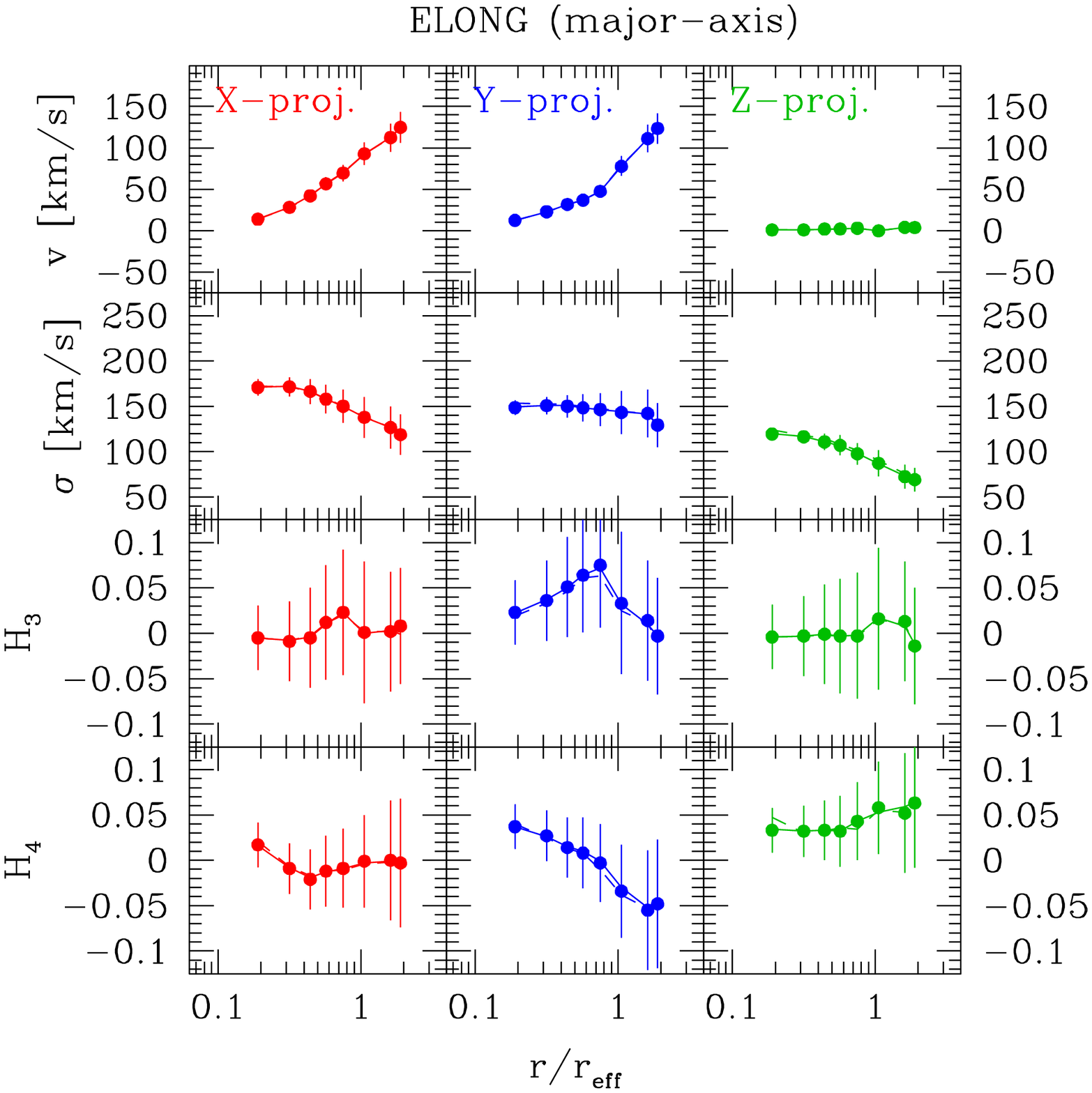}
\includegraphics[width=84mm,angle=0]{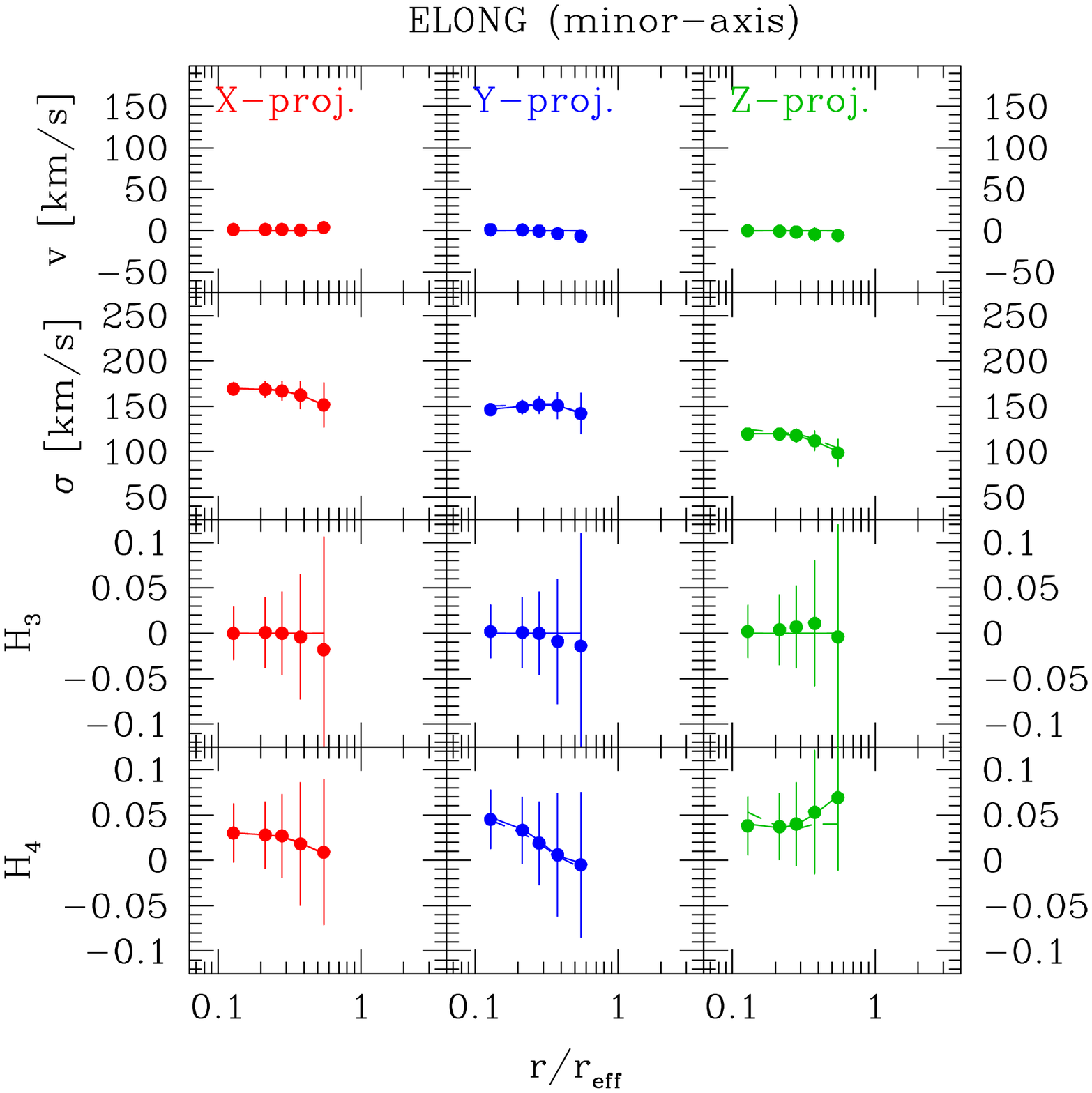}
\caption{As Fig.~\ref{11mcs05:kinematics}, but for the ELONG target.
$\mlrat=0.9,\,0.6,\,0.5$ (X,Y and Z-model, respectively).}
\label{31mcs29:kinematics}
\end{figure}

\begin{figure}
\includegraphics[width=84mm,angle=0]{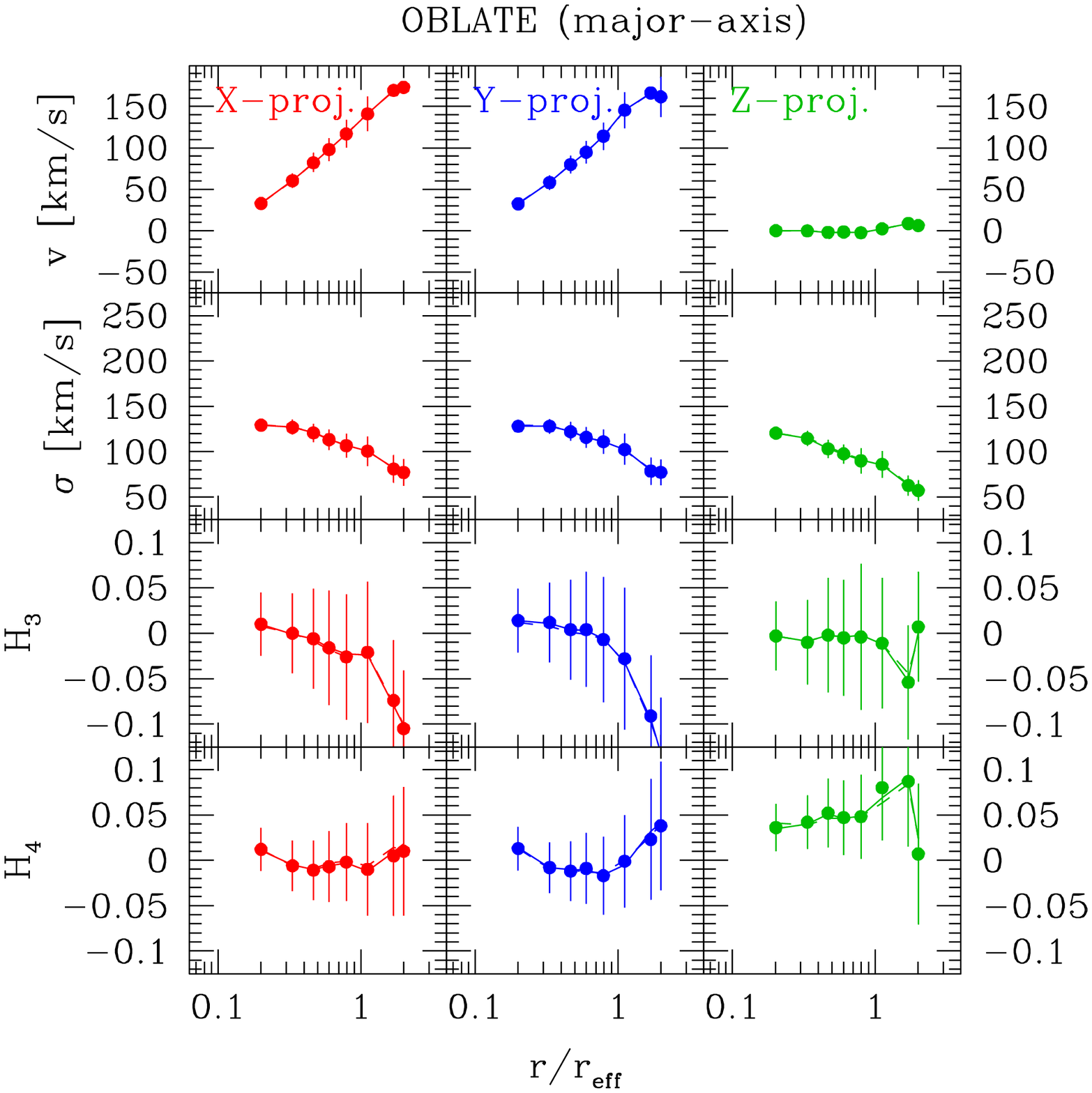}
\includegraphics[width=84mm,angle=0]{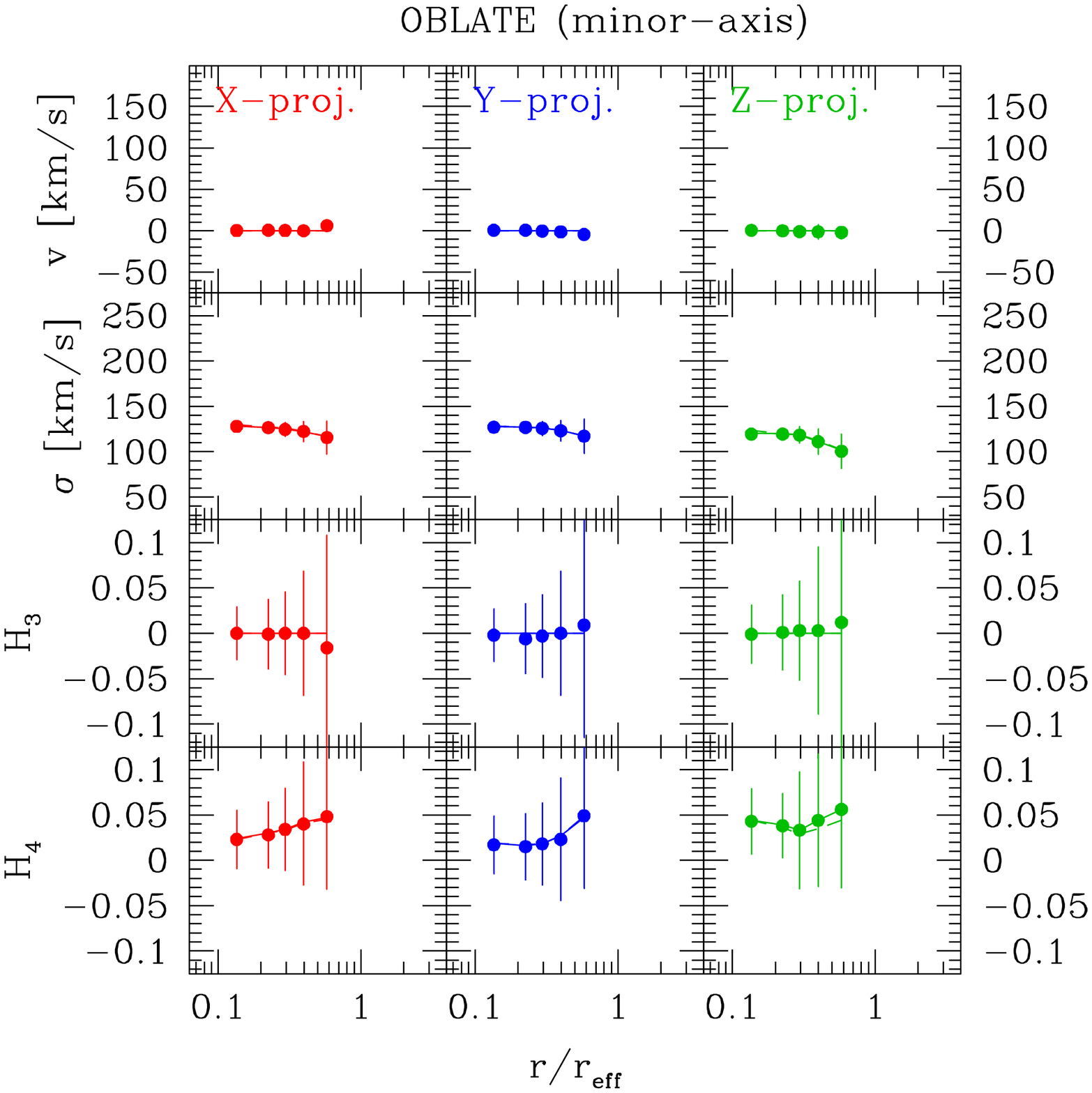}
\caption{As Fig.~\ref{11mcs05:kinematics}, but for the OBLATE target.
$\mlrat=0.7,\,0.9,\,0.7$ (X,Y and Z-model, respectively).}
\label{41mcs11:kinematics}
\end{figure}

Figs.~\ref{11mcs05:kinematics} - \ref{41mcs11:kinematics} survey the 
kinematics of the individual targets (dots with error bars) and the model fits
(lines). For each target the bestfit model (solid; best-fit luminous mass-to-light ratios
$\mlrat$ are given in the captions and summarised in Tab.~\ref{tab:mlvals}) 
as well as the bestfit with
$\Upsilon \equiv \mlin$ (dashed) are shown. Tab.~\ref{tab:mlvals} also lists the
accuracy of reconstructed total masses at $\reff$.

\begin{table}
\begin{center}
\begin{tabular}{lcccccc}
remnant & \multicolumn{3}{c}{$\mlrat$} & \multicolumn{3}{c}{$\Delta M$ [\%]}\\
& X & Y & Z & X & Y & Z\\
\hline
(1) & (2) & (3)    & (4)          &  (5)  & (6) & (7)\\
\hline 
TRIAX   &  0.9   & 0.7    &  0.4 &  $-2.3$   & $-14.9$   &  $-45.8$ \\
PROLATE &  0.9   & 0.6    &  0.8 &  $3.7$   & $-18.2$    &  $-3.8$ \\
ROUND   &  0.9   & 0.8    &  0.7 &  $9.3$   & $10.0$    &  $-13.1$ \\
FLAT    &  0.9   & 0.7    &  0.6 &  $5.1$   & $-4.4$    &  $-45.6$ \\
ELONG   &  0.9   & 0.7    &  0.5 &  $10.0$   & $-9.4$    &  $-47.8$ \\
OBLATE  &  0.7   & 0.9    &  0.7 &  $2.8$   & $0.4$    &  $-34.5$ \\
\hline 
\end{tabular}
\caption{Accuracy of reconstructed masses. (1) Merger remnant; (2-4) reconstructed 
$\mlrat$ for X, Y and Z-models, respectively; (5-7) fractional error 
$\Delta M \equiv (M\fit-M\inp)/M\inp$ of reconstructed 
(luminous + dark) mass inside $\reff$ for X, Y and Z-models, respectively. 
\label{tab:mlvals}}
\end{center}
\end{table}

%%%%%%%%%%%%%%%%%%%%%%%%%%%%%%%%%%%%%%%%%%
% 11mcs05
%%%%%%%%%%%%%%%%%%%%%%%%%%%%%%%%%%%%%%%%%%

%%%%%%%%%%%%%%%%%%%%%%%%%%%%%%%%%%%%%%%%%%
% 11mcs07
%%%%%%%%%%%%%%%%%%%%%%%%%%%%%%%%%%%%%%%%%%

%%%%%%%%%%%%%%%%%%%%%%%%%%%%%%%%%%%%%%%%%%
% 21mcs12
%%%%%%%%%%%%%%%%%%%%%%%%%%%%%%%%%%%%%%%%%%

%%%%%%%%%%%%%%%%%%%%%%%%%%%%%%%%%%%%%%%%%%
% 21mcs17
%%%%%%%%%%%%%%%%%%%%%%%%%%%%%%%%%%%%%%%%%%

%%%%%%%%%%%%%%%%%%%%%%%%%%%%%%%%%%%%%%%%%%
% 31mcs29
%%%%%%%%%%%%%%%%%%%%%%%%%%%%%%%%%%%%%%%%%%

%%%%%%%%%%%%%%%%%%%%%%%%%%%%%%%%%%%%%%%%%%
% 41mcs11
%%%%%%%%%%%%%%%%%%%%%%%%%%%%%%%%%%%%%%%%%%

\label{lastpage}
\end{document}